\documentclass{aa}  

\usepackage{graphicx}
\usepackage{epstopdf}

\usepackage{txfonts}

\usepackage{subfigure}
\usepackage{amsmath}
\usepackage{amssymb}
\usepackage{pdflscape}
\usepackage{color}

\begin{document}

   \title{Multi-phase feedback processes in the Sy2 galaxy NGC\,5643}

   \author{I. Garc\'ia-Bernete\inst{1}\fnmsep\thanks{E-mail: igbernete@gmail.com},
          A. Alonso-Herrero\inst{2}, S. Garc\'ia-Burillo\inst{3}, M. Pereira-Santaella\inst{4}, B. Garc\'ia-Lorenzo\inst{5,6}, F. J. Carrera\inst{7}, D. Rigopoulou\inst{1}, C. Ramos Almeida\inst{5,6}, M. Villar Mart\'in\inst{4}, O. Gonz\'alez-Mart\'in\inst{8}, E. K. S. Hicks\inst{9}, A. Labiano\inst{2}, C. Ricci\inst{10,11,12} \and S. Mateos$^7$\\}
   \institute{$^1$Department of Physics, University of Oxford, Oxford OX1 3RH, UK \\
$^2$Centro de Astrobiolog\'ia, CSIC-INTA, ESAC Campus, E-28692, Villanueva de la Ca\~nada, Madrid, Spain\\
$^3$Observatorio Astron\'omico Nacional (OAN-IGN)-Observatorio de Madrid, Alfonso XII, 3, 28014 Madrid, Spain\\
$^4$Centro de Astrobiolog\'ia, CSIC-INTA, Carretera de Torrej\'on a Ajalvir, E-28880 Torrej\'on de Ardoz, Madrid, Spain\\
$^5$Instituto de Astrof\'isica de Canarias, Calle v\'ia L\'actea, s/n, E-38205 La Laguna, Tenerife, Spain\\
$^6$Departamento de Astrof\'isica, Universidad de La Laguna, E-38205 La Laguna, Tenerife, Spain\\
$^7$Instituto de F\'isica de Cantabria (CSIC-UC), Avenida de los Castros, E39005 Santander, Spain\\
$^8$Instituto de Radioastronom\'ia y Astrof\'isica (IRyA-UNAM), 3-72 (Xangari), 8701, Morelia, Mexico\\
$^9$Department of Physics \& Astronomy, University of Alaska Anchorage, AK 99508-4664, USA\\
$^{10}$N\'ucleo de Astronom\'ia de la Facultad de Ingenier\'ia, Universidad Diego Portales, Av. Ej\'ercito Libertador 441, Santiago, Chile\\
$^{11}$Kavli Institute for Astronomy and Astrophysics, Peking University, Beijing 100871, China\\
$^{12}$George Mason University, Department of Physics \& Astronomy, MS 3F3, 4400 University Drive, Fairfax, VA 22030, USA\\}
%\titlerunning{Multi-phase feedback processes of NGC\,5643}
\authorrunning{Garc\'ia-Bernete et al.}

   \date{}

% \abstract{}{}{}{}{} 
% 5 {} token are mandatory

  \abstract
   {We study the multi-phase feedback processes in the central $\sim$3 kpc of the barred Seyfert 2
galaxy NGC\,5643. We use observations of the cold molecular gas (ALMA CO(2-1) transition) and ionized gas (MUSE IFU optical emission lines). We study different regions along the outflow zone which extends out to $\sim$2.3 kpc in the same direction (east-west) as the radio jet, as well as nuclear/circumnuclear regions in the host galaxy disk. The CO(2-1) line profiles of regions in the outflow and spiral arms show two or more different velocity components, one 
associated with the host galaxy rotation and the others with out/inflowing material. In the outflow region, the [O\,{\sc iii}]$\lambda$5007$\AA$ emission lines have two or more components: the narrow component traces rotation of the gas in the disk and the others are related to the ionized outflow. The deprojected outflowing velocities of the cold molecular gas (median $V_{\rm central}\sim$189~km s$^{-1}$) are generally lower than those of the outflowing ionized gas, which reach deprojected velocities of up to 750~km~s$^{-1}$ close to the AGN, and their spatial profiles follow those of the ionized phase. This suggests that the outflowing molecular gas in the galaxy disk is being entrained by the AGN wind. We derive molecular and ionized outflow masses of $\sim$5.2$\times$10$^7$~M$_{\odot}$ ($\alpha_{\rm CO}^{\rm Galactic}$) and 8.5$\times$10$^4$~M$_{\odot}$ and molecular and ionized outflow mass rates of $\sim$51~M$_{\odot}\,{\rm yr}^{-1}$ ($\alpha_{\rm CO}^{\rm Galactic}$) and 0.14~M$_{\odot} \,{\rm yr}^{-1}$, respectively. Therefore, the molecular phase dominates the outflow mass and outflow mass rate, while the outflow kinetic power and momentum are similar in both phases. However, the wind momentum load ($\dot{P}_{out}/\dot{P}_{AGN}$) for the molecular and ionized outflow phases are $\sim$27--5 ($\alpha_{\rm CO}^{\rm Galactic}$ and $\alpha_{\rm CO}^{\rm ULIRGs}$) and $<1$, which suggests that the molecular phase is not momentum conserving while the ionized one most certainly is. The molecular gas content {{($M_{\rm east}\sim$1.5$\times$10$^7$~M$_\sun$; $\alpha_{\rm CO}^{\rm Galactic}$)}} of the eastern spiral arm  is approximately 50-70\% of the content of the western one. We interpret this as destruction/clearing of the molecular gas produced by the AGN  wind impacting in the eastern side of the host galaxy (negative feedback process). The increase of the molecular phase momentum implies that part of the kinetic energy from the AGN wind is transmitted to the molecular outflow. This suggest that in Seyfert-like AGN such as NGC~5643, the radiative/quasar and the kinetic/radio AGN feedback modes coexist and may shape the host galaxies even at kpc-scales via both positive and (mild) negative feedback.}

      
   \maketitle

%________________________________________________________________

\section{Introduction}

The impact of the energy released by Active Galactic Nuclei (AGN) in the form of radiation and/or mechanical outflows in the host galaxy interstellar medium has been proposed as a key mechanism responsible for regulating star formation in galaxies. In cosmological simulations, AGN feedback is needed to reproduce the observed number of massive galaxies via the quenching of star formation 
(e.g. \citealt{Bower06,Croton06,Bongiorno16}), although other less significant heating sources such as supernovae could also play a role (e.g. \citealt{Silk12}). Additionally, recent studies found observational evidence of positive AGN feedback \citep[e.g.,][]{Klamer04, Norris09, Cresci15, Maiolino17,Shin19} which favours compression and subsequent collapse in molecular gas leading to an enhanced star formation rate.

\begin{figure*}
\centering
\includegraphics[width=12.5cm, angle=90]{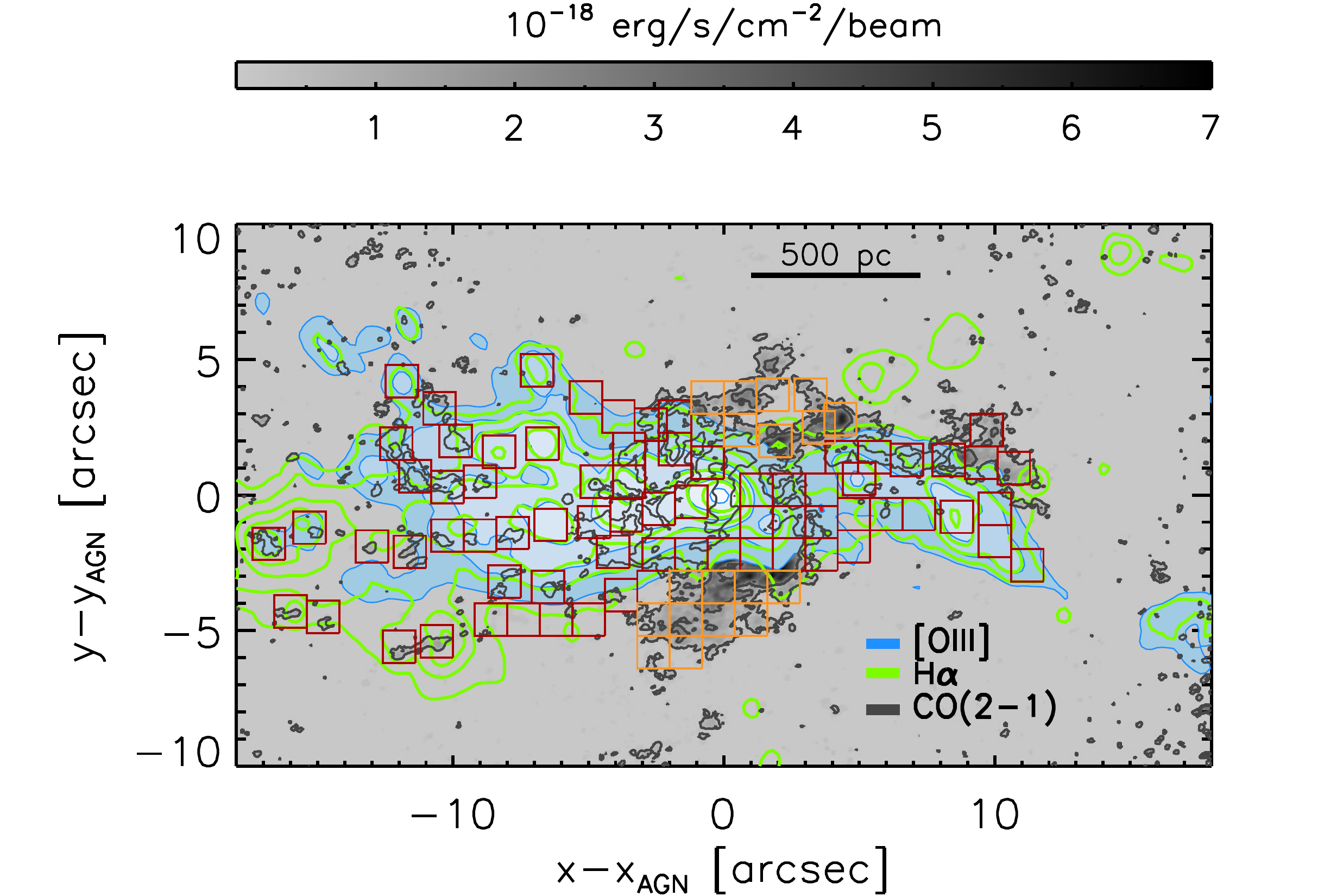}
\caption{In a grey-scale we show the ALMA CO(2-1) integrated intensity map of NGC\,5643 produced from the natural-weight data cube in a linear scale.  The black contours are the CO(2-1) emission in a logarithmic scale with the first contour at 8$\sigma$ and the last contour at 8.7$\times$10$^{-18}$~erg/s/cm$^{-2}$/beam. Blue filled and green contours are the MUSE [O\,{\sc iii}]$\lambda$5007$\AA$ and H$\alpha$ emission maps (see Section \ref{optical_fit}), respectively. Red and orange regions correspond to the outflow and nuclear spiral selected zones, respectively. North is up and east is left, and offsets are measured relative to the AGN.}
\label{fig1}
\end{figure*}

AGN-driven winds appear to be ubiquitous in AGN, although their effect on the host galaxies is unclear. Moreover, AGN-driven outflows are detected in several phases from the extremely hot X-ray gas to the ionized phase to the cold molecular and neutral phases (e.g. \citealt{Morganti17}). It is thus essential to quantify the overall mass, momentum and energy budget of each phase (e.g. \citealt{Cicone18}). Recently,  \citet{Fiore17}  compiled observations for a sample of 94 AGN with  massive winds. They found a correlation between the molecular and ionized mass outflow rates and the AGN bolometric luminosity. These authors also reported that the outflow velocity is larger in the ionized phase but the mass outflow rate is dominated by the molecular one, which becomes especially important at low to intermediate luminosities (L$_{\rm bol}$=10$^{42}$-10$^{45}$~erg~s$^{-1}$). However, this work employed a heterogeneous sample of AGN (0.003$<$z$<$6.4) that not always has multiphase observations for the same galaxy. Except for a few  nearby AGN, e.g.,  NGC\,1068 by \citealt{garcia-burillo14,garcia-burillo19} and NGC\,5728 by \citealt{Shimizu19}, there is a lack of detailed multiphase studies using various regions within the same source. Nearby Seyfert galaxies are intermediate luminosity AGN (L$_{\rm X}$=10$^{42}$-10$^{44}$~erg~s$^{-1}$) and afford the necessary physical resolution to study the multiphase outflow properties in different regions. 

This pilot study aims to investigate the AGN-driven outflow in the cold molecular and ionized gas phases of the Seyfert 2 galaxy NGC\,5643. It is a barred Sy2 galaxy at $\sim$16.9\,Mpc seen almost face-on (i$\sim$-27$^\circ$; \citealt{deVaucouleurs76}). It presents ionization cones detected in [O\,{\sc iii}]$\lambda$5007$\AA$ (hereafter [O\,{\sc iii}]) and H${\alpha}$ at both sides of the nucleus with an elongated morphology along the east-west direction (e.g. \citealt{Schmitt94, Simpson97}). Using VLT/MUSE data \citet{Cresci15} showed that the kinematics of this double-sided ionization cone was consistent with outflows, based on the detection of a blueshifted asymmetric wing of the [O\,{\sc iii}] emission line. In the same orientation, VLA radio observations show emission at both sides of this radio-quiet galaxy ($\log(\rm L_{1.4 \rm GHz})=$37.3 erg~s$^{-1}$; \citealt{melendez10}) related with the emission of a radio jet \citep{Morris85, Leipski06}. This morphology is also approximately coincident with a large scale stellar bar \citep{Mulchaey97}. We refer the reader to \citet{Cresci15} to further details on the stellar and radio morphology. The nuclear H$_{2}~1-0$S(1) velocity field shows also evidence of non-circular motions in the hot molecular gas \citep{Davies14}.  Recently, \citet{Herrero18} presented high angular resolution CO(2-1) line and 232 GHz continuum observations obtained with ALMA. They showed that in the outflow region the CO(2-1) emission has generally two kinematic components, one associated to rotation in the disk of the galaxy and another due to the interaction of the AGN outflow with the molecular gas. Besides, there is observational evidence of positive AGN feedback in this galaxy \citep{Cresci15}.

The paper is organized as follows. Section \ref{observations} describes the observations and data compilation, respectively. The spectral analysis is presented in Section \ref{data_analysis}. In Section \ref{distr_kinematic}, we compare the molecular and ionized gas distribution and kinematics. In Section \ref{connection_kinematic}, we study the connection between the molecular and ionized gas phases and in Section \ref{Simulation} we compare the ionized wind observational data with outflow wind simulations. Finally, in Section \ref{conclusions} we summarize the main conclusions of this work. 

\begin{figure*}
\centering
\par{
\includegraphics[width=7.3cm, angle=90]{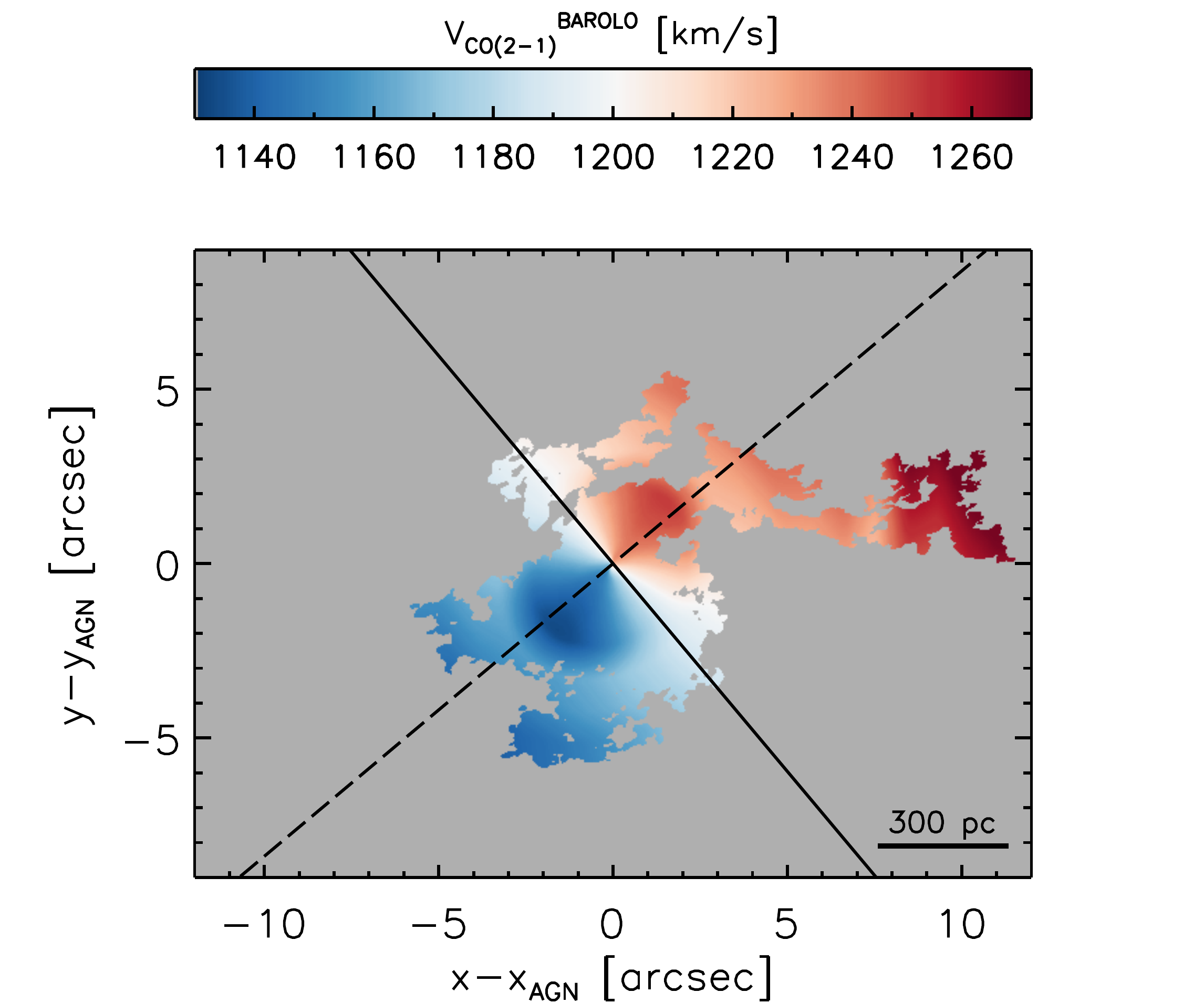}
\includegraphics[width=7.5cm, angle=90]{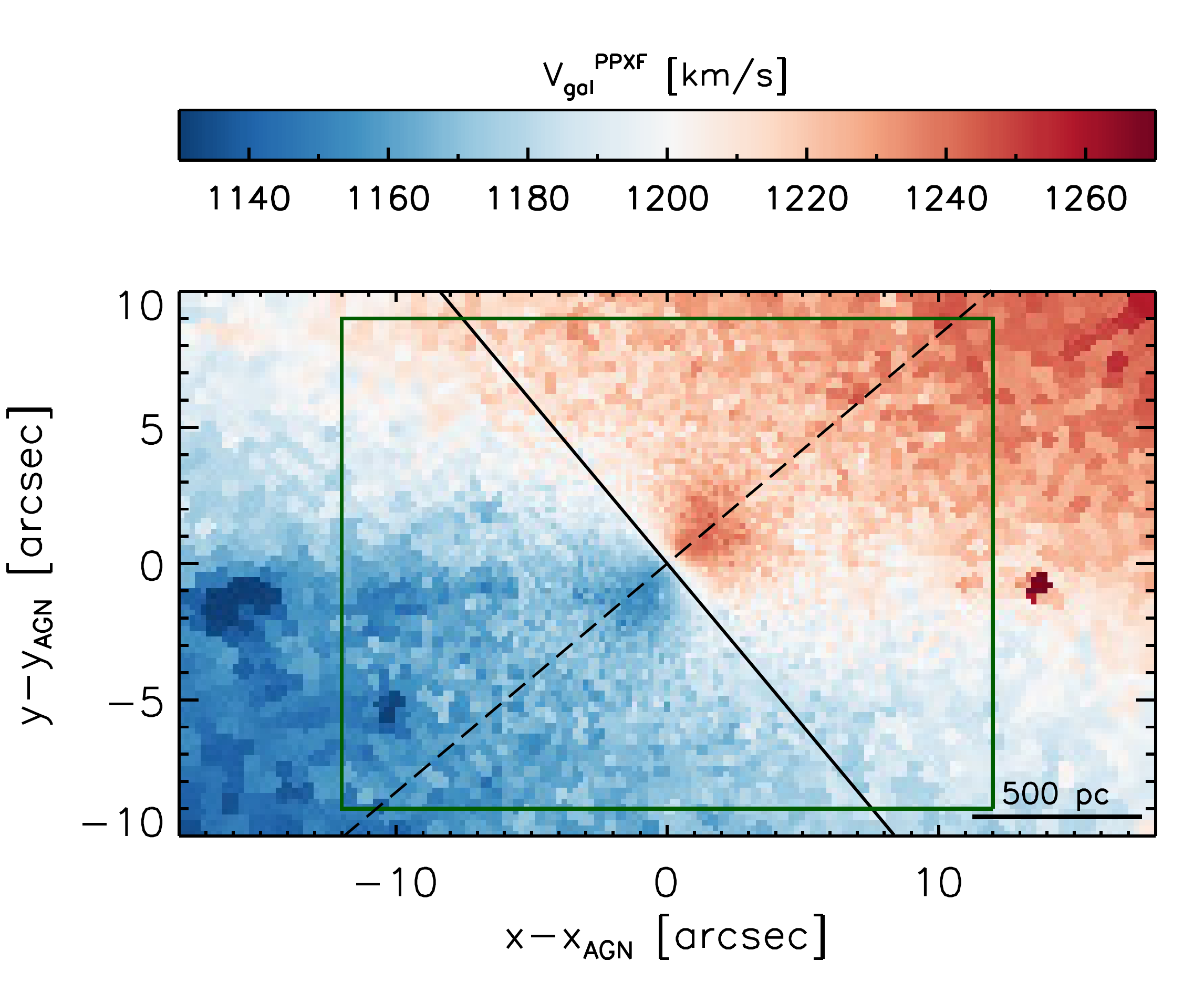}
\par}
\caption{Left panel: $^{3D}$BAROLO model of the mean velocity field of NGC\,5643 using a rotating disk model fitted to the natural-weight CO(2-1) data cube. Right panel: penalised pixel fitting (pPXF) model of the stellar kinematics from the Ca\,II triplet lines using $\lambda>$8000$\AA$ MUSE data and pixels binned to S/N$=$40. The solid and dashed black lines correspond to the kinematic minor- and major-axes, respectively. The green box in right panel shows the FoV of the left panel. The north-eastern (south-western) region corresponds with the far (near) side. North is up and east is left, and offsets are measured relative to the AGN.}
\label{fig2}
\end{figure*}

\section{Observations}
\label{observations}

\subsection{ALMA data}
\label{alma}

NGC\,5643 was observed in Band 6 using the 12-m ALMA array with a compact (baselines between 15 and 492 m) and an extended configuration (baselines between 17 and 3700~m). The data were obtained as part of the project 2016.1.00254.S (PI: A. Alonso-Herrero) and the on-source integration times were 11 and 36 minutes for the compact and extended configuration, respectively. The original field of view (FoV) of this data is $\sim$40$\arcsec$ $\times$ 20$\arcsec$ ($\sim$3.3~kpc $\times$ 1.6~kpc) with spectral resolution of 15~km~s$^{-1}$ and angular resolution of 0.26$\arcsec$ $\times$ 0.17$\arcsec$ at a beam position angle (PA$_{beam}$) of -69.6$^\circ$. The fully reduced and clean CO(2-1) natural weight data cube was taken from \citet{Herrero18}.  Fig.~\ref{fig1} shows the central $\sim$2.9~kpc $\times$ 1.6~kpc FoV where the outflow region dominates.

\subsection{Archival VLT/MUSE integral field spectroscopy}
\label{muse}

Optical integral field spectroscopy of NGC\,5643 was taken using the Multi Unit Spectroscopic Explorer (MUSE, \citealt{Bacon10}) on the 8.2~m Very Large Telescope (VLT). These data were observed as part of the program 095.B-0532(A) (PI: Carollo). Note that this dataset is presented in \citet{Erroz19}. We downloaded the fully reduced and calibrated science data cube from the ESO data archive\footnote{http://archive.eso.org}. We remark that this MUSE data cube, first presented in \citet{Herrero18}, was observed under significant better seeing conditions than the one presented in \citet{Cresci15}, i.e. $\sim$0.5\,\arcsec vs $\sim$0.88\arcsec.

In this work we are only interested in the emission line properties, thus we first subtract the stellar continuum. We used STARLIGHT \citep{Cid05} to model the stellar continuum. This code combines library spectra (STELIB; \citealt{Bruzual03}) of various ages and metallicities  to reproduce the input spectrum. After the continuum subtraction, we fit the emission lines in each spaxel with amplitude-over-noise (AoN; note that the amplitude is the peak of the emission line) ratio is larger than three with Gaussian functions to derive the 2-D maps of the brightest emission lines.  Fig. \ref{fig1} shows the resulting [O\,{\sc iii}] map (blue image and blue contours) as well as the H$\alpha$ emission (green contours).

\subsection{Archival Spitzer data}
\label{spitzer}

We downloaded the fully reduced and extracted low-resolution MIR Spitzer/IRS (\citealt{Houck04}) spectrum from the Cornell Atlas of Spitzer/IRS Source (CASSIS\footnote{https://cassis.sirtf.com/}; \citealt{Lebouteiller11}). CASSIS uses an optimal extraction to get the best signal-to-noise ratio. The spectra were obtained using the staring mode and the low-resolution (R$\sim$60--120) IRS modules: the short-low (SL; 5.2--14.5~$\mu$m) and the long-low (LL; 14--38~$\mu$m). Finally, we only needed to apply a small offset to stitch together the different modules, taking the shorter wavelength module (SL2; 5.2–7.6~$\mu$m) as the basis, which has associated a slit width of 3.7~arcsec.

\section{Spectral Analysis}
\label{data_analysis}

\subsection{Region selection}
\label{Regions}

To study the kinematics of the ionized and molecular gas of NGC~5643, we selected 91 different regions. The outflow region, as probed by the ionized gas emission, extends out to 2.3 kpc in the 
east-west direction (as the radio emission) as well as  in the north-south direction approximately 
$\sim$800~pc to the east of the AGN and  $\sim$500 pc to the west. The selected regions include  bright CO(2-1) emission along the outflow zone, the borders of the outflow at both sizes of the galaxy (delimited by the [O\,{\sc iii}]/H$\alpha$ bipolar nebulosity) and  clumps with bright [O\,{\sc iii}] and H$\alpha$ emission. These regions  are marked with 
red boxes in Fig.~\ref{fig1}. We also selected regions in the northern and southern parts of the nuclear/circumnuclear spiral structure seen in the CO(2-1)  map, where the ionized gas emission is not strong. We marked these as orange boxes in Fig.~\ref{fig1}. To extract the CO(2-1) and ionized gas emission from these regions, we chose 1.2$\arcsec\times$1.2$\arcsec$ square apertures which are $\sim$2.5 times the MUSE FWHM (Section \ref{observations}).

Since the selected regions do not cover completely the ionized gas emission,  we also extracted several larger slices (see Fig.~\ref{stacking_slices} in Appendix \ref{stacking}) along the outflow region.  We defined slices with widths of 2.4\arcsec and  
heights  ranging from $\sim$6\arcsec ~to $\sim$10\arcsec depending on the extent of the [O\,{\sc iii}] emission. We used the location of the AGN as the reference, and then we applied the same offsets from the nucleus along the east-west direction. We constructed the slices by stacking 
spectra extracted with  1.2$\arcsec\times$1.2$\arcsec$ square apertures and shifted them to a  common rest-frame wavelength reference. We show the stacked [O\,{\sc iii}] and CO(2-1) spectra for each slice in Section \ref{stacking}. This stacking technique improves greatly the S/N of the extracted spectra and thus allows us to detect weak kinematic components which will be used to  derive the outflow properties in a uniform way (see Section \ref{connection_kinematic}).

\subsection{CO(2-1) data}
\label{molecular_fit}

\subsubsection{Rotating disk model}
\label{rotating_disk}
\cite{Herrero18} showed that the ALMA CO(2-1) emission shows different kinematic components (see also Figs.~\ref{line_fits_co1} and \ref{line_fits_co2} of Appendix \ref{stacking}). The main component traces galaxy rotation. Following \cite{Herrero18}, we fitted  a rotating disk model to the CO(2-1) kinematics using the $^{3D}$BAROLO code \citep{Teodoro15} but covering a larger FoV. We fixed the following parameters: inclination i$=35^\circ$, the position angle of the major kinematic axis PA$=320^\circ$ and the systemic velocity v$_{sys}=1994$~km~s$^{-1}$, as derived by \cite{Herrero18}. $^{3D}$BAROLO produces the observed and best-fit model maps of the velocity-integrated intensity (0$^{th}$ moment), mean-velocity field (1$^{st}$ moment) and velocity dispersion (2$^{nd}$ moment). Figure \ref{fig2} (left panel) shows the resulting $^{3D}$BAROLO model for the central $\sim$10\arcsec $\times$10\arcsec .

\subsubsection{Individual regions}
\label{individual_co}

For the majority of the selected regions, the CO(2-1) emission shows clearly deblended velocity components, except in regions near the AGN (see Figs.~\ref{line_fits_co1} and \ref{line_fits_co2}).  We fit the various kinematic components to accurately measure their velocities and fluxes. Therefore, only when necessary (significant residuals of the fit; $>$3 times the standard deviation) we used two or three Gaussians for the fit. We used the {\textit{PySpecKit}} code \citep{Ginsburg11}, which produces the best fit using the minimization of $\chi^2$. 

In the region modelled by the rotating galaxy disk, there is a good agreement between the predicted velocity from the model and the brightest CO(2-1) component. Since we could not model the large-scale faint emission with BAROLO, we extrapolated the velocity field when possible to identify the galaxy disk emission line in the fitted kinematic components. In those cases, we considered the brightest feature as the galaxy disk component. Finally, we also checked that the velocities of the identified CO(2-1) rotation disk component were in agreement within the errors with those of the stellar kinematics (see Section \ref{stellar} and right panel of Fig. \ref{fig2}, and Appendix \ref{stacking}). When we detected more  than one kinematic component, the fainter were assumed to trace non-circular motions (i.e., outflowing or inflowing material, see below).

\subsubsection{Radial profile along the outflow}
\label{outflow_co}

As discussed in Section \ref{Regions}, we also extracted several larger slices (see Fig.~\ref{stacking_slices}) along the outflow region. Then, we fit all the emission lines with AoN$>$3 for each slice spectrum using the same method as in Appendix \ref{individual_co} (see Fig. \ref{line_fits_co1} and \ref{line_fits_co2} in Appendix \ref{stacking}).

\begin{figure*}[ht!]
\centering
\par{
\includegraphics[width=9cm]{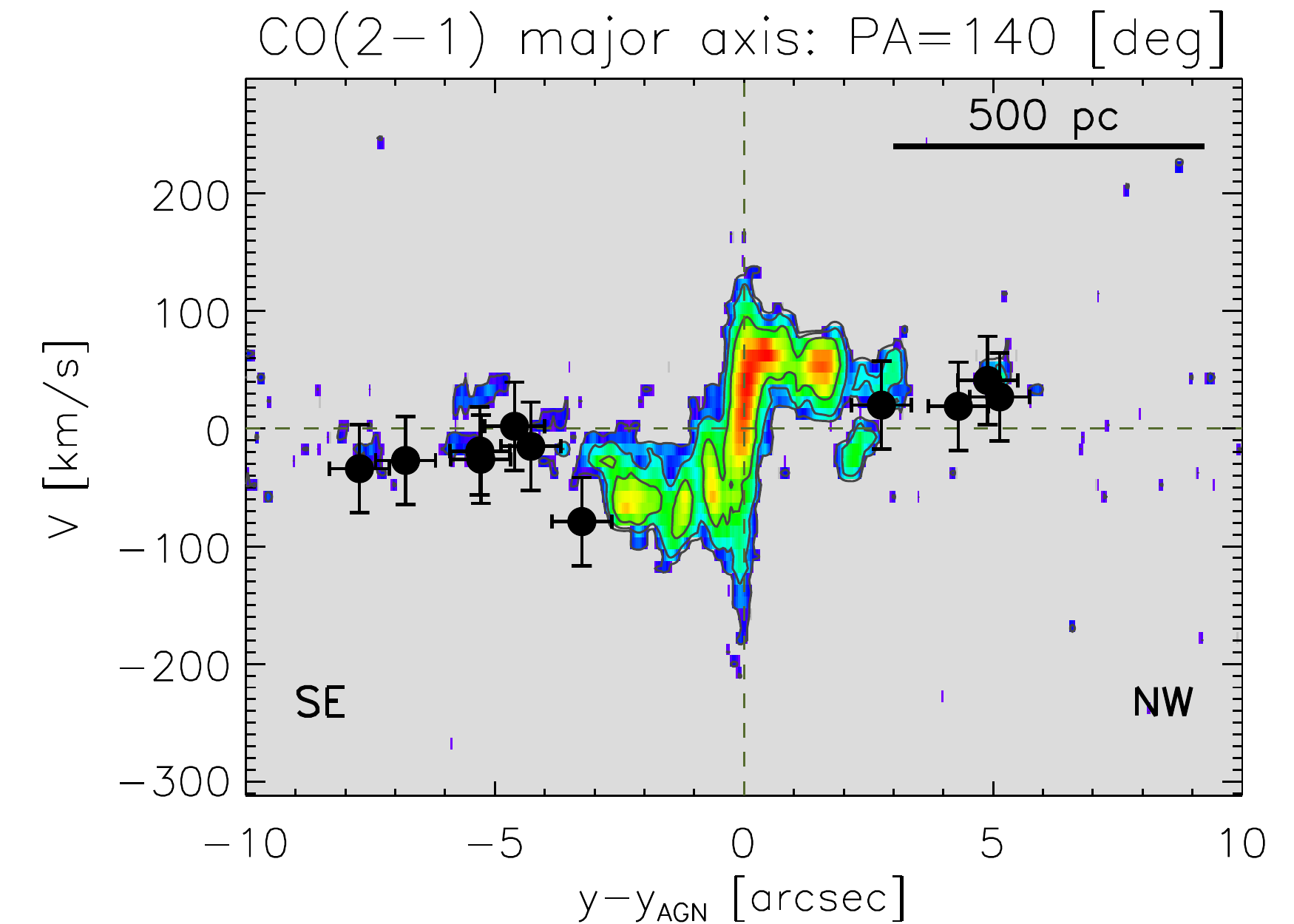}
\includegraphics[width=9cm]{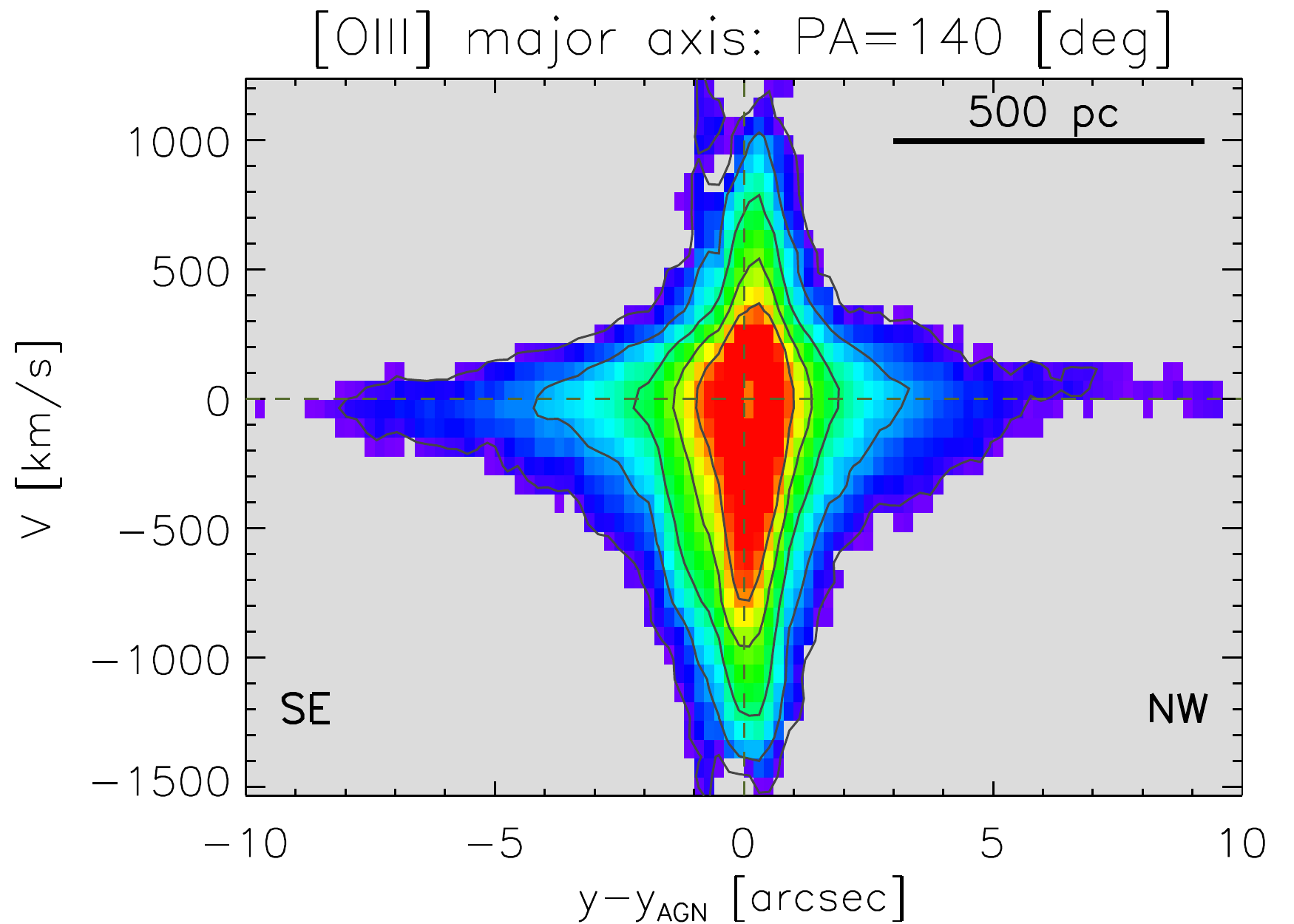}
\par}
\caption{Left panel: Observed CO(2-1) {\textit{p-v}} diagram along the kinematic major-axis (fluxes above 3$\sigma$). The black filled circles correspond to the velocities of the [O\,{\sc iii}] narrow components along the kinematic major-axis. Right panel: same as in left panel, but for the [O\,{\sc iii}] emission. The horizontal dashed line indicates the zero velocity and the vertical line the AGN position.}
\label{major_axis}
\end{figure*}

\subsection{MUSE data}
\subsubsection{Stellar kinematics}
\label{stellar}

We derived the stellar kinematics from the Ca\,II triplet lines using MUSE data at $\lambda>$8000$\AA$  and the penalised pixel fitting (pPXF) method \citep{Cappellari04,Cappellari17}. To achieve the best spatial resolution possible with good S/N, we used the Voronoi binning algorithm \citep{Cappellari03} for S/N$=$40. The right panel of Figure \ref{fig2} shows that the NGC\,5643 MUSE data used in this work has enough S/N in the full $\sim$2.9~kpc $\times$ 1.6~kpc, and that the stellar kinematics is in fairly good agreement within the errors with the CO(2-1) rotating disk model (see left panel of Figure \ref{fig2}).

\subsubsection{Individual regions}
\label{individual_o3} 

To derive the kinematics of the ionized gas and in particular the properties of the outflow, we used the [O\,{\sc iii}] emission line. The [O\,{\sc iii}] line emission in all the selected regions shows complex and asymmetrical profiles. This suggests the presence of different kinematic components that require more than one Gaussian for the fit  (see Fig. \ref{line_fits_o31} and \ref{line_fits_o32} of Appendix \ref{stacking}). To perform the emission line fitting we used the {\textit{PySpecKit}} code as in Section \ref{molecular_fit}. Then, after the stellar continuum subtraction, we fit the [O\,{\sc iii}] doublet emission lines (when the [O\,{\sc iii}]$\lambda$5007$\AA$ emission line has AoN$>$3) simultaneously with the same velocity and widths, which are always larger than the instrumental one (i.e. MUSE nominal spectral resolution FWHM$\sim$2.5$\AA$). We also fixed the [O\,{\sc iii}] doublet to its theoretical value (1/3; \citealt{Osterbrock06}). For those fits with single Gaussians which show residuals in the [O\,{\sc iii}] region above 3 times the standard deviation, we included a second kinematic component. If  residuals above 3 times the standard deviation were still present, we included a third kinematic component. Finally, we only considered for the analysis those fits using two/three components when the residuals improved considerably  (i.e., $\rm{standard\:deviation}^{\rm final}_{residual}\ll\rm{standard\:deviation}^{\rm initial}_{residual}$).

As expected, due to the presence of an outflow in NGC\,5643, we detected two or more components in regions along the outflow region. The main difference between these various kinematic components is their widths and amplitudes. We found median values for the narrow component, which has the highest amplitudes ($\sim$10 times higher on average), of $\sigma_{\rm narrow}^{\rm [O\,III]}=$97$\pm$20~km s$^{-1}$ and the broad component has $\sigma_{\rm broad}^{\rm [O\,III]}=$233$\pm$75~km s$^{-1}$, where the $\sigma$ values are corrected by the instrumental FWHM. Note that, hereafter, we will refer to these components as narrow and broad, respectively. However, this broad component has smaller line widths than the classical broad line region of AGNs (FWHM$>$1000~km s$^{-1}$). The mean velocities of the narrow component are consistent within the errors with those derived from the stellar kinematics. Therefore, we assumed that the narrow component of the [O\,{\sc iii}] line traces the  galaxy disk rotation (see also Section~\ref{pvdiagrams}). 

\subsubsection{Radial profile along the outflow}
\label{outflow_o3}

Following the same methodology as in Section \ref{outflow_co} and \ref{individual_o3}, for each slice along the outflow we fit all the ionized gas emission lines with AoN$>$3. In addition, for the slices we fit the 
H$\alpha$, H$\beta$ and [S\,II]$\lambda$6718,6732$\AA$ emission lines using the same velocity-widths to estimate the extinction and the electron density. As for [O\,{\sc iii}], we  used the same criteria  to include one or two Gaussians in each individual fit. Note that we only used a maximum of two Gaussians for the fits. We also fixed the [N\,II] doublet ratios to their theoretical values (e.g. \citealt{Osterbrock06}). For the  extinction correction,  we assumed an intrinsic ratio of H$\alpha$/H$\beta=$2.86 and the \citet{Calzetti00} attenuation law (R$_V=$3.12). In addition, we estimated the electron density using the [S\,II]$\lambda$6716,6730 doublet ratio (e.g. \citealt{Osterbrock06}). To do so, we use the PyNeb\footnote{http://www.iac.es/proyecto/PyNeb/} \citep{Luridiana15}  task \textit{temden}, which computes the electron density from diagnostic line ratios, under the assumption of the typical narrow line region (NLR) gas temperature on Seyfert galaxies (10$^4$~K; e.g. \citealt{Vaona12}). In agreement with previous works on
Seyfert galaxies (e.g. \citealt{Bennert06a,Bennert06b}), the observed gas density of the NLR gas in NGC\,5643 increases with decreasing distance from the AGN. In Table \ref{tab_o3} of Appendix \ref{stacking}, we show the electron density results for each slice.

On the other hand, we remark that in \citet{Davies20} the authors found that this method estimate significantly lower electron densities in AGN photoionized gas than using auroral and transauroral lines. However, the latter has also the limitation that the auroral and transauroral lines are generally weak. We refer the reader to \citet{Davies20} for further discussion on the various method to derived the electron density in AGNs.

\begin{figure*}
\centering
\par{
\includegraphics[width=8.75cm]{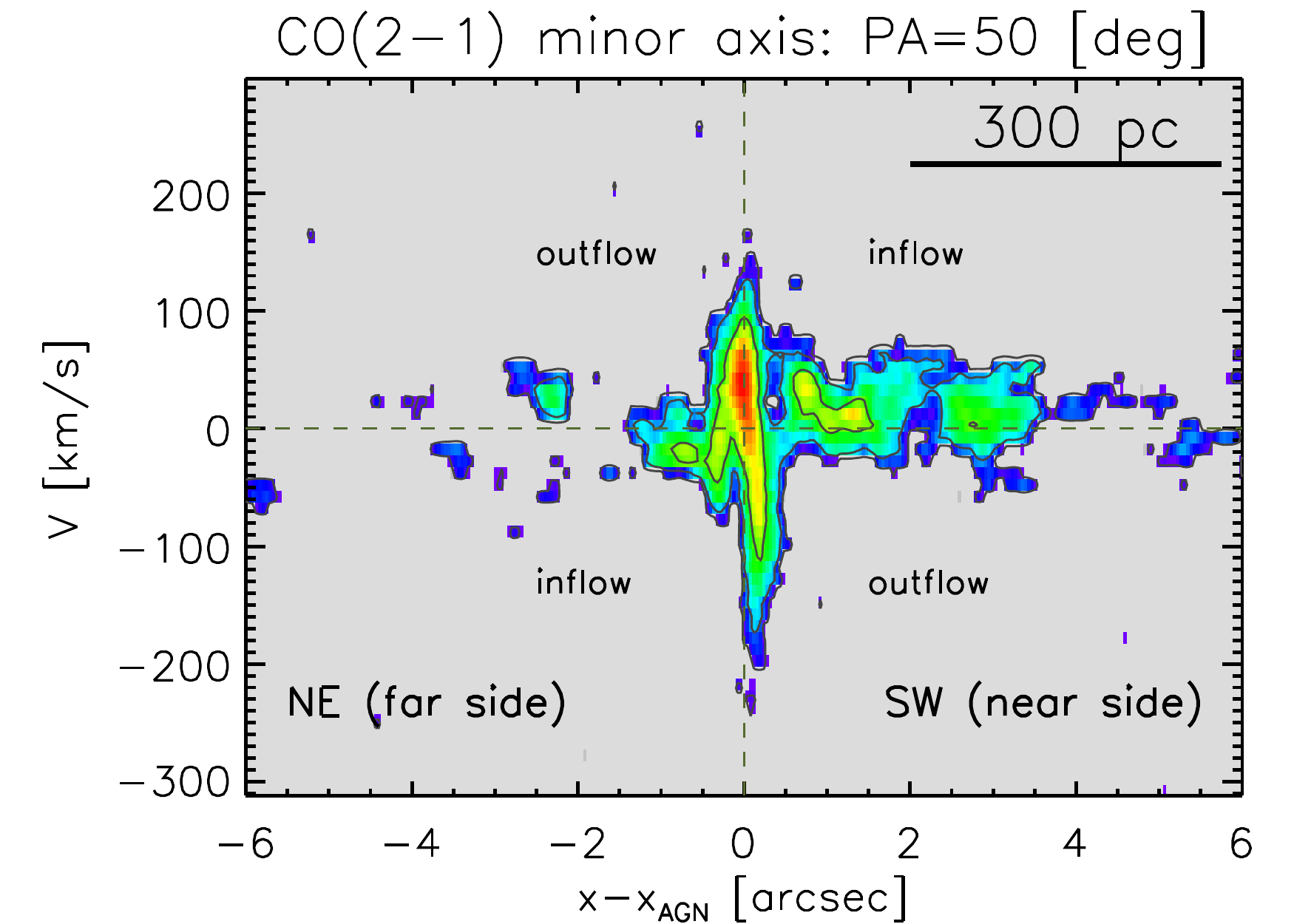}
\includegraphics[width=8.75cm]{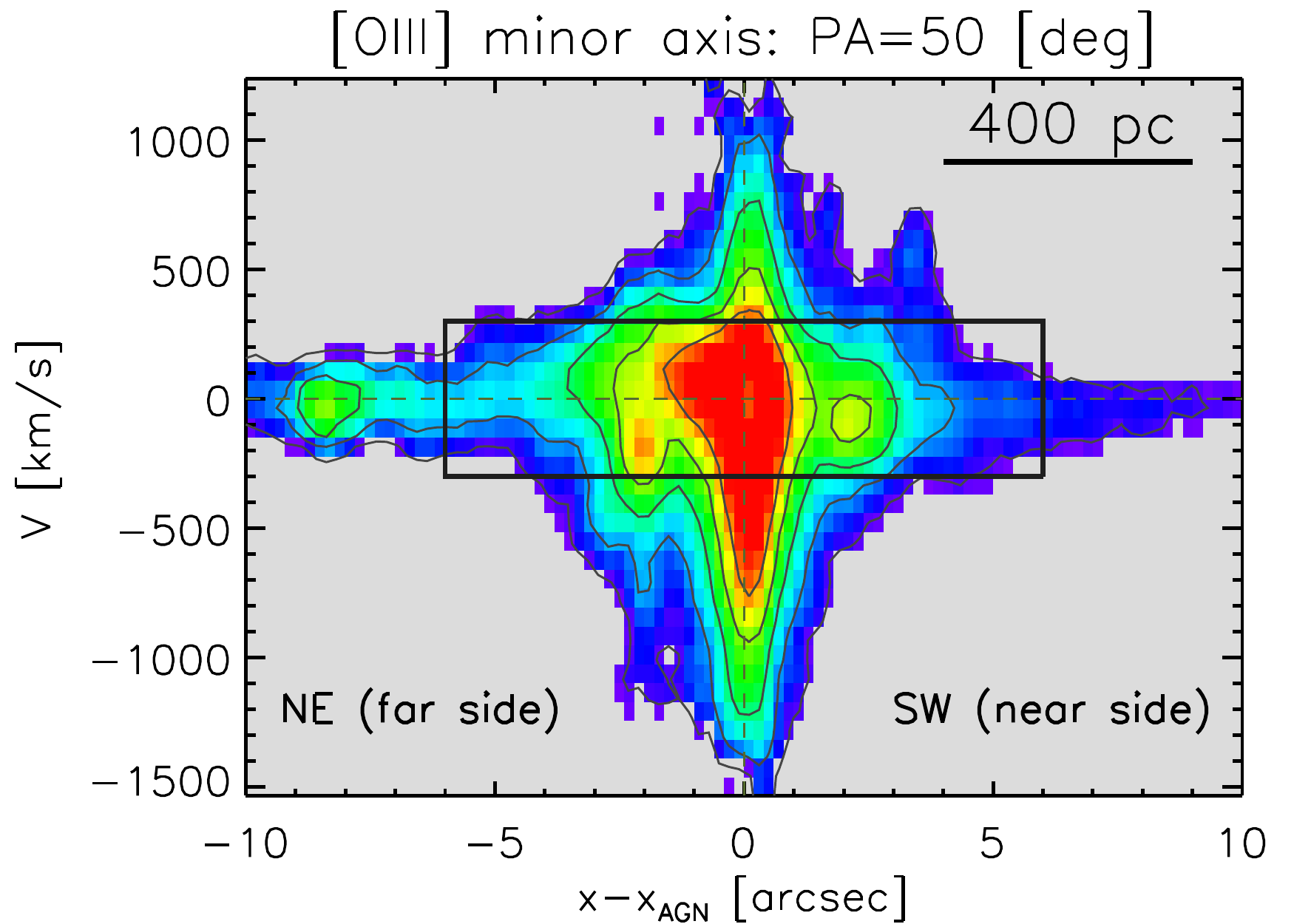}
\par}
\caption{Left panel: Observed CO(2-1) {\textit{p-v}}diagram along the kinematic minor-axis (fluxes above 3$\sigma$). Right panel: same as in the left panel, but for the [O\,{\sc iii}] emission. The black box in right panel shows the FoV of the left panel. The horizontal dashed lines indicate the zero velocity and the vertical lines the AGN position.}
\label{minor_axis}
\end{figure*}

\section{Distribution and kinematics of the circumnuclear molecular/ ionized gas}
\label{distr_kinematic}

\label{optical_fit}

\subsection{Morphology}
\label{Morphology}

As previously discussed by \citet{Herrero18}, the brightest CO(2-1) emission comes from the nuclear region that is clearly connected with the two-arm spiral (see Fig. \ref{fig1}). This spiral structure extends out to $\sim$12\arcsec ($\sim$1~kpc) at both sides of the galaxy and is oriented practically in the same direction (east-west) as the radio emission and the large-scale bar. This morphology could be explained by the canonical gas response to a large-scale bar ($\sim$5.5~kpc; \citealt{Mulchaey97}) with an inner Lindblad resonance (ILR). The gas shows an asymmetric two-arm spiral stretching along the leading edges of the bar (PA=85$^\circ$; \citealt{Mulchaey97}). The spiral does not end in a clear ring in the nuclear region that may correspond to the ILR. Rather, a significant molecular gas concentration around the AGN on scales of 10-50~pc forms a compact disk/torus (see \citealt{Herrero18}). There is also a region of molecular gas in the east side of the spiral approximately 5\arcsec \, from the AGN \citep[see figure~1 in][]{Herrero18}, with a deficit of CO(2-1) emission (see estimates in Section~\ref{wind_entrained}). There are also other two-spiral arms oriented from the northeast to the southwest of the nuclear region (i.e. the nuclear spiral).

The ionization cone traced by the [O\,{\sc iii}] emission line extends for $\sim$28\arcsec ($\sim$2.3 kpc) in the east-west direction and is weaker in the west side due to the host galaxy obscuration (see Fig. \ref{fig1}). As also discussed by \citet{Herrero18}, the region in the  east spiral arm with a CO(2-1) deficit is coincident with bright [O\,{\sc iii}] emission. This could be related to the destruction/clearing of the molecular gas produced by the AGN wind impacting on the eastern side of the host galaxy (see Section \ref{wind_entrained}). The dust extinction map (see e.g. Fig. A1 of \citealt{Mingozzi19}) reveals a curved structure with higher extinction values in the western part of the galaxy, which could be connected to the dust lane in the large scale bar \citep{Cresci15}. We also found that there is a good match between H$\alpha$ emission and the faint CO(2-1) clumps detected in the ionization cone at the eastern part of the galaxy. This suggests that these clumps are star forming regions (see also \citealt{Cresci15}).

\subsection{Kinematics}
\label{kinematic}

\subsubsection{Position-velocity diagrams}\label{pvdiagrams}
To investigate the overall kinematics of the ionized and the molecular gas, as well as deviations from pure circular motions, we produced position-velocity ({\textit{p-v}}) diagrams. The  {\textit{p-v}} diagram of the CO(2-1) emission extracted along the major kinematic axis shows a clear rotation pattern (left panel of Fig.~\ref{major_axis}). We also plotted in this figure the [O\,{\sc iii}] narrow component velocities (black circles) along the same axis, which are in excellent agreement with the CO(2-1) galaxy rotation curve.  This together with the fact that the [O\,{\sc iii}] narrow component velocities are consistent with those derived from the stellar kinematics (see Section \ref{individual_o3}) supports our assumption that this [O\,{\sc iii}] component traces the galaxy rotation and thus the  broad components the non-circular movements (i.e. the outflow). 

\begin{figure}
\centering
\includegraphics[width=8.cm]{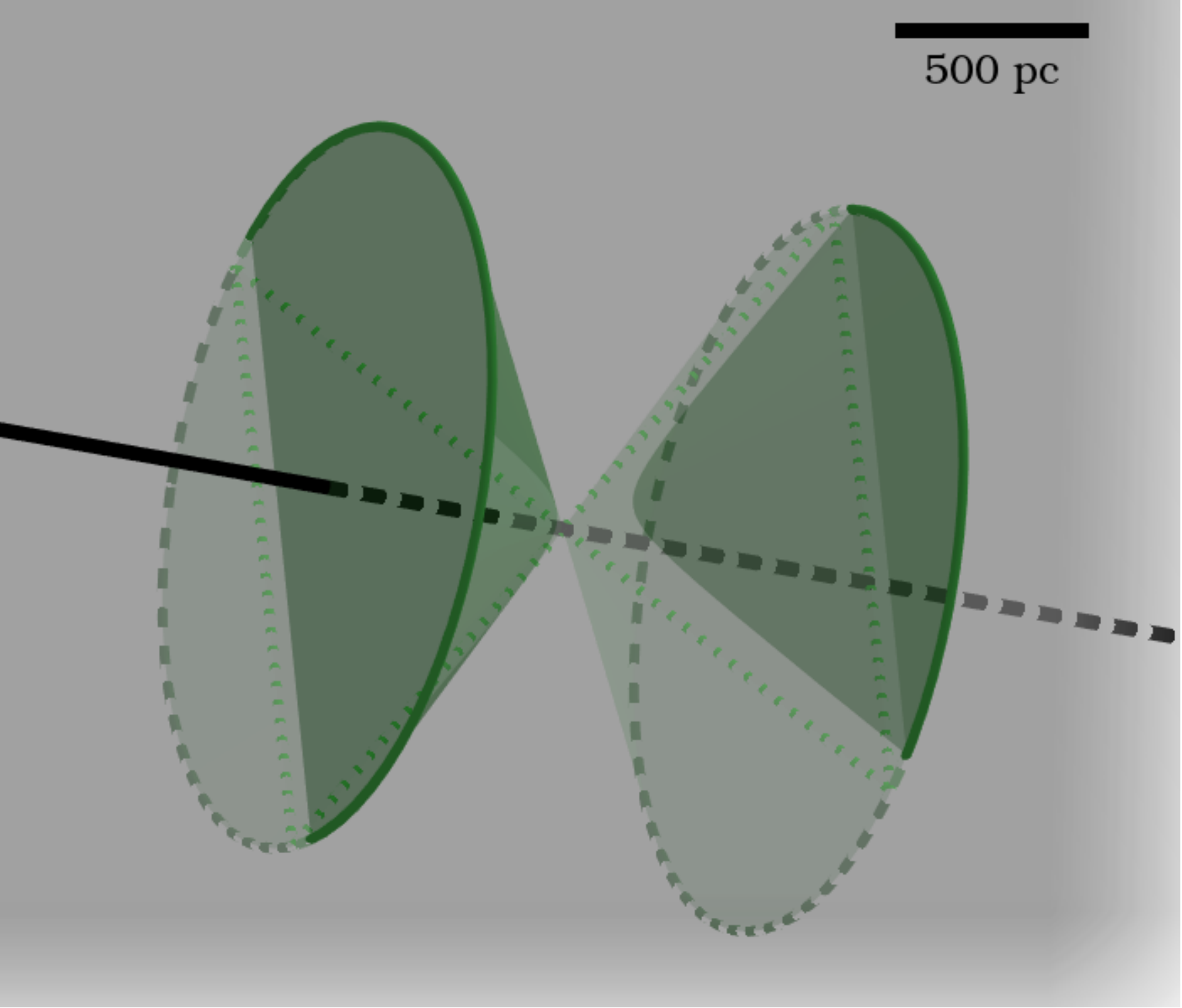}
\caption{3D scheme for the geometry of NGC\,5643 derived from the NLR modelling by \citet{Fischer13}. The green bicone indicates the AGN outflow and its axis is illustrated as a black line. The grey plane corresponds to the disk of the host galaxy. Dark green 
and light green shaded regions are in front of and behind the galaxy disk, respectively. North is up and east is left.}
\label{3d_scheme}
\end{figure}

The left panel of Fig.~\ref{minor_axis} shows the {\textit{p-v}} diagram along the kinematic minor-axis for the molecular gas, which reveals  non-circular motions. Leaving aside the nuclear region (inner $\sim 1\arcsec$), which was discussed in detail by \cite{Herrero18}, the CO(2-1) non-circular motions are observed both to the northeast and southwest of the AGN extending for several arcseconds. The typical amplitudes of the non-circular motions are $\sim$50~km s$^{-1}$ and are blueshifted and redshifthed on both sides of the AGN. This indicates that outflowing and inflowing motions are present out to distances of $\sim$6\arcsec (500~pc) from the AGN. To interpret the CO(2-1) non-circular motions, we first discard density wave-driven inflows. We took into account the  orientation of the stellar bar (PA=85$^\circ$) and the orientation of the galaxy (see Fig.~\ref{fig2}). Assuming that the molecular gas in the disk of the galaxy is rotating counter clockwise, the bar induced streaming motions (local inflow) would be redshifted to the southwest and blueshifted to the northeast along the kinematic minor axis (see Fig.\ref{minor_axis}, left panel). However, it is not expected an extra-planar outflow component since generally the non-circular velocity is below the escape velocity ($\sim$200~km/s at 5\arcsec), and any extra-planar component would fall back onto the disk. Therefore, for the molecular gas outflowing motions we  make the reasonable assumption that they take place in the plane of the galaxy. Taking into account the geometry model derived by \citet{Fischer13} for the ionized gas outflow (see Fig.~\ref{3d_scheme}) to the east (far side of the disk) the redshifted velocities should trace the molecular outflow whereas to the west (near side of the disk) we expect blueshifted velocity excess. 

The [O\,{\sc iii}] {\textit{p-v}} diagram along the kinematic major- and minor-axis is more complex due to the larger number of kinematic components (see right panels of Fig. \ref{major_axis} and Fig. \ref{minor_axis}). A visual inspection  shows that the maximum observed outflow velocities reach $\sim$1500~km s$^{-1}$. Indeed the high velocity gas region ($\sim$1000-1500~km s$^{-1}$) is spatially resolved. The emission appears blueshifted and redshifted at both sides of the galaxy. This confirms the 3-D nature of the ionized emission of the AGN-driven outflow, first inferred by \citet{Fischer13} with long-slit spectroscopy. It is clear that the western components are weaker due to the larger extinction derived from the H$\alpha$/H$\beta$ ratio in this side of the galaxy (see e.g. Fig. A1 of \citealt{Mingozzi19}).

\subsubsection{Spatially resolved kinematics}

As discussed in the previous sections, the majority of the selected regions show two or more different velocity components in each of the gas phases studied here (i.e. molecular and ionized). One is associated with the host galaxy rotating disk and the others with non-circular movements. To derive the spatially resolved kinematics of the in/outflows in this galaxy we need to subtract the galaxy rotation  from the ionized and molecular gas velocity profiles. Once we identified the rotation disk components, we subtract the corresponding velocity of the galaxy disk component in each selected region from the other non-circular velocity components.

\begin{figure*}
\centering
\par{
\includegraphics[width=9.55cm, angle=90]{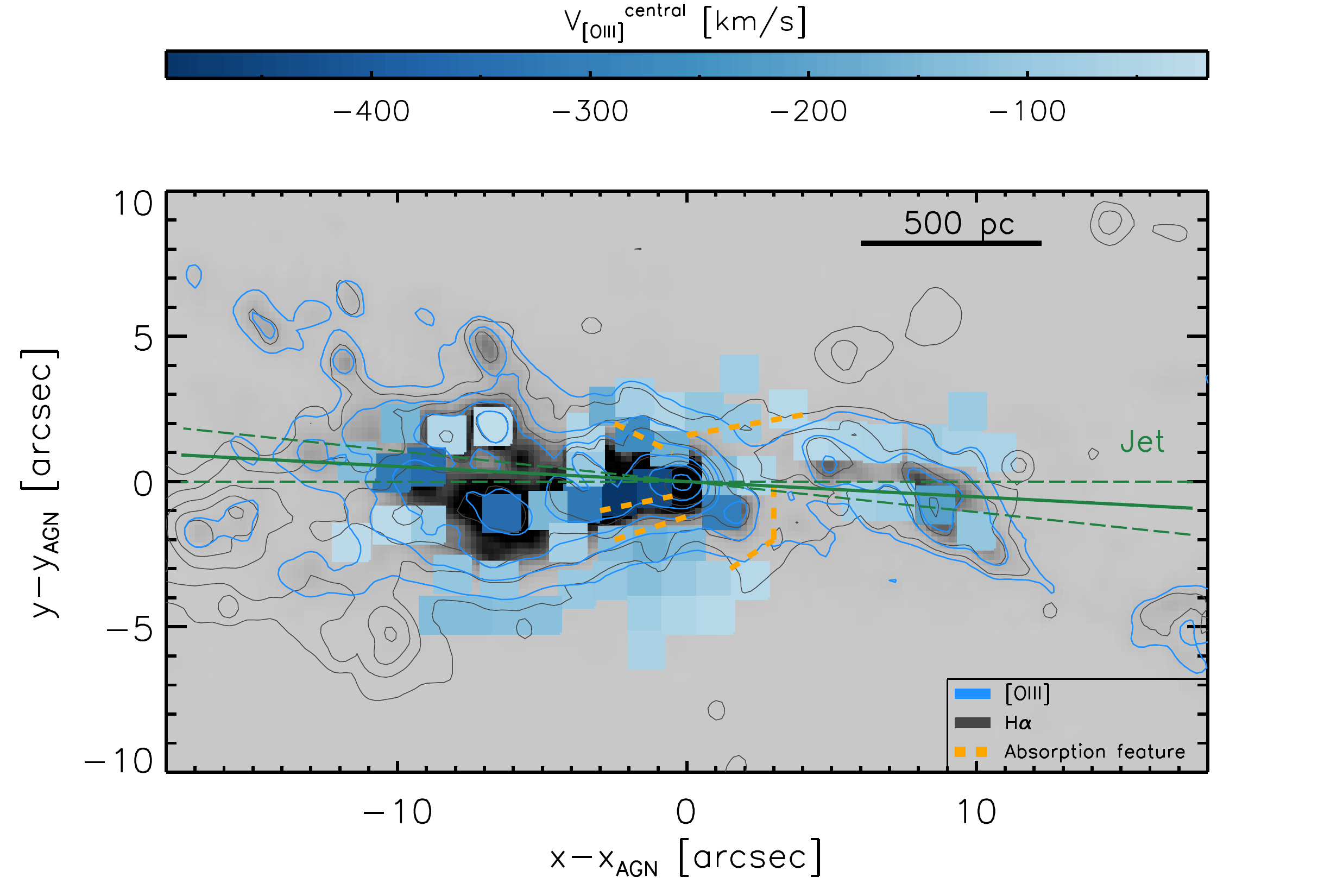}
\includegraphics[width=9.55cm, angle=90]{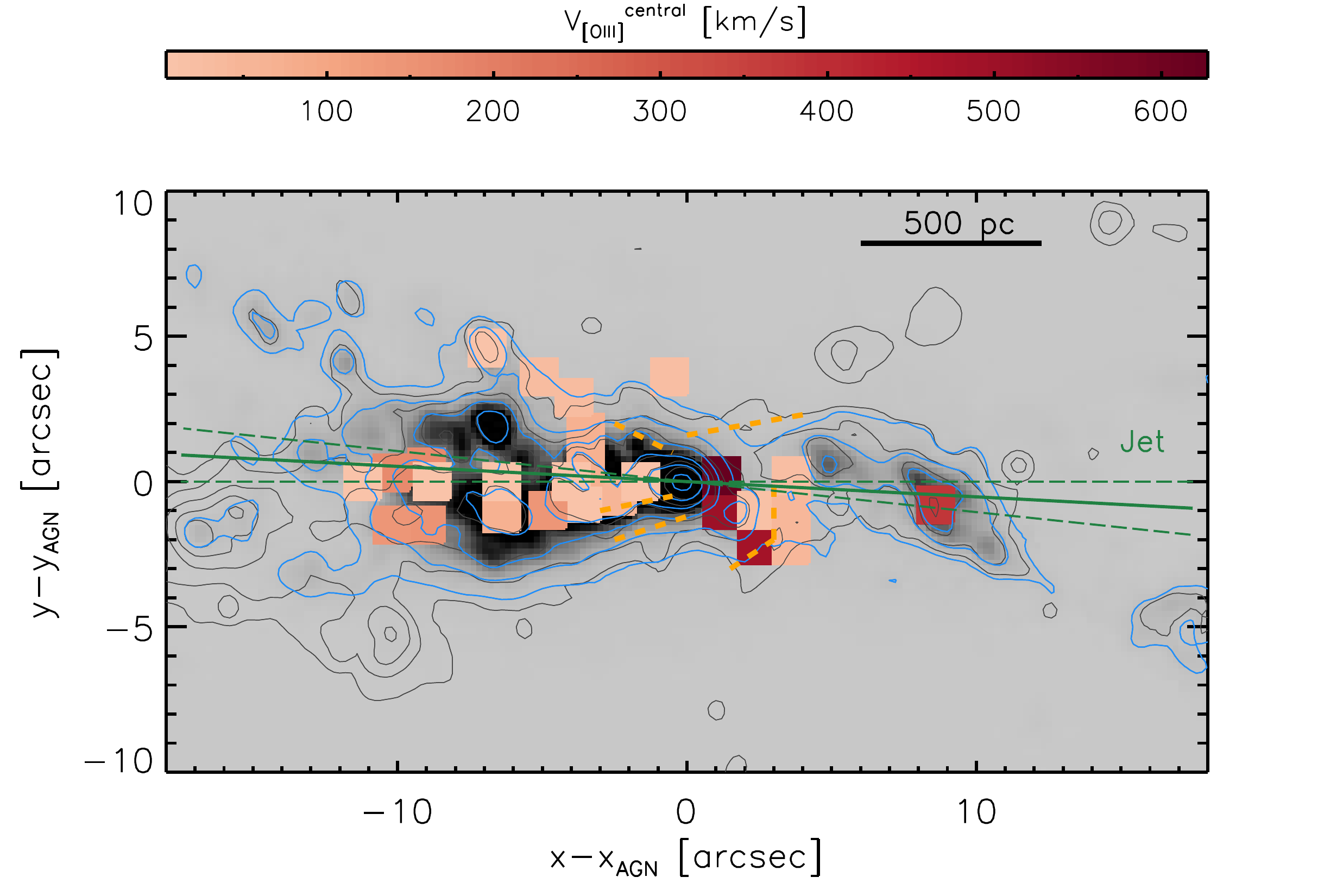}
\par}
\caption{The grey scale image is the MUSE [O\,{\sc iii}]$\lambda$5007$\AA$  integrated intensity map and blue and brown contours to the [O\,{\sc iii}]$\lambda$5007$\AA$ and H$\alpha$ emission (see Section \ref{optical_fit}). The square regions are color-coded according to the central (projected) outflow velocity in the ionized gas phase. The top panel are the blueshifted [O\,{\sc iii}] outflow $v_{\rm central}$ (projected) and the bottom panel redshifted velocities. The solid and dashed green lines correspond to the direction of the radio jet.  The orange dashed lines trace the three main absorption features in the dust structure map of \citet{Davies14}. The feature to the south-west may be associated with outflow rather than inflow. }
\label{blue_o3_vel}
\end{figure*}

\begin{figure*}
\centering
\par{
\includegraphics[width=9.55cm, angle=90]{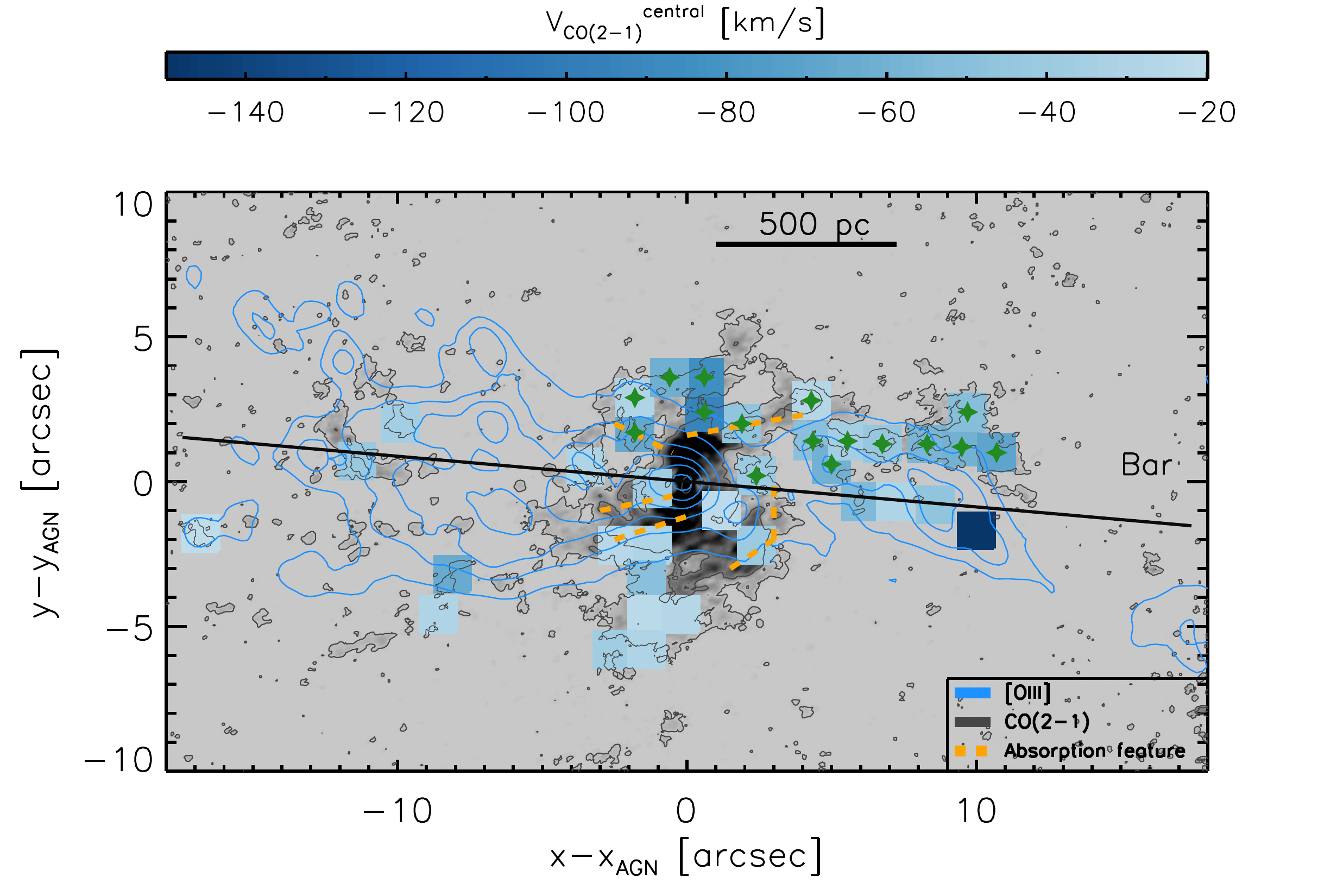}
\includegraphics[width=9.55cm, angle=90]{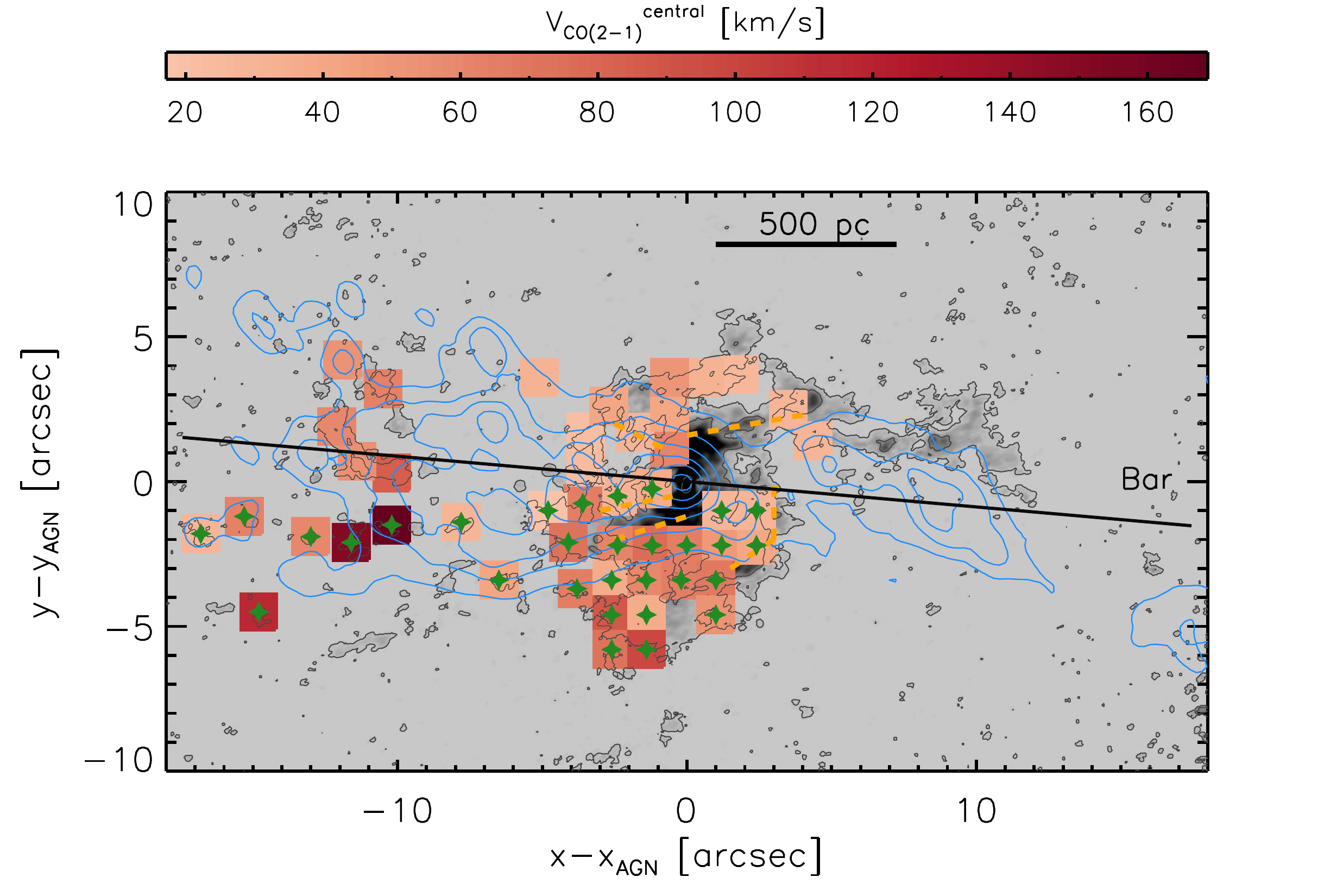}
\par}
\caption{The grey scale image is the ALMA CO(2-1) integrated intensity map and blue contours to the [O\,{\sc iii}]$\lambda$5007$\AA$ emission  (see Section \ref{optical_fit}). The black contours are the CO(2-1) emission in a logarithmic scale with the first contour at 6$\sigma$ and the last contour at 2.2$\times$10$^{-17}$~erg/s/cm$^{-2}$/beam. The square regions are color-coded according to the central (projected) non-circular velocity in the cold molecular gas phase. The top panel shows blueshifted CO(2-1) non-circular $V_{\rm central}$ (projected) and the bottom panel redshifted velocities.  The solid black line  shows the orientation of the large scale stellar bar \citep{Mulchaey97}. Filled green stars  mark regions with the expected local inflow regions due to the bar (see Section \ref{kinematic}). The orange dashed lines are as in Fig.~\ref{blue_o3_vel}. }
\label{blue_co_vel}
\end{figure*} 

The median value (projected) for central velocity of the molecular phase non-circular motions is 44$\pm$29~km s$^{-1}$ with most regions showing velocities below 100~km s$^{-1}$. The only exceptions are three eastern regions corresponding to CO(2-1) clumps located at projected distances of $\sim$10-16\arcsec ($\sim$0.9-1.3~kpc) and one in the western region at $\sim$10\arcsec ($\sim$0.8~kpc). The ionized gas outflow normally shows larger velocities, reaching central velocities of $\sim$720~km~s$^{-1}$ (projected) in regions close to the AGN.

Figure~\ref{blue_o3_vel}  shows the spatial distribution of $V_{\rm central}$ (projected) of the [O\,{\sc iii}] outflowing kinematic components, both blueshifted (top) and redshifted (bottom). According to the ionization cone modelling of \citet{Fischer13}, the ionization cone has an inclination angle of $\sim$40$^{\circ}$ with respect to the host galaxy disk (see Fig. \ref{3d_scheme}) and an opening angle of $\sim$110$^{\circ}$. Then, blueshifted velocities are expected at both sides of the galaxy for the ionized phase, as first noticed in the minor-axis [O\,{\sc iii}] {\textit{p-v}} diagram. Indeed, the outflow zone is mainly dominated by blueshifted [O\,{\sc iii}] velocities  (Fig. \ref{blue_o3_vel}, top). However, there are also  redshifted [O\,{\sc iii}] velocities in the north-eastern and south-western borders of the outflow, and the central-eastern part of the outflow (Fig.~\ref{blue_o3_vel}, bottom). This could be explained by a hollow ionization cone \citep{Fischer13}, and, therefore, we are observing the internal wall of the hollow ionization cone in the north-eastern part. In the case of the south-western border and the central-eastern region of the outflow the material is likely outflowing behind the galaxy disk. These findings together with the geometry derived by \citet{Fischer13} suggest that the outflow wind and the radio jet should be impacting the galaxy disk. 

Figure~\ref{blue_co_vel} shows the spatial distribution for the non-circular motions of the CO(2-1) emission. As already seen in 
the minor-axis {\textit{p-v}} diagram (Fig.~\ref{minor_axis}), some are associated with streaming motions linked to the gas response to the bar along the leading edges of the  spiral arms. Thus, we interpret the CO(2-1) blueshifted (reshifted) non-circular motions in the northern (southern) part of the spiral arms as local inflow. We marked these regions with green crosses for reference. Indeed, such CO(2-1) streaming motions associated with the presence of bars are detected in other Seyfert galaxies (e.g. \citealt{Herrero19,Shimizu19}; and Dominguez-Fernandez et al. 2020). Thus, the CO(2-1) non-circular motions in the remaining regions in the east-west [OIII] outflow region can be explained as molecular gas entrained by the AGN wind and outflowing in the plane of the galaxy \citep[see also][]{Herrero18}. We finally note that the velocities of the outflowing and inflowing molecular gas are similar (see also \citealt{Dominguez20}), and thus, detailed studies such as this, with a good understanding of the geometry of galaxy and presence of a bar, are necessary to be able to disentangle them.

\begin{figure*}[ht!]
\centering
\par{
\includegraphics[width=8.0cm]{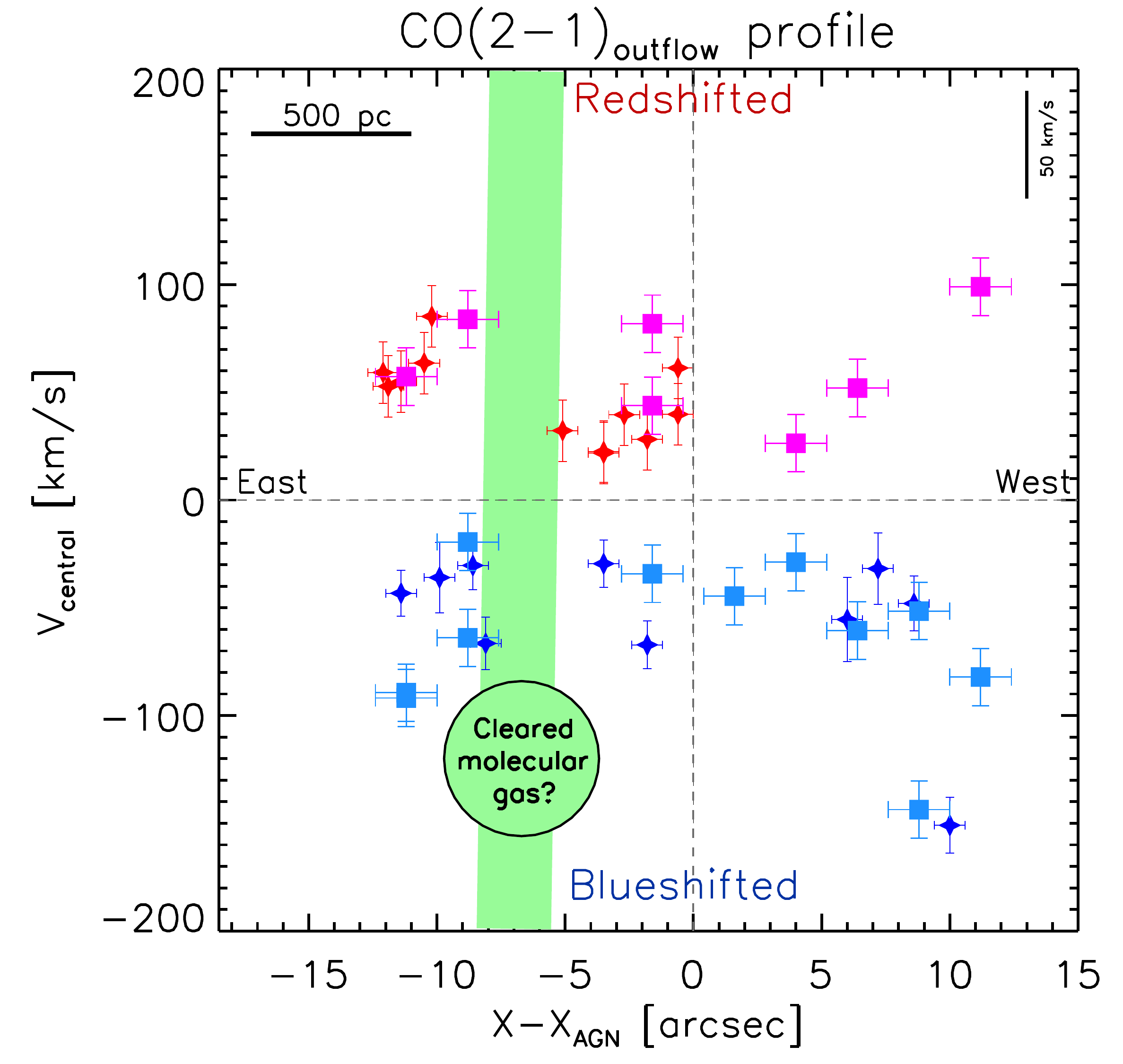}
\includegraphics[width=8.0cm]{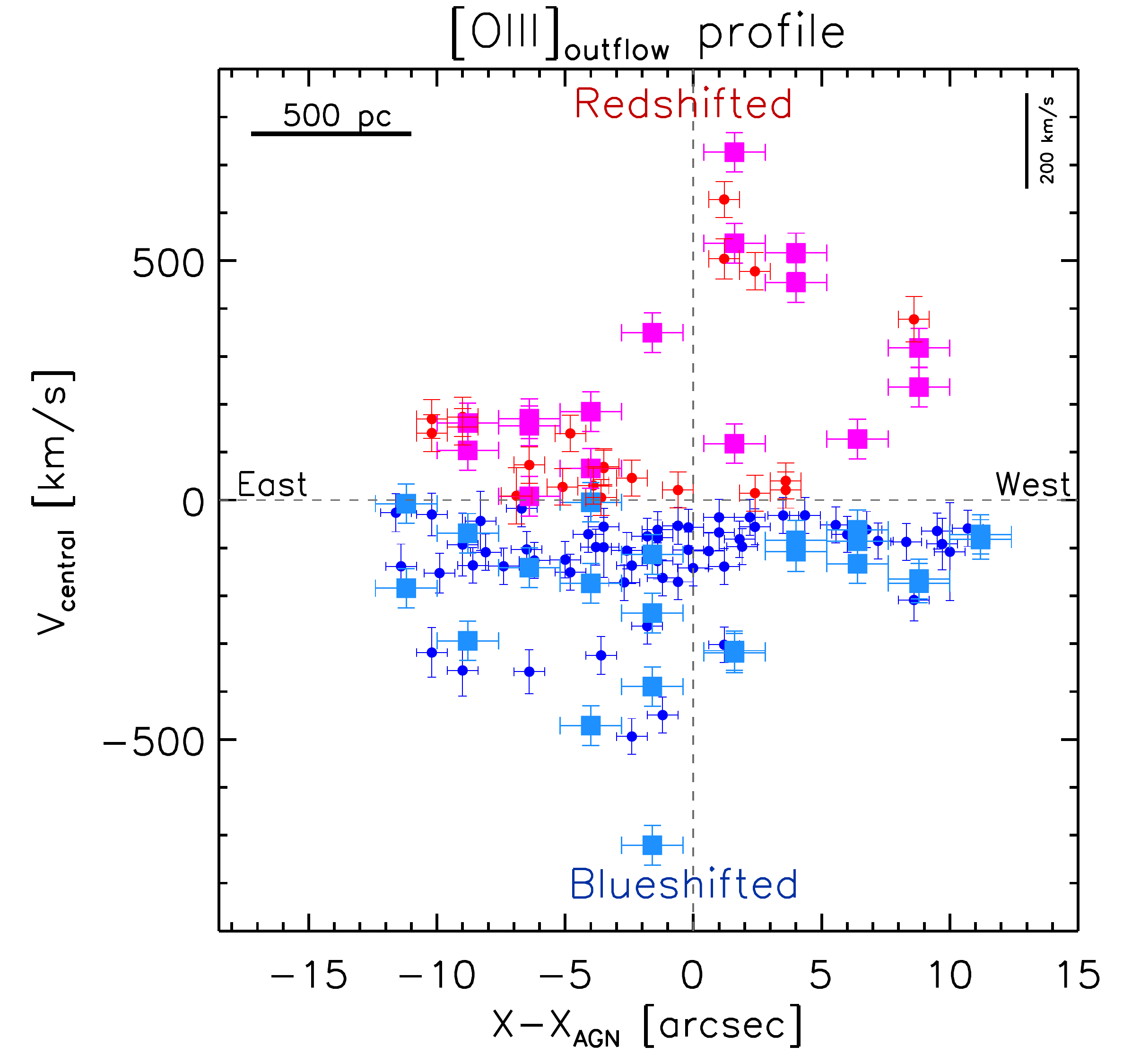}
\par}
\caption{Spatially-resolved profile of the central outflow velocities (projected) for CO(2-1) (left) and for [O\,{\sc iii}] (right). Blue and red symbols correspond to blueshifted and redshifted velocities. Note that, for comparison, we also show the central velocities of the slices (pale blue and magenta squares; see Sections~\ref{molecular_fit} and \ref{optical_fit}, and Appendix \ref{stacking}). The black circle indicates the cleared molecular gas region.} 
\label{fig7}
\end{figure*}

\section{Multiphase outflow connection}
\label{connection_kinematic}

\subsection{Spatially resolved outflow properties}
\label{Velocity_profiles}
The relation between the various phases  of AGN-driven outflows has been previously studied in active galaxies from the integrated approach and generally not for the same galaxies (e.g. \citealt{Fiore17}). In this section we examine the spatially resolved properties of the outflow of NGC~5643 in the molecular and ionized phases. Fig.~\ref{fig7} shows the spatial profile of the outflowing V$_{\rm central}$ (projected) for the ionized phase (right panel) and the central outflowing velocities (projected) for the molecular gas (left panel) phase along the east-west direction. The ionized outflow shows the largest outflow velocities relatively close (projected distances of approximately 400~pc) to the AGN, with the $V_{\rm central}$ (projected) reaching $\sim$720~km s$^{-1}$ blueshifted to the east of the AGN and redshifted to the west. 
On the eastern side of the galaxy, the outflowing ionized gas is decelerated rapidly from $\sim$720~km s$^{-1}$ to 100-200~km s$^{-1}$ (projected velocities) at 5--6\arcsec (projected distances of $\sim$400--500~pc). The electron density map (see e.g. Fig. B1 of \citealt{Mingozzi19}) reveals higher values of the electron density in this region, which could be related with the compression effect of the outflow-induced shocks (e.g. \citealt{VillarMartin14,VillarMartin15,Arribas14}). This region also coincides with the region depleted of molecular gas, first noticed in the integrated CO(2-1) map. This suggests that this is the region where the AGN-driven wind impacted the host galaxy. See Section \ref{wind_entrained} for further discussion.

Given that the signatures of the gas outflow are not strong in the individual regions, to derive the outflow properties we used of the slices analyzed in Sections~\ref{molecular_fit} and \ref{optical_fit} (see also Appendix \ref{stacking}). Since we assume that the molecular outflow is taking place in the galaxy disk, we can calculate their deprojected non-circular velocities as:

\begin{equation}
v_{\rm out}=v_{\rm central}/(\rm sin{\rm~i} ~\rm sin~\Psi)
\label{mass_o3}
\end{equation}

\noindent where i is the inclination angle of the galaxy disk and ~$\Psi$ is the phase angle measured in the galaxy plane from the receding side of the line of nodes. The median deprojected outflowing velocities of the cold molecular gas is $\sim$189~km s$^{-1}$. 

As previously discussed, the nature of the ionized phase of NGC\,5643 is 3D with a hollow ionization cone. Therefore, we use the outflow geometry of NGC\,5643 derived from the NLR modelling by \citet{Fischer13} to correct the ionized outflow velocities. We deprojected the central velocities of the ionized outflow taking the inclination angles of each wall of the cone. Following \citet{Fischer13}, the inclination angle of the bicone with respect to the plane of the sky is 25$^{\circ}$ and the hollow ionization bicone has internal and external semi-opening angles of 50$^{\circ}$ and 55$^{\circ}$, respectively. Therefore, the correction factor is more important for redshifted (blueshifted) velocities in the eastern (western) side of the cone (angle of 27.5$^{\circ}$). However, the applied correction factor for the blueshifted (redshifted) velocities in the eastern (western) side of the cone is small (angle of 77.5$^{\circ}$). The deprojected outflowing velocities of the ionized gas reach velocities (blueshifted and redshifted) of up to $\sim$750~km~s$^{-1}$ at both sides of the galaxy.

To calculate the mass outflow rate of the ionized phase in each slice we used the extinction-corrected  [O\,{\sc iii}] luminosities of the outflowing component and derived electron densities using the same methodology as in \citet{Fiore17}:

\begin{equation}
\rm M_{\rm \rm out}^{\rm [O\,III]} =  4.0\times 10^{7}~\rm M_\odot \left(\frac{C}{10^{\rm O/H}} \right)\left(\frac{\rm L_{[O\,III]~{\rm out}}}{10^{44}~{\rm erg/s}} \right)\left(\frac{n_{\rm e}}{1000~{\rm cm^{-3}}} \right)^{-1}
\label{mass_o3}
\end{equation}

\noindent where $L_{\rm [O\,III]~out}$ is the  [O\,{\sc iii}] emission line luminosity in units of 10$^{44}$~erg~s$^{-1}$, n$_{\rm e}$ is the electron density in the ionized gas clouds in units of 10$^3$~cm$^{-3}$ (see Table \ref{tab_o3} of Appendix \ref{stacking}), 10$^{\rm [O/H]}$ is the oxygen abundance in solar units ([O/H]$_\odot\sim$8.86; \citealt{Centeno08}) and C is the \textit{condensation factor}. This last parameter can be approximated as C$=$1 under the assumption that all ionizing gas clouds have the same density (see e.g. \citealt{Cano12} for further details). For the oxygen abundance we used the median value of [O/H]$=$9.0 derived for the NLR of NGC\,5643 \citep{Storchi-Bergmann98}. Note that this estimate also assumes a fully ionized gas with an electron temperature of 10$^4$~K. We finally assumed that the total ionized gas mass is $3\times M_{\rm [O\,III]}$ \citep[see][]{Fiore17}. The largest source of uncertainty in this measure is related to the electron density (see e.g. \citealt{Harrison18}). 

For the outflowing molecular gas mass we use the CO(2-1) fluxes. Since the CO-to-H$_2$ conversion factor is poorly constrained for Seyfert galaxies, we took the Galactic conversion factor ($\alpha_{\rm CO}=$M$_{gas}$/L$_{CO}^{'}$=4.35~M$_{\sun}$[K~km~s$^{-1}$~pc$^2$]$^{-1}$; \citealt{Bolatto13}) and a the CO(1-0)/CO(2-1) brightness temperature ratio of one. Note that the molecular gas mass depends on the assumed conversion factor, $\alpha_{\rm CO}$, and CO(1-0)/CO(2-1) ratio. Some previous works also used the conversion factor of ULIRGs ($\alpha_{\rm CO}$=0.8~M$_{\sun}$[K~km~s$^{-1}$~pc$^2$]$^{-1}$; \citealt{Bolatto13,Cicone14,Lutz20}), therefore, we discuss the results using $\alpha_{\rm CO}^{\rm Galactic}$ and $\alpha_{\rm CO}^{\rm ULIRGs}$ conversion factors. The latter would reduce our estimation in a factor of $\sim$5, and, therefore, the results do not change significantly (see Section \ref{appendixD}). We used the methodology by \citet{Solomon05} to estimate the molecular gas mass that includes a correction factor (36\%) to include the He mass.  The assumed uncertainty of the molecular gas mass using the Galactic conversion factor is $\sim$30\% (e.g. \citealt{Bolatto13}).

The mass outflow rate for both molecular and ionized phases can be calculated as: 

\begin{equation}
\dot{\rm M}_{\rm out} =  \left(\frac{\rm M_{\rm out}~v_{\rm out}}{\Delta D} \right)
\label{mass_rate}
\end{equation}

\noindent where M$_{out}$ is the outflowing mass, $v_{out}$ is the outflowing velocity and $\Delta$D is the size of our slices (2.4\arcsec$\sim$192~pc). The outflow kinetic power and momentum are:

\begin{figure*}[ht!]
\centering
\par{
\includegraphics[width=8.0cm]{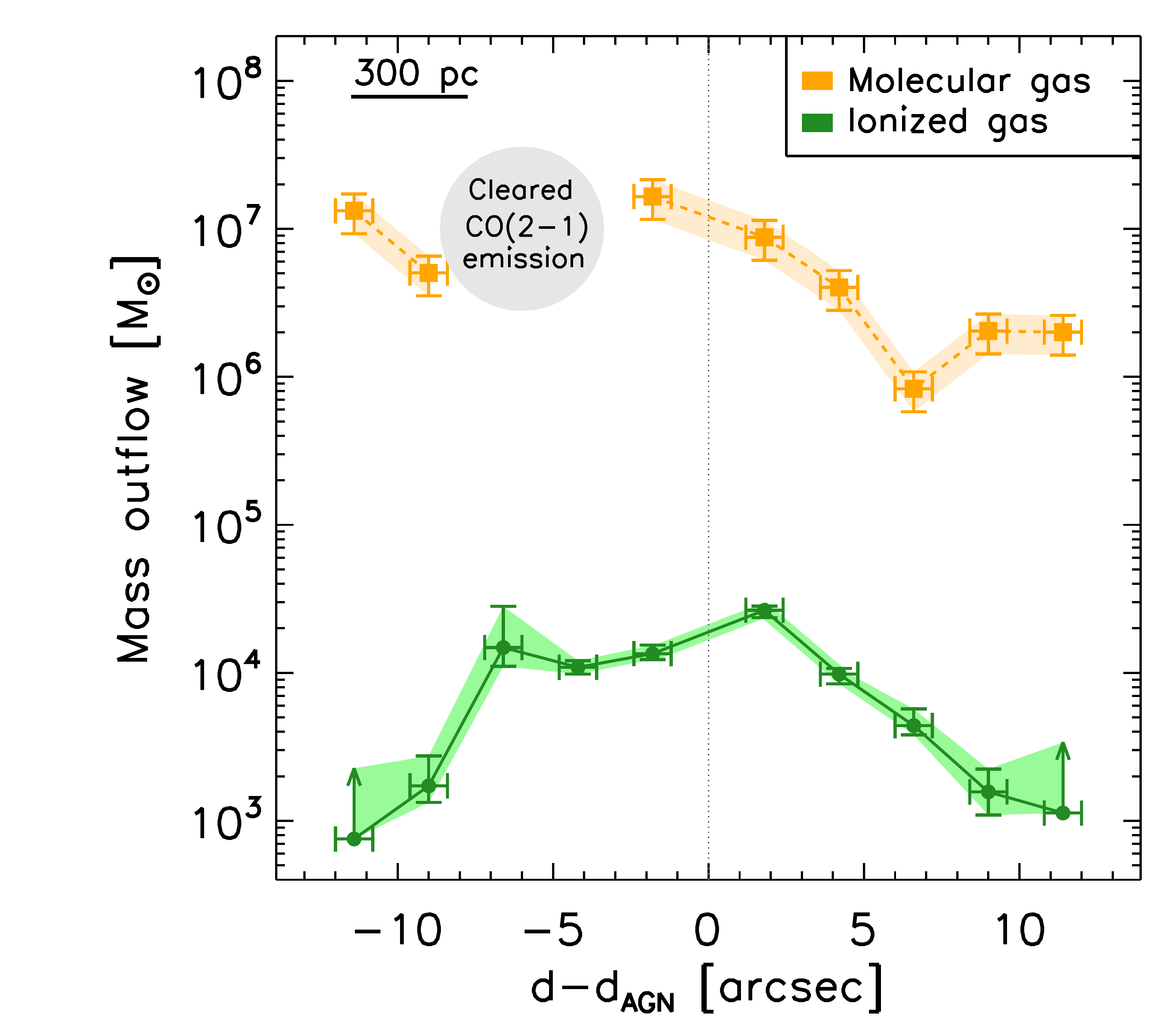}
\includegraphics[width=8.0cm]{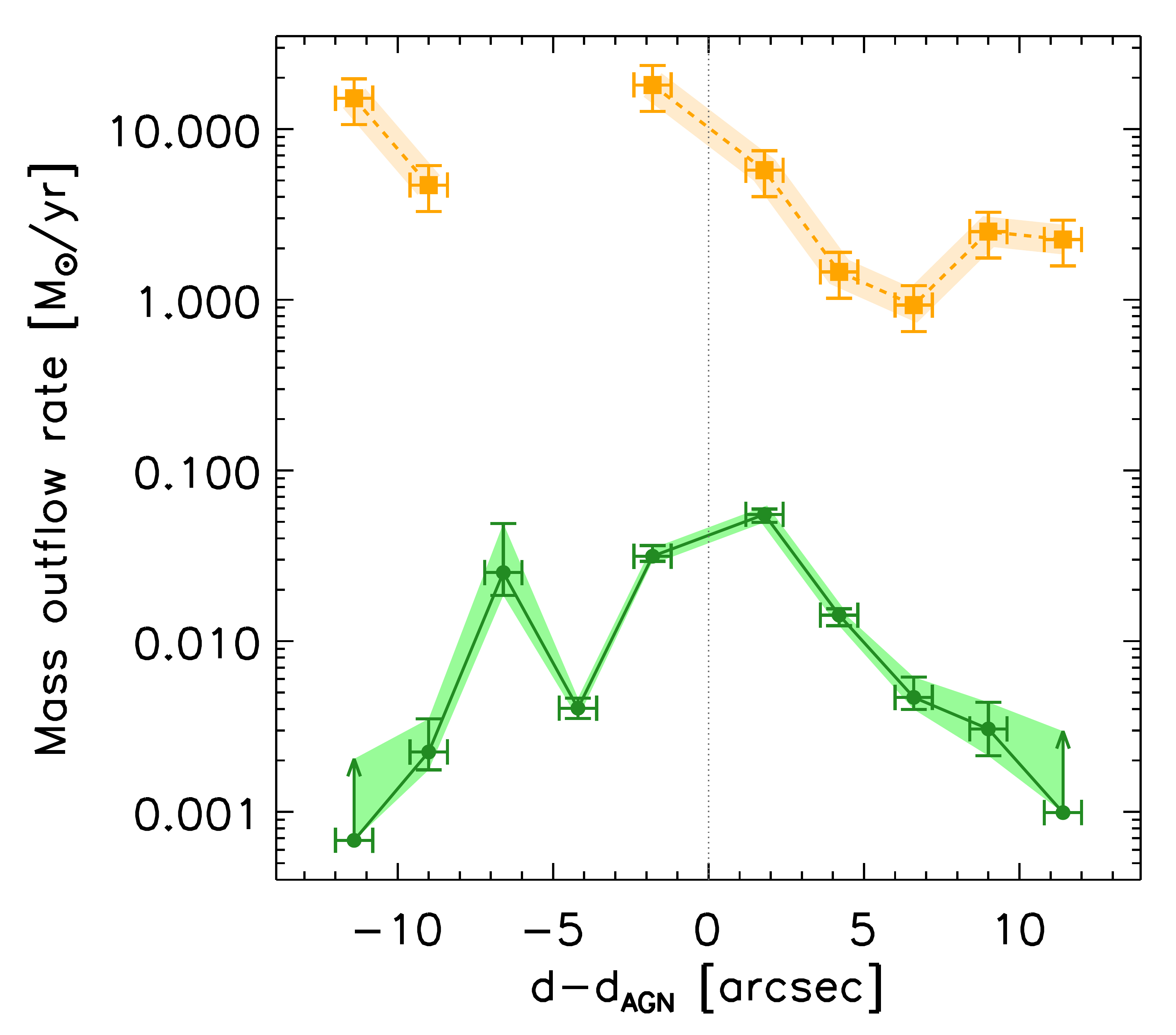}
\includegraphics[width=8.0cm]{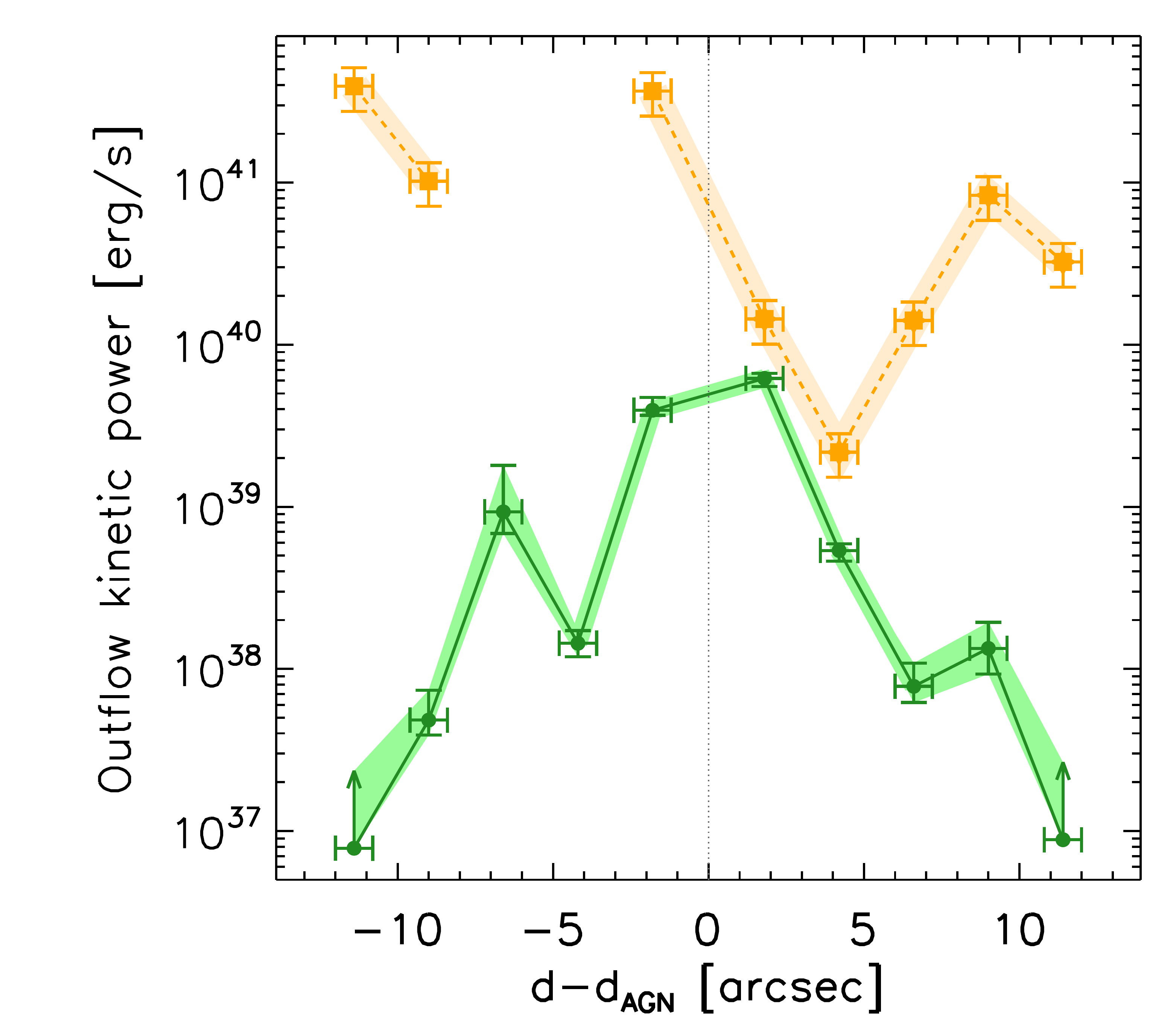}
\includegraphics[width=8.0cm]{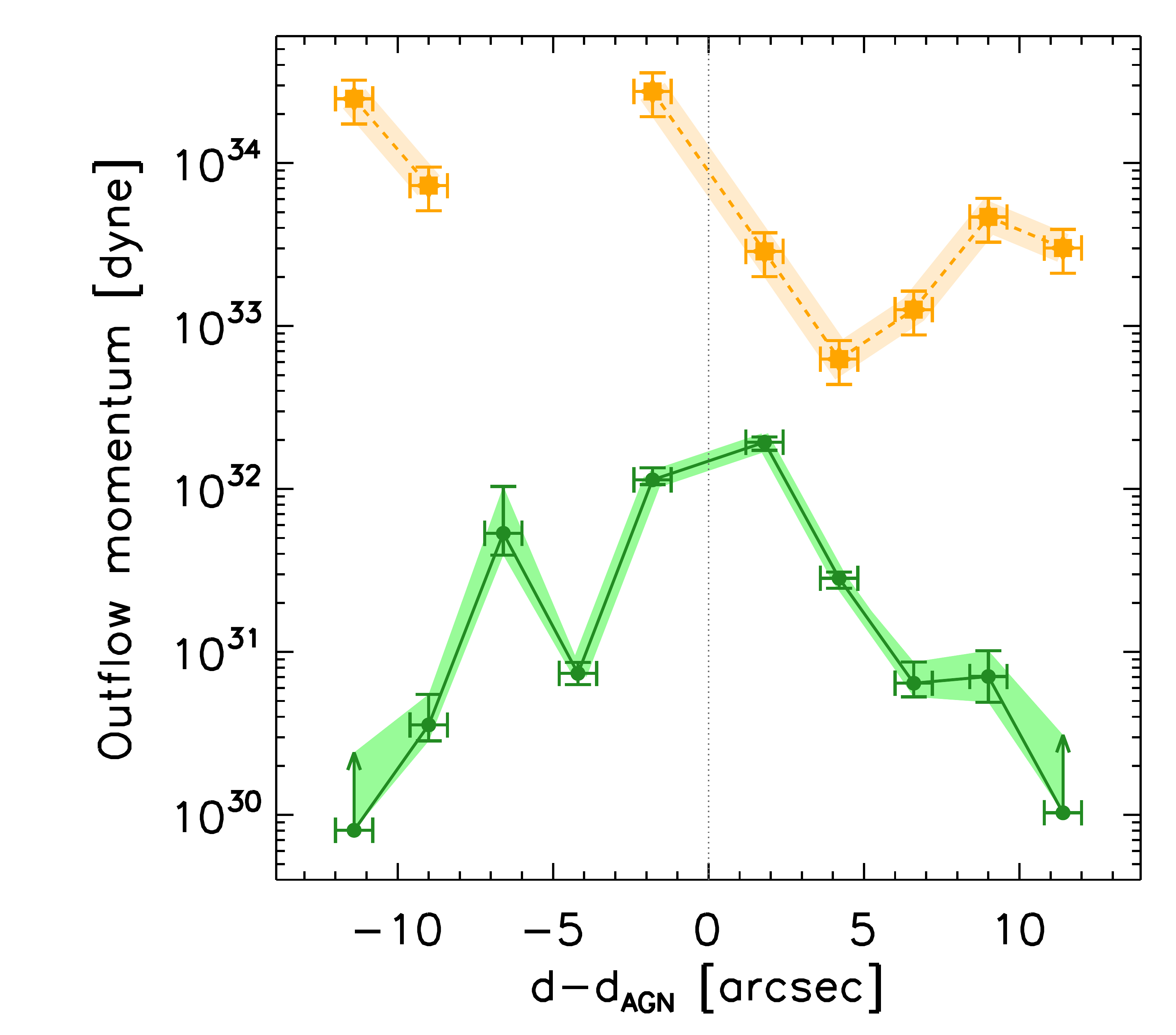}
\par}
\caption{Spatially resolved properties of the ionized (green circles) and molecular (orange squares) phases of the outflow:  outflowing mass (top left), 
mass outflow rate (top right), outflow kinetic power (center left), outflow momentum (center right). Orange squares and green circles correspond to the molecular and ionized outflow gas, respectively. Note that the two regions to the east of the AGN (negative d-d$_{AGN}$ values) with no molecular outflow derived properties correspond to regions with cleared CO(2-1) emission.} 
\label{fig10}
\end{figure*}

\begin{figure*}
\centering
\par{
\includegraphics[width=8.6cm]{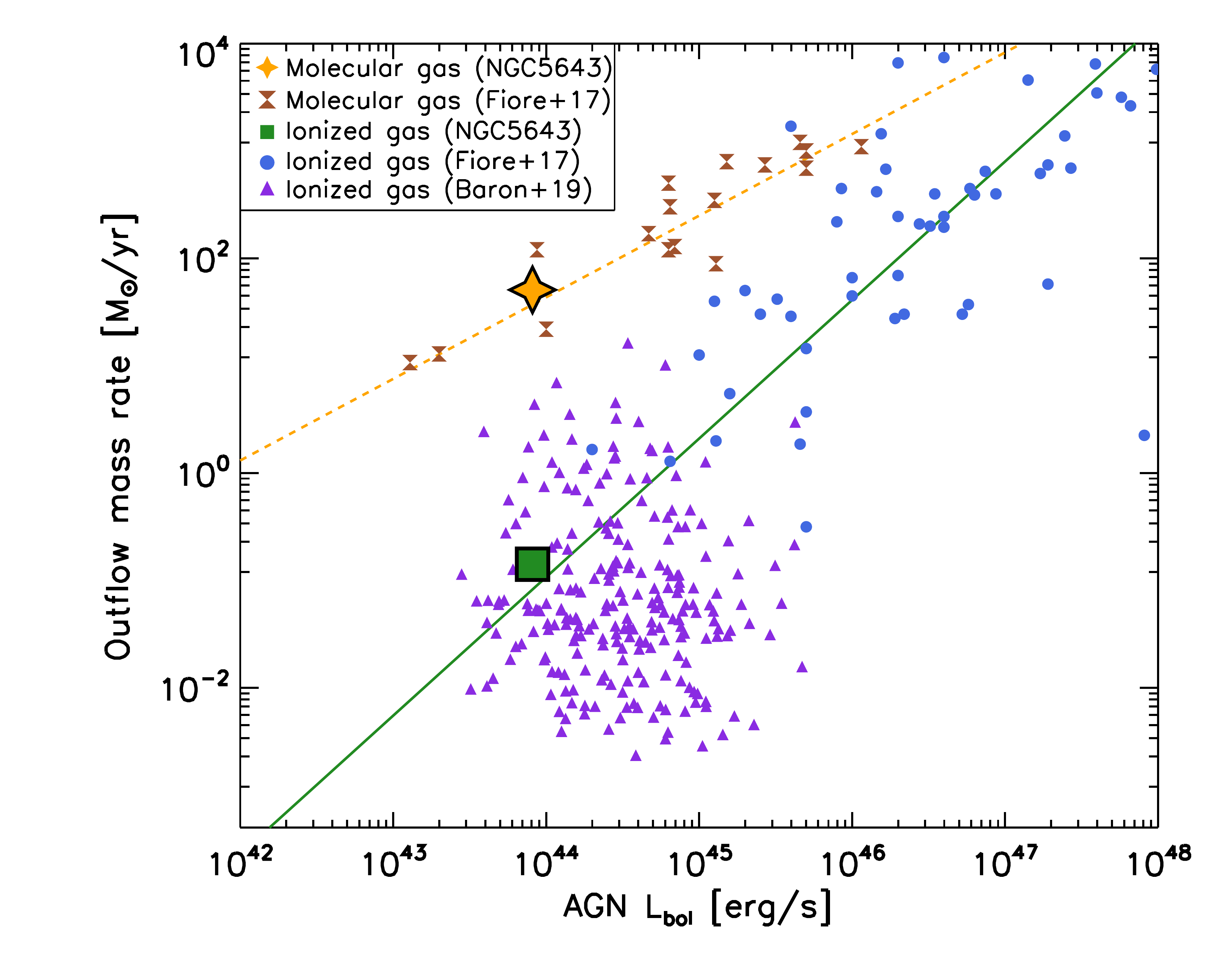}
\includegraphics[width=8.6cm]{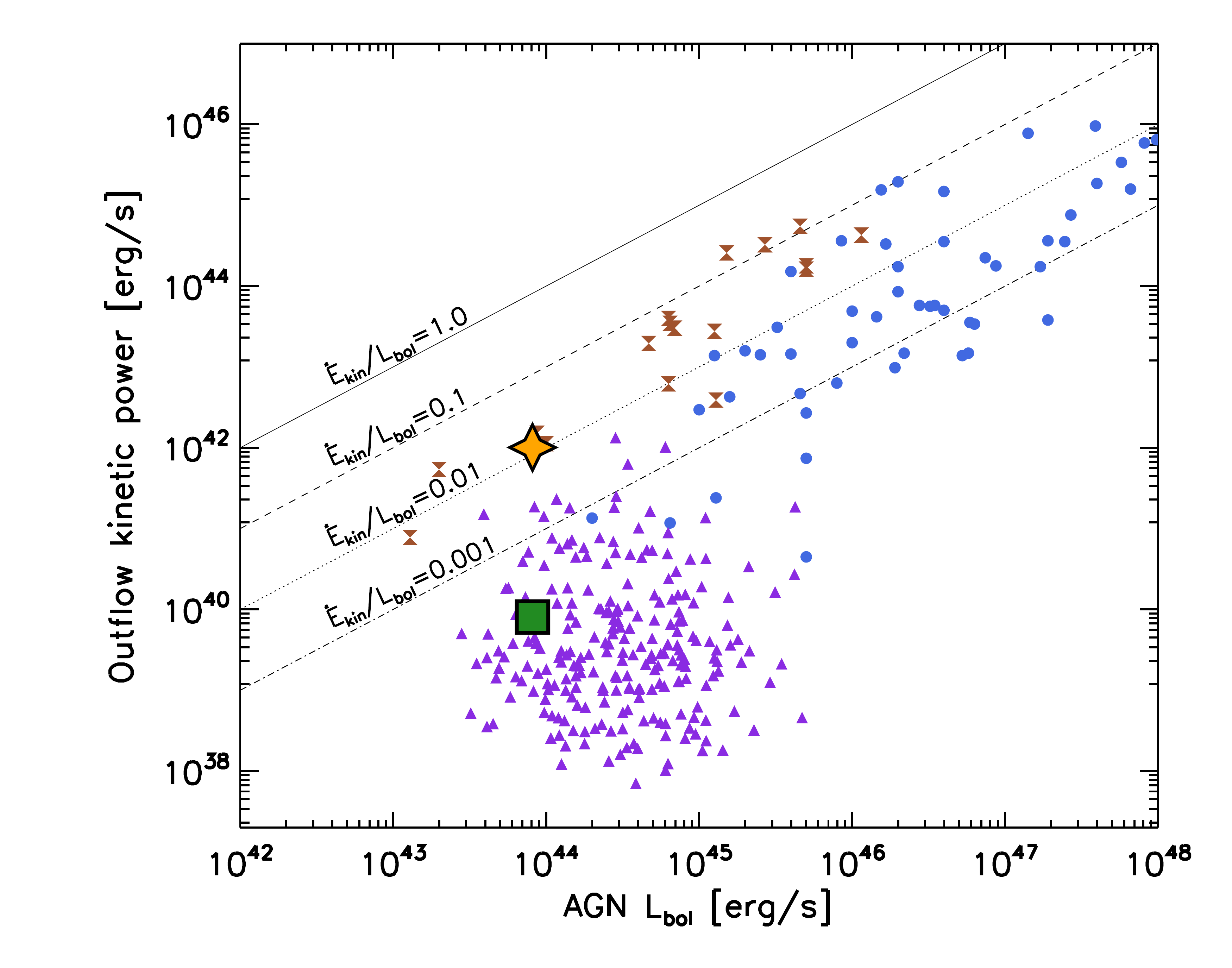}
\includegraphics[width=8.6cm]{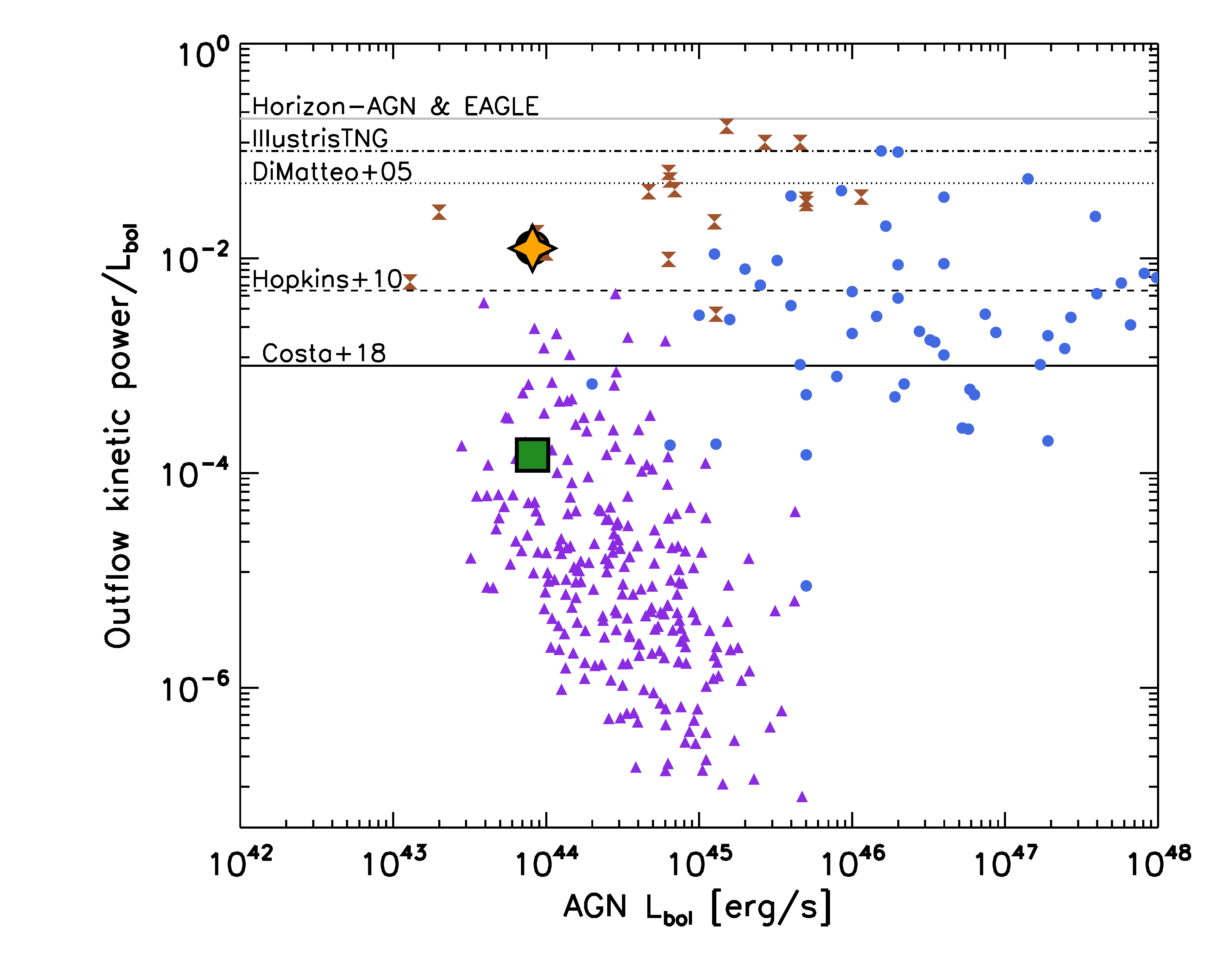}
\par}
\caption{Top left panel: outflow mass rate as a function of the AGN luminosity. The dashed orange and solid green lines are the best fit correlations derived by \citet{Fiore17} for the molecular and ionized gas, respectively. Top right panel: same as  top left panel, but for the outflow kinetic power. Solid, dashed, dotted and dash-dotted lines represent  $\dot{E}_{kin}=$1.0,0.1,0.01,0.001~L$_{bol}$. Bottom right: kinetic coupling efficiencies. The various horizontal lines correspond to theoretical values (\citealt{Costa18,DiMatteo05,Dubois14,Hopkins10,Schaye15,Weinberger17}). Orange stars and green squares represent the values derived in this work for the molecular and ionized phase of NGC\,5643, respectively. The black circle is the total kinetic coupling efficiencies for both (ionzed and molecular) gas phases.  Brown hourglass and blue circles are from \citet{Fiore17} for the molecular and ionized phase, respectively, and the purple triangles from \citet{Baron19} for ionized outflows. Note that we have consistently applied the same methodology as in \citet{Fiore17} for  the total ionized gas mass, as $3\times  M_{\rm out}^{\rm [O\,III]}$.}
\label{fig15}
\end{figure*}

\begin{equation}
\dot{\rm E}^{\rm kin} =  \frac{1}{2}~\dot{\rm M}_{\rm out}~v^2_{\rm out}= 3.2\times10^{35}  \left(\frac{\dot{\rm M}_{\rm out}}{\rm M_{\odot}~yr^{-1}} \right) \left(\frac{v_{\rm out}}{\rm km~s^{-1}} \right)^2~{\rm erg~s^{-1}}
\label{kinet}
\end{equation}

\begin{equation}
\dot{\rm P} =  \dot{\rm M}_{\rm out}~v_{\rm out}= 6.3\times10^{30}  \left(\frac{\dot{\rm M}_{\rm out}}{\rm M_{\odot}~yr^{-1}} \right) \left(\frac{v_{\rm out}}{\rm km~s^{-1}} \right)~{\rm dyne}
\label{momentum}
\end{equation}

Figure~\ref{fig10} shows the spatially resolved properties of the outflow on both sides of the AGN for the ionized and the molecular phases. The spatial profiles of the outflowing mass for both phases are clearly different. Indeed, the maximum outflowing mass in the eastern side for the ionized phase coincides with the minimum (or lack) of outflowing molecular gas on the eastern side of the AGN. The outflowing gas masses are however larger for the molecular phase than for the ionized, with a ratio between them of $\sim$600 (110 for a ULIRG $\alpha_{\rm CO}$), which is larger than the uncertainties. Similarly, the molecular gas mass outflow rate profile has larger values than those of the ionized phase (see top right panel of Fig. \ref{fig10}), but in this case the ratio between the ionized and molecular phase is lower $\sim$360 (70 for a ULIRG $\alpha_{\rm CO}$).

The profiles of the outflow kinetic power and momentum (see central panels of Fig. \ref{fig10}) of both phases show a decrease at $\sim$5-10\arcsec ($\sim$400--800~pc). However, the kinetic power and momentum profiles in the ionized phase show a local minimum in the eastern side at $\sim$5\arcsec$ (\sim$400~pc) where the AGN wind is probably impacting the galaxy strongly. We find tentative evidence of clearing of the molecular gas due to the interaction of the AGN wind with molecular gas in the galaxy (see Section \ref{wind_entrained}) including the outflowing component coinciding with the location where the AGN wind is decelerated. Indeed, the profile of the eastern part shows a lack of outflowing molecular mass in the cleared gas region. Interestingly, in this side of the outflow, the farthest slice regions show higher values of the outflowing molecular mass content which is comparable to those in the inner region. This is in agreement with the clearing and pushing gas scenario proposed here. 

\begin{table*}[ht]
%\small 
\centering
\begin{tabular}{lccc}
\hline
Integrated outflow properties		&Eastern &Western & Total\\
			&side    &side	  &\\
 (1)&(2)&(3)& (4)\\	
			
\hline
Molecular gas outflow mass (M$_{\odot}$)			& (3.5$\pm1.0$)$\times$10$^7$ & (1.8$\pm0.5$)$\times$10$^7$&(5.2$\pm1.6$)$\times$10$^7$\\
 \noalign{\smallskip}
Molecular gas outflow mass rate (M$_{\odot}$ yr$^{-1}$)	& 38.0$\pm11.4$				&	12.9$\pm3.9$& 50.9$\pm15.3$\\
 \noalign{\smallskip}
Molecular gas outflow kinematic power (erg s$^{-1}$)& 
(8.6$\pm2.6$)$\times$10$^{41}$&(1.5$\pm0.4$)$\times$10$^{41}$ &(1.0$\pm0.3$)$\times$10$^{42}$ \\
 \noalign{\smallskip}
Molecular gas outflow momentum (dyne)& 
(6.0$\pm1.8$)$\times$10$^{34}$&(1.2$\pm0.4$)$\times$10$^{34}$ &(7.2$\pm2.2$)$\times$10$^{34}$ \\
 \noalign{\smallskip}
Ionized gas outflow mass  (M$_{\odot}$)			&  (4.2$\pm_{0.4}^{1.4}$)$\times$10$^4$ & (4.3$\pm_{0.3}^{0.2}$)$\times$10$^4$&(8.5$\pm_{0.7}^{1.6}$)$\times$10$^4$\\
 \noalign{\smallskip}
Ionized gas outflow mass rate  (M$_{\odot}$ yr$^{-1}$)	& 0.06$\pm_{0.01}^{0.02}$				&	0.08$\pm0.01$    &0.14$\pm_{0.01}^{0.03}$\\
 \noalign{\smallskip}
Ionized gas outflow kinematic power (erg s$^{-1}$)& (5.1$\pm_{0.4}^{1.2}$)$\times$10$^{39}$&(6.9$\pm_{0.7}^{0.5}$)$\times$10$^{39}$ &(1.2$\pm_{0.1}^{0.2}$)$\times$10$^{40}$ \\
 \noalign{\smallskip}
Ionized gas outflow momentum (dyne)& 
(1.8$\pm_{0.2}^{0.5}$)$\times$10$^{32}$&(2.4$\pm0.2$)$\times$10$^{32}$ &(4.2$\pm_{0.4}^{0.7})\times$10$^{32}$ \\
 \noalign{\smallskip}
%\hline
%SFR (M$_{\odot}$ yr^{-1}) & 0.41$\pm0.10$ & 0.25$\pm0.06$ & 0.66$\pm0.12$  \\ 
 %\noalign{\smallskip}
\hline
\end{tabular}						 
\caption{Summary of the molecular and ionized outflow properties. Note that the molecular outflow properties reported here were calculated using the Galactic CO-to-H$_2$ conversion factor. In Appendix \ref{appendixD} we also report these properties using the ULIRGs CO-to-H$_2$ conversion factor.}
\label{tab1}
\end{table*}

\subsection{Integrated Outflow Properties}
\label{integrated_outflow}
Table~\ref{tab1} reports the main integrated outflow properties on both sides of the nucleus. The molecular gas outflowing mass of the eastern side, ${\rm M}_{\rm out,molecular} \sim3.5\times10^7$~M$_{\odot}$
 ($0.6\times10^7$~M$_{\odot}$ for a ULIRG $\alpha_{\rm CO}$), is
roughly twice that of the western one. However, in the ionized phase the outflowing mass in the eastern region (${\rm M}_{\rm out,ionized} \sim5.1\times10^4~M_{\odot}$) is comparable to that in the western side (see top left panel of Fig. \ref{fig10}). Note that in Appendix  \ref{appendixC} we also present the results for the observed values (i.e. without using the extinction correction). 

The molecular gas outflow kinetic power and momentum are $\sim$80 and $\sim$170 ($\sim$15 and $\sim$30 for a ULIRG $\alpha_{\rm CO}$) times higher than the values of the ionized phase. Moreover, the eastern part of the galaxy appears to be more affected by the AGN-driven outflow based on the lack of molecular gas in this side of the galaxy (see Section \ref{Morphology} and \ref{Velocity_profiles}). Interestingly, \citet{Cresci15} also found in this part of the galaxy evidence of positive AGN feedback in the star forming clumps at $\sim$6--8\arcsec \, ($\sim$500-650~pc). All this suggests that outflows in Seyfert galaxies as that studied here for NGC\,5643 can produce both positive and negative (mild) feedback processes. 

Finally, in Figure \ref{fig15} we present a comparison between the total derived properties for NGC~5643 and the properties for other AGN (e.g. \citealt{Fiore17} and \citealt{Baron19}). For the AGN luminosity of NGC~5643 (L$_{bol}$=8.14$\times10^{43}$~erg~s$^{-1}$; \citealt{Ricci17})\footnote{We calculated the bolometric luminosity using the intrinsic 14-195~keV Swift/BAT luminosity from \citet{Ricci17} and a fixed bolometric correction (L$_{\rm bol}$/L$_{14-195~{\rm keV}}$=7.42; \citealt{Bernete19}).}, the derived outflow mass and kinematic rates for both gas phases are compatible with the values derived for other AGN. The total momentum rates for the molecular and ionized gas are $\dot{P}_{out}\sim$7.2$\times$10$^{34}$ ($\sim$1.3$\times$10$^{34}$ for a ULIRG $\alpha_{\rm CO}$) and~$\sim$4.2$\times$10$^{32}$ dyne, respectively. These values are similar to the radiation momentum rate expected for the AGN in NGC\,5643 ($\dot{P}_{AGN}=$L$_{bol}$/c$=$2.72$\times$10$^{33}$ dyne). Therefore, the wind momentum load ($\dot{P}_{out}/\dot{P}_{AGN}$) for the molecular and ionized outflow phases are $\sim$27 ($\sim$5 for a ULIRG $\alpha_{\rm CO}$) and $<1$, which are consistent with the derived values of other AGN of similar luminosity \citep{Fiore17}. This indicates that the outflow molecular phase is not momentum conserving, while the ionized one most certainly is ($\dot{P}_{out}/\dot{P}_{AGN}<$1). This increase of the molecular phase momentum implies that part of the kinetic energy from the AGN wind is transmitted to the molecular outflow. These results suggest that radiative and kinetic AGN feedback modes coexist in NGC\,5643. 

In addition, the presence of a radio jet in NGC\,5643 could play a role injecting outflow power. Following the methodology presented in \citet{Birzan08} using the monochromatic luminosity at 1.4~GHz of the NGC\,5643 jet \citep{melendez10} we find that the jet power is  $\sim$2--10$\times$P$_{\rm kin}^{ \rm \:molecular}$ and $\sim$160$\times$P$_{\rm kin}^{ \rm \:ionized}$. This indicates that the jet of NGC\,5643 can contribute to driving the molecular and ionized outflows efficiently.

On the other hand, the bottom panel of Figure \ref{fig15} shows the kinetic coupling efficiencies for molecular and ionized outflows. Although the various assumptions and methods employed in the literature might be responsible for some of the scatter, there is a large range of values for a fixed AGN bolometric luminosity in this plot ($\sim$eight orders of magnitude). It is of interest to compare the observed kinetic powers from those derived from AGN feedback models (see e.g. \citealt{Harrison18}). Indeed, the values derived in this work (see also Appendix \ref{appendixD}) for the molecular and ionized outflow of NGC\,5643 are in good agreement with those theoretical values reported by \citet{Hopkins10} and \citet{Costa18}. 

\begin{figure}[ht!]
\centering
\includegraphics[width=9.1cm]{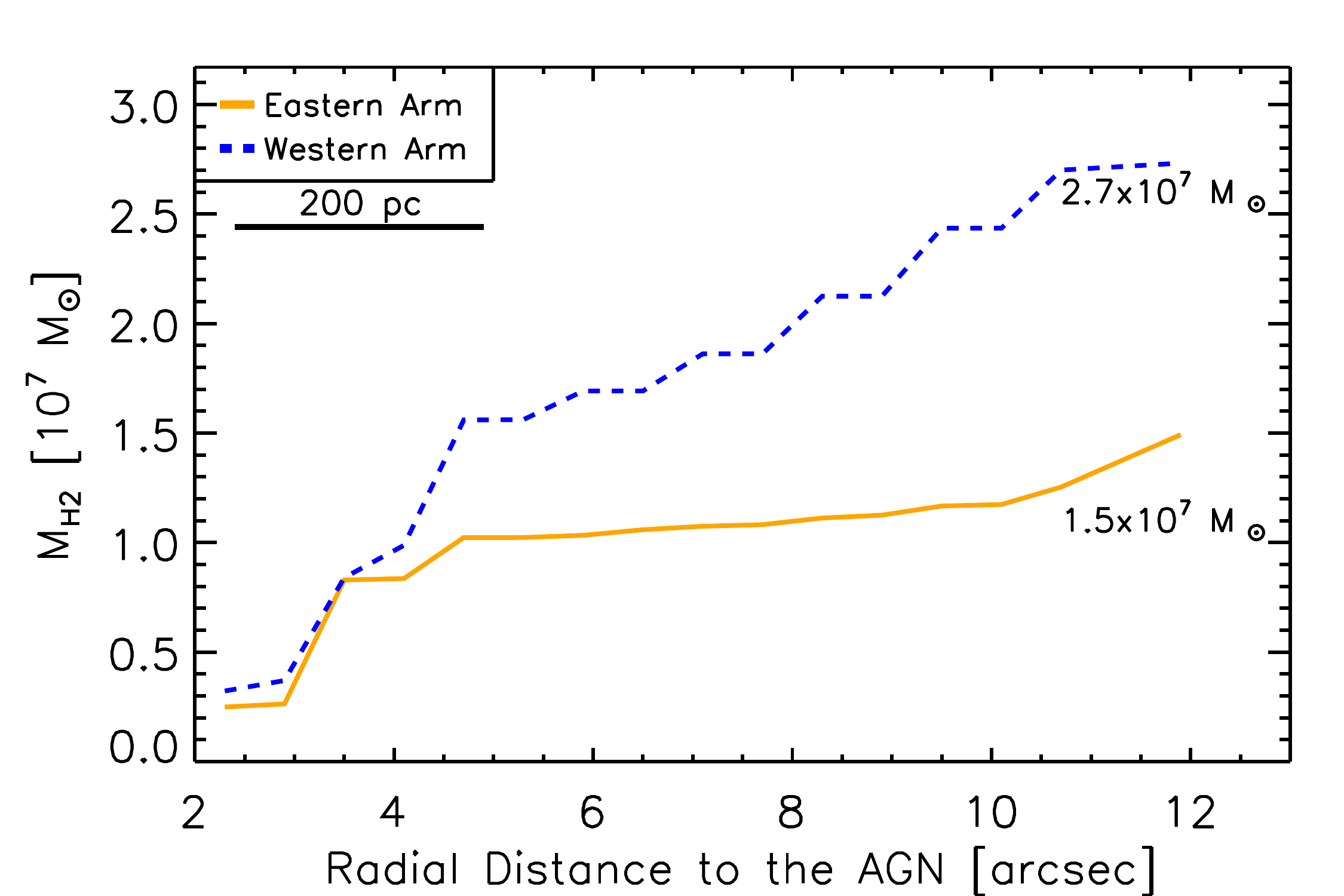}
\caption{Cumulative molecular mass profiles using the symmetric selected regions in the spiral arms on both sides of the nucleus (see Appendix \ref{spiral_arms}). Eastern and western sides are the orange solid and blue dashed lines, respectively.}
\label{fig12}
\end{figure}

\subsection{Destruction/Clearing of the Molecular Gas}
\label{wind_entrained}
As previously discussed in \citet{Herrero18}, see also Section \ref{Morphology}, there is a deficit of CO(2-1) emission on the eastern side of the spiral. This could be related to the destruction/clearing of the molecular gas produced by the AGN wind impacting on the host galaxy. Therefore, to estimate the possible impact of the outflow wind in the destruction/clearing of the molecular gas, we measure the total gas mass in both spiral arms. To do so, using the fully reduced optical HST/F606W image of NGC\,5643 from the ESA Hubble Legacy Archive\footnote{http://archives.esac.esa.int/hst/}, we define the region of the two main spiral arms. Then, using the CO(2-1) data we extract several regions following both spiral arms (see Fig. \ref{fig11}). See Appendix \ref{spiral_arms} for further details on the spiral arms region selection.

\begin{figure*}[ht!]
\centering
\includegraphics[width=14.0cm]{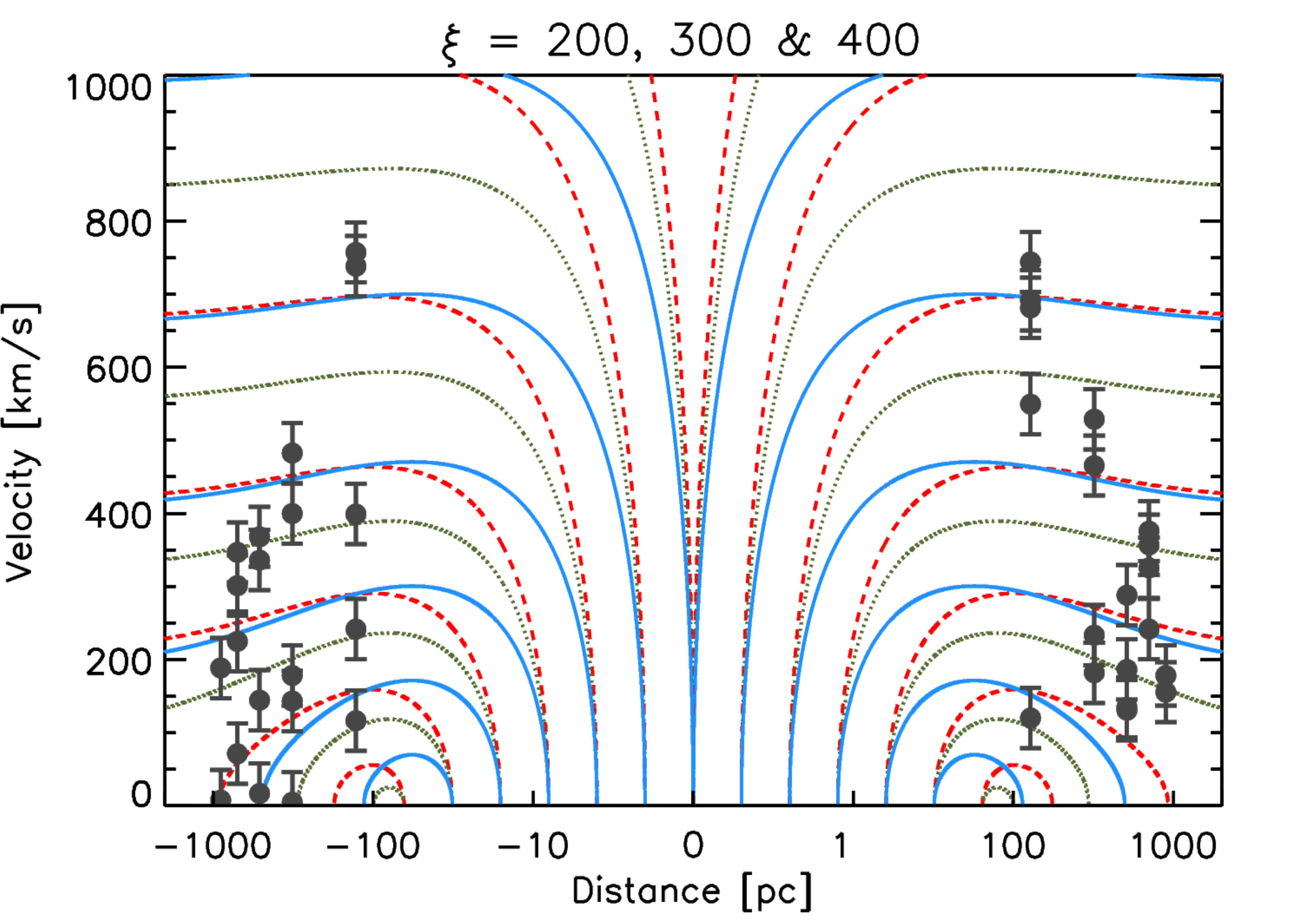}
\caption{Outlfow wind analytic model: velocity profiles for various force multipliers ($\xi=$200, 300 \& 400) and launch radii (r$_1$=1, 2, 4, 8, 16, 32 and 64~pc). Lines correspond to the various force multipliers used ($\xi$=200 -solid blue-, 300 -dotted green-, 400 -dashed red-). Black circles correspond to the ionized outflow deprojected velocities. Negative (positive) distances indicate eastern (western) radial distances to the AGN.}
\label{fig9}
\end{figure*}

In Fig. \ref{fig12} shows the cumulative molecular gas mass profile, which confirms the molecular gas deficit in the eastern part of the galaxy reveals by the CO(2-1) emission map. Indeed, the molecular gas mass of the eastern spiral arm using two different approaches (see Appendix \ref{spiral_arms}) is $M_{\rm east}\sim$(1.5-2.0)$\times$10$^7$~M$_\sun$ ($\alpha_{\rm CO}^{\rm Galactic}$). These are only 50-70\% of the molecular gas content in the western spiral arms. This should be related to the destruction/clearing of the molecular gas produced by the AGN wind strongly impacting on the eastern side of the host galaxy. The molecular gas content in the spiral arms is comparable to the outflowing molecular mass in the outflow regions. However, note that for the spiral arms we did not include regions in the nuclear regions which accounts for the bulk of the molecular gas mass both in the disk and in the molecular outflow. We also note that we calculate the deficit of molecular gas in the eastern part of the galaxy under the assumption of symmetry in the surface brightness. However, it is possible that intrinsic variations in the surface brightness due to asymmetry could affect this measurement.

\section{Outlfow Wind Analytic Model}
\label{Simulation}

NGC\,5643 is an intermediate luminosity AGN with a large outflow extending $\sim$1~kpc on each side of the nucleus. Previous works studied how the radial extent of the NLR in scales with AGN luminosity. The expected radial extent of the ionized phase depends on the relation employed. Comparing the [O\,{\sc iii}]] luminosity of NGC\,5643 ($\log$(L$_{[O\,III]}$)=40.4~erg~s$^{-1}$; \citealt{melendez08}) with Fig. 10 of \citet{Fischer18} the expected radial extent is ranging from $\sim$100--500~pc. However, this value is much larger (ranging from 0.5--2.5~kpc) for the relation with the 8~$\mu$m luminosity\footnote{The 8~$\mu$m luminosity ($\log$(L$_{8~\mu m}$)=42.2~erg~s$^{-1}$) was calculated by taking the average of a 1~$\mu$m window centred at 8~$\mu$m in the {\textit{Spitzer} spectrum (see Section \ref{spitzer}).}} presented in Fig. 3 of \citet{Hainline14}. Since the observed outflow properties are not well constrained, it is of interest to compare them (e.g. size and velocity) with those estimated from a theoretical point of view. To do so, we follow the analytic model proposed by \citet{Das07} which considers the radiation-driven wind and the gravitational drag as:

\begin{equation}
a(r) = \frac{L_{\rm bol} \sigma_T \xi}{4 \pi r^2 cm_p}-\frac{G M_{\rm tot}(r)}{r^2}
\label{ar}
\end{equation}

\noindent where L$_{\rm bol}$ is the bolometric luminosity of the AGN, $\sigma_T$ is the Thomson scattering cross section for the electron, $\xi$ is the force multiplier, r is the distance, c is the speed of light, m$_p$ is the mass of the proton, G is the universal gravitational constant and M(r) is the total enclosed mass within r. The force multiplier ($\xi$) is primarily a function of the ionization parameter ({\textit{U}}) for a given spectral energy distribution (see \citealt{Das07} and references therein for further discussion). In the latter work, the authors found from photoionization models values of the force multiplier ranging from $\xi\sim$500-6000 for the Sy2 galaxy NGC\,1068. This approach takes into account the drag forces from gravity stopping and turning back the gas outflow but not the possible deceleration of the NLR outflow due to the resistance of the ISM.

Using a bulge mass of M$_{\rm bulge}$=5.48$\times10^{9}$~M$_{\sun}$ and effective radius of 0.46~kpc (\citealt{Weinzirl09}) and a black hole mass M$_{\rm BH}$=2.75$\times10^{6}$~M$_{\sun}$ ( \citealt{Goulding10}), and following Eqs. (5), (6) and (7) of \citet{Das07}, the enclosed mass as a function of the distance from the central SMBH) is:

\begin{equation}
M_{\rm tot}(r) = 2.75 \times 10^{6} +2.37 \times 10^{9} \left(\frac{r_{pc}}{r_{pc}+345} \right)^{1.5}
\label{mr}
\end{equation} 

Then, integrating equation \ref{ar} and setting the initial velocity to zero we obtain the following velocity of the model:

\begin{equation}
v(r) = \sqrt{\int_{r1}^{r} \left[6840 \frac{L_{\rm bol}}{10^{44}} \frac{\xi}{t^2} - 8.6 \times 10^{-3} \frac{M_{\rm tot}(t)}{t^2} \right] dt}
\label{vr}
\end{equation}

Note that this model assumed a spherical symmetry, for the sake of simplicity. We used the same value of the bolometric luminosity as in Section \ref{integrated_outflow}. 

In Fig.~\ref{fig9} we show the resulting velocity profiles of NGC\,5643 for various launch radii (r$_1$=1, 2, 4, 8, 16, 32 and 64~pc) and force multiplier values ($\xi=$200, 300 \& 400), which reproduce well the ionized phase observations (see black circles in Fig. \ref{fig9}). Although we are not considering here other drag forces such as the ISM, the value of the force multiplier needed to fit the observational data is quite small compare with the maximum value derived for NGC\,1068. This means that, even when including other drag forces, it be possible to reproduce the observations by increasing the force multiplier.

As can be seen from Figure \ref{fig9}, up to a launch radius of r$_1$=4~pc the velocity of the gas quickly reaches terminal velocity, even for the smaller force multiplier (i.e., $\xi$=200). This means that it does not slow down significantly with the radii for small launch distances.  However, from larger launch radii the gas velocity starts to turn over and returns to the systemic velocities at r$\sim$100~pc for the smaller force multiplier. These results suggest that in Seyfert galaxies with intermediate AGN luminosities, bulge and BH masses similar to those of NGC\,5643, the ionized outflow can reach kiloparsec scales as observed in this galaxy.

\section{Conclusions}
\label{conclusions}
We presented a detailed study of the multiphase feedback processes in the central $\sim$3 kpc of the barred Seyfert 2 galaxy NGC\,5643. We used observations of the cold molecular gas (ALMA CO(2-1) transition) and ionized gas (MUSE IFU optical emission lines). We studied different regions along the outflow zone which extends out to $\sim$2.3 kpc in the same direction (east-west) as the radio jet, as well as nuclear/circumnuclear regions in the host galaxy disk. The main results are as follows: \\ 

   \begin{enumerate}
\item  The [O\,{\sc iii}]$\lambda$5007$\AA$ emission lines in regions in the outflow have two or more components. The narrow component  traces the rotation of the gas in the disk. The others, which have broader profiles (median $\sigma_{broad}^{[O\,III]}=$233$\pm$75~km s$^{-1}$), are related to the ionized gas outflow. The projected ionized outflow velocities in this phase reach 
$\sim720\,{\rm km \, s}^{-1}$ in the inner 830~pc ($\sim$10\arcsec), and then get decelerated at further distances from the AGN.

\item The CO(2-1) line profiles of regions in the outflow and in the nuclear/circumnuclear spiral also show two or more different velocity components. One is associated with the host galaxy rotating disk and the others with inflowing and outflowing motions. The streaming motions (local inflow) are due to the presence of a large-scale stellar bar.

\item The deprojected outflowing velocities of the cold molecular gas (median $V_{\rm central}\sim$189~km s$^{-1}$) are generally lower than those of the outflowing ionized gas, which reach deprojected velocities of up to 750~km~s$^{-1}$ (both blueshifted and redshifted) close to the AGN, and their spatial profiles follow those of the ionized phase., although the molecular phase velocity profile follows that of the ionized phase. This suggests that molecular gas in the host galaxy disk is being entrained by the AGN wind in these regions. \\

\item   The molecular and ionized outflow masses are $\sim$5.2$\times$10$^7$~M$_{\odot}$ ($\sim$1.0$\times$10$^7$~M$_{\odot}$ for a ULIRG $\alpha_{\rm CO}$) and 8.5$\times$10$^4$~M$_{\odot}$, respectively. Furthermore, the derived molecular and ionized outflow mass rates are $\sim$51 (9)~M$_{\odot}$ yr$^{-1}$ and 0.14~M$_{\odot}$ yr$^{-1}$. Thus, the molecular phase dominates the outflow mass and outflow mass rate, with ratios between 600 (110) and 360 (70) over the ionized one.\\ 

\item The outflow kinematic power for both outflow gas phases are similar. Interestingly the kinematic power shows an important decrease at projected distances $\sim$400~pc east and west from the AGN, probably where the greatest effect of the AGN wind impacting the host galaxy is taking place. The wind momentum load ($\dot{P}_{out}/\dot{P}_{AGN}$) for the molecular and ionized outflow phases are $\sim$27 (5) and $<1$, which suggests that the molecular phase is not momentum conserving while the ionized one most certainly is. \\

\item  We estimated the molecular gas mass in the two main spiral arms in the circumnuclear region of NGC\,5643. The molecular gas mass of the eastern spiral arm, $M_{\rm east}\sim$1.5$\times$10$^7$~M$_\sun$, is 50-70\% the content of the western one. This should be related, at least in part, with the destruction/clearing of the molecular gas produced by the AGN  wind impacting on the eastern side of the host galaxy.\\

\item  Using a simple analytic model of a radiation-driven AGN wind and gravitational drag, we reproduced the $\sim$2~kpc scale outflow observed in NGC\,5643. This suggests that for AGN luminosities, bulge and BH masses as in this galaxy ionized outflows can reach large scales.\\

   \end{enumerate}
   
   All of these results suggest that outflows in Seyfert-like AGN as in NGC\,5643 can produce both positive \citep{Cresci15} and negative (mild) feedback processes. In particular, in the Seyfert galaxy NGC\,5643 the molecular phase is not momentum conserving while the ionized one most certainly is. This suggests that the radiative/quasar and the kinetic/radio AGN feedback modes coexist and may shape the host galaxies even at kpc-scales. Although to date there have been only few detailed multiphase studies using various regions within the same source, it is essential to quantify the overall outflow properties of each phase. Future observations of large samples, combining high-resolution optical and submillimetre observations, will be able to characterize the typical effect of AGN feedback on the host galaxy.
   
\begin{acknowledgements}
IGB, AAH and FJC acknowledge financial support through grant PN AYA2015-64346-C2-1-P (MINECO/FEDER), funded by the Agencia Estatal de Investigaci\'on, Unidad de Excelencia Mar\'ia de Maeztu. IGB and DR also acknowledge support from STFC through grant ST/S000488/1.
DR acknowledges support from the University of Oxford John Fell Fund. AAH, SGB and MVM also acknowledge support through grant PGC2018-094671-B-I00 (MCIU/AEI/FEDER,UE). AAH, MPS, MVM and AL work was done under project No. MDM-2017-0737 Unidad de Excelencia ``Mar\'ia de Maeztu'' - Centro de Astrobiolog\'ia (INTA-CSIC). MPS acknowledges support from the Comunidad de Madrid, Spain, through Atracci\'on de Talento Investigador Grant 2018-T1/TIC-11035 and PID2019-105423GA-I00 (MCIU/AEI/FEDER,UE). BG-L acknowledges support from the State Research Agency (AEI) of the Spanish Ministry of Science, Innovation and Universities (MCIU) and the European Regional Development Fund (FEDER) under grant with reference AYA2015-68217-P. FJC and SM acknowledge financial support from the Spanish Ministry MCIU under project RTI2018-096686-B-C21 (MCIU/AEI/FEDER/UE), cofunded by FEDER funds and from the Agencia Estatal de Investigaci\'on, Unidad de Excelencia Mar\'ia de Maeztu, ref. MDM-2017-0765. CRA acknowledges support from the Spanish Ministry of Science, Innovation and Universities (MCIU), the Agencia Estatal de Investigaci\'on (AEI) and the Fondo Europeo de Desarrollo Regional (EU FEDER) under project AYA2016-76682-C3-2-P  and PID2019-106027GB-C42. CRA also acknowledges support from the MCIU under grant RYC-2014-15779. AL acknowledges the support from Comunidad de Madrid through the Atracción de Talento grant 2017-T1/TIC-5213. CR acknowledges support from the Fondecyt Iniciacion grant 11190831.

This paper makes use of the following ALMA data: ADS/JAO.ALMA$\#$2016.1.00254.S. This work is based [in part] on archival data obtained with the Multi Unit Spectroscopic Explorer (MUSE) on the Very Large Telescope (VLT) under ESO programme 095.B-0532(A), and the Spitzer Space Telescope, which is operated by the Jet Propulsion Laboratory, California Institute of Technology under a contract with NASA. This research has also made use of the NASA/IPAC Extragalactic Database (NED), which is operated by the Jet Propulsion Laboratory, California Institute of Technology under a contract with NASA.

Finally, we are extremely grateful to the anonymous referee for useful comments. 

\end{acknowledgements}

%-------------------------------------------------------------------

%\Online

\begin{appendix} %First online appendix
\section{Stacking regions}
\label{stacking}

\begin{figure*}[ht!]
\centering
\includegraphics[width=13.1cm,angle=90]{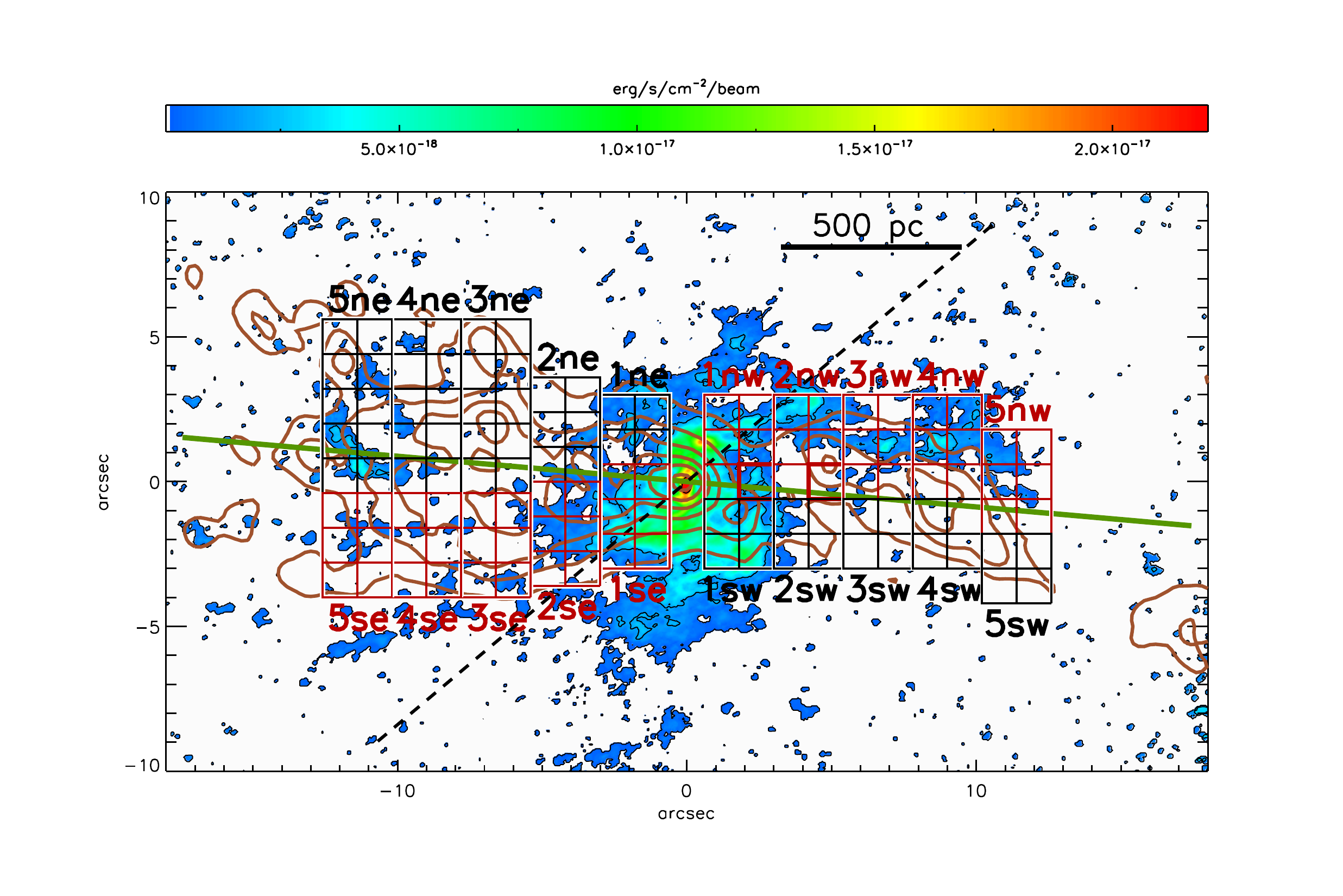}
\caption{ALMA CO(2-1) integrated intensity map of NGC\,5643 produced from the natural-weight data cube. The map is shown in a linear color scale. Black contours corresponds to the CO(2-1) emission map, which are shown in a logarithmic scale with the first contour at 3$\sigma$ and the last contour at 2.2$\times$10$^{-17}$~erg/s/cm$^{-2}$/beam. Brown contours correspond to the [O\,{\sc iii}]$\lambda$5007$\AA$ emission map (see Section \ref{optical_fit}).The dashed black line indicates the kinematic major-axis. The green solid line corresponds with the direction of the large scale stellar bar \citep{Mulchaey97}. Red and black boxes correspond to regions where streaming motions (local inflow) are and are not expected, respectively. For instance, region {\textit{5ne}} consists of 2$\times$5 black boxes (1.2$\arcsec\times$1.2$\arcsec$ square apertures), and the {\textit{eastern 5}} full slice corresponds to {\textit{5ne$+$5se}}.}
\label{stacking_slices}
\end{figure*}

\begin{figure*}
\centering
\par{
\includegraphics[width=4.5cm]{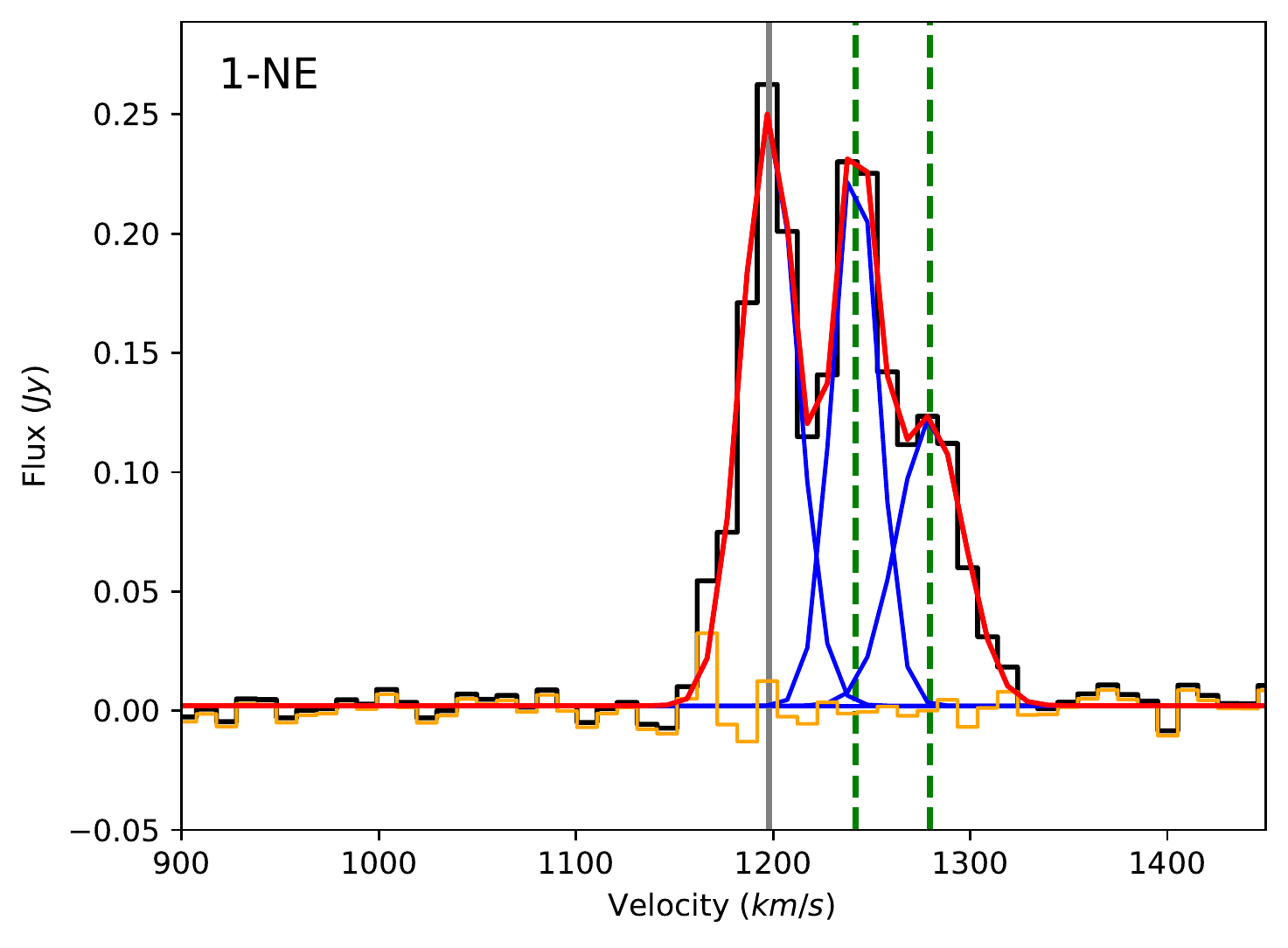}
\includegraphics[width=4.5cm]{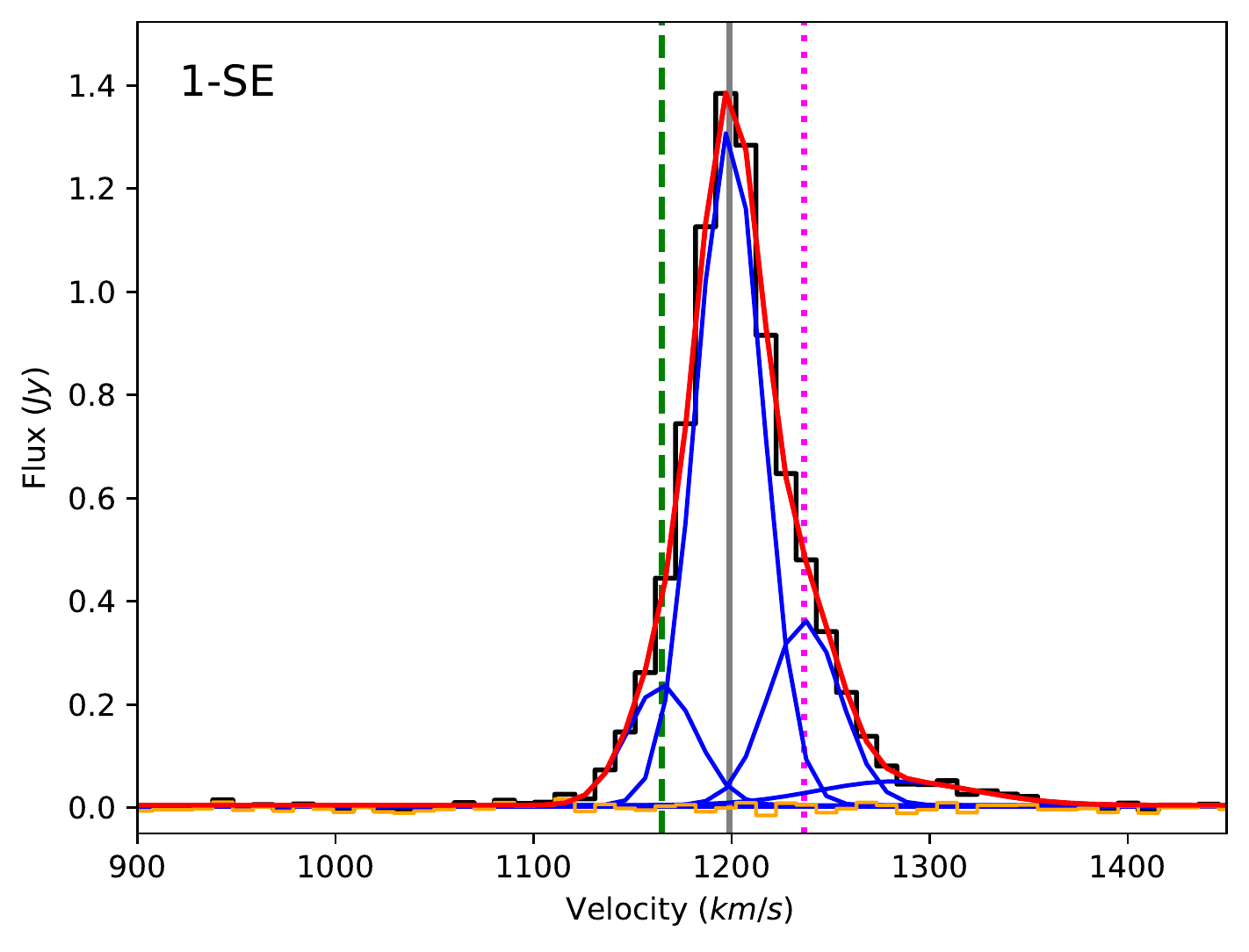}
\includegraphics[width=4.5cm]{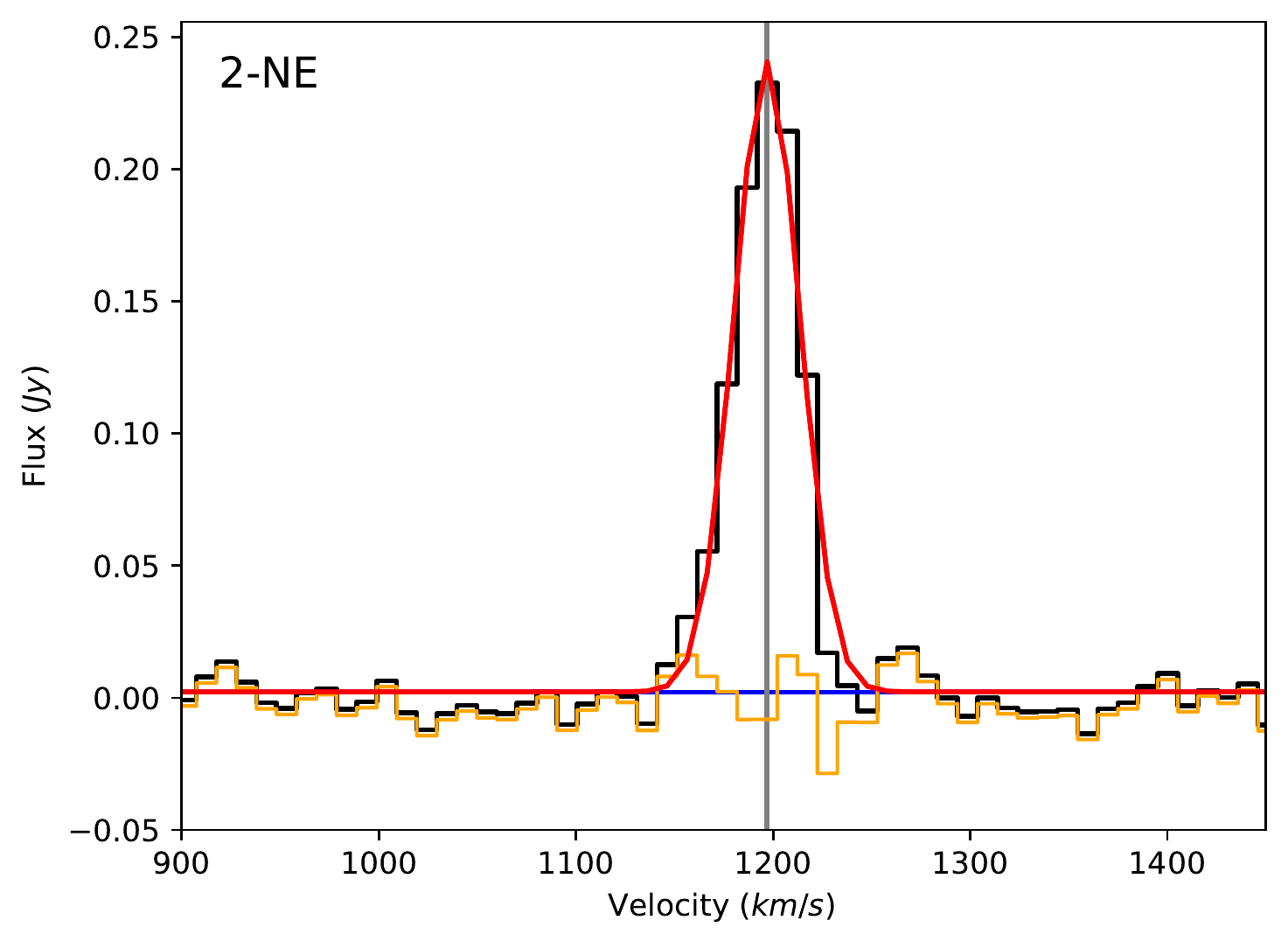}
\includegraphics[width=4.5cm]{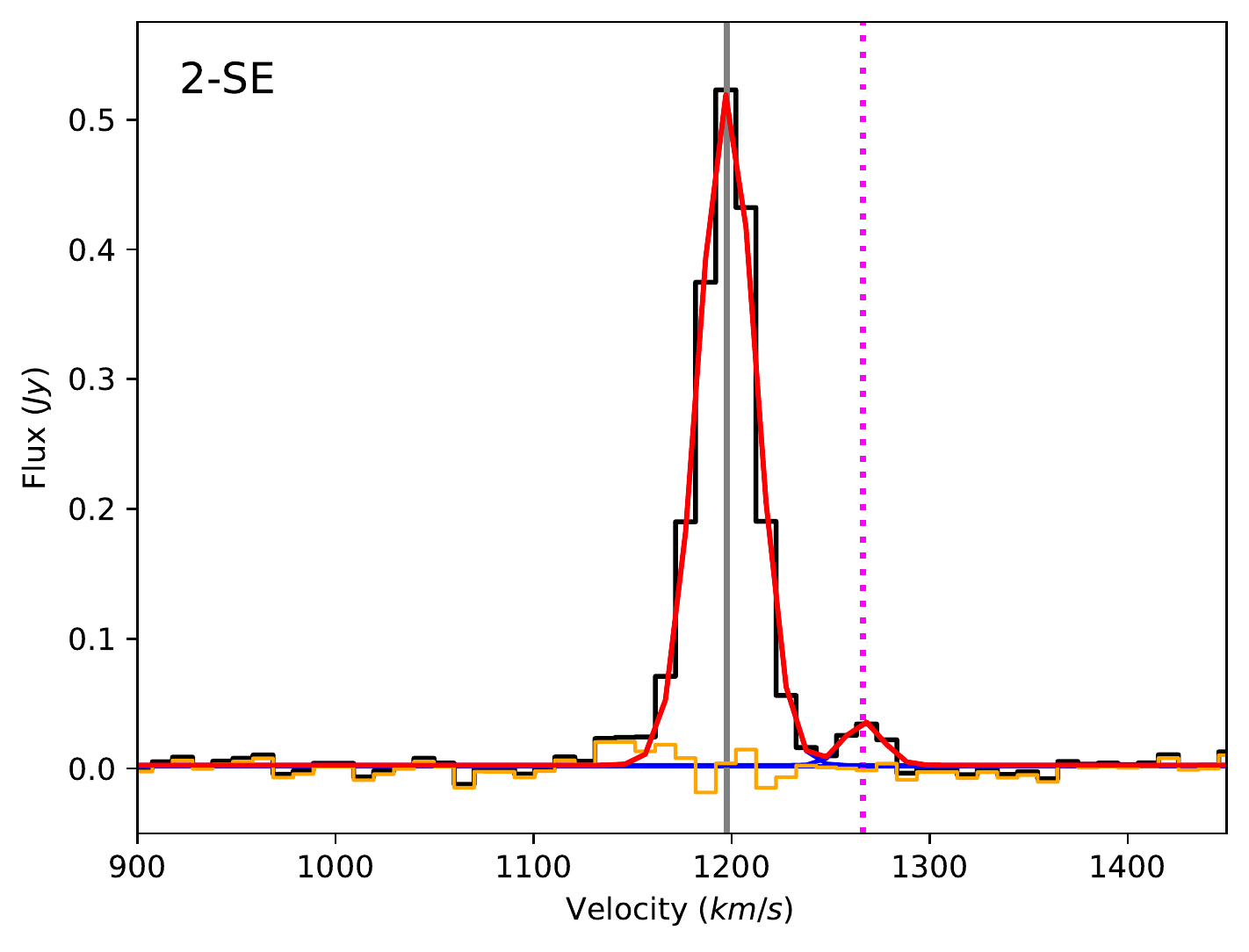}
\includegraphics[width=4.5cm]{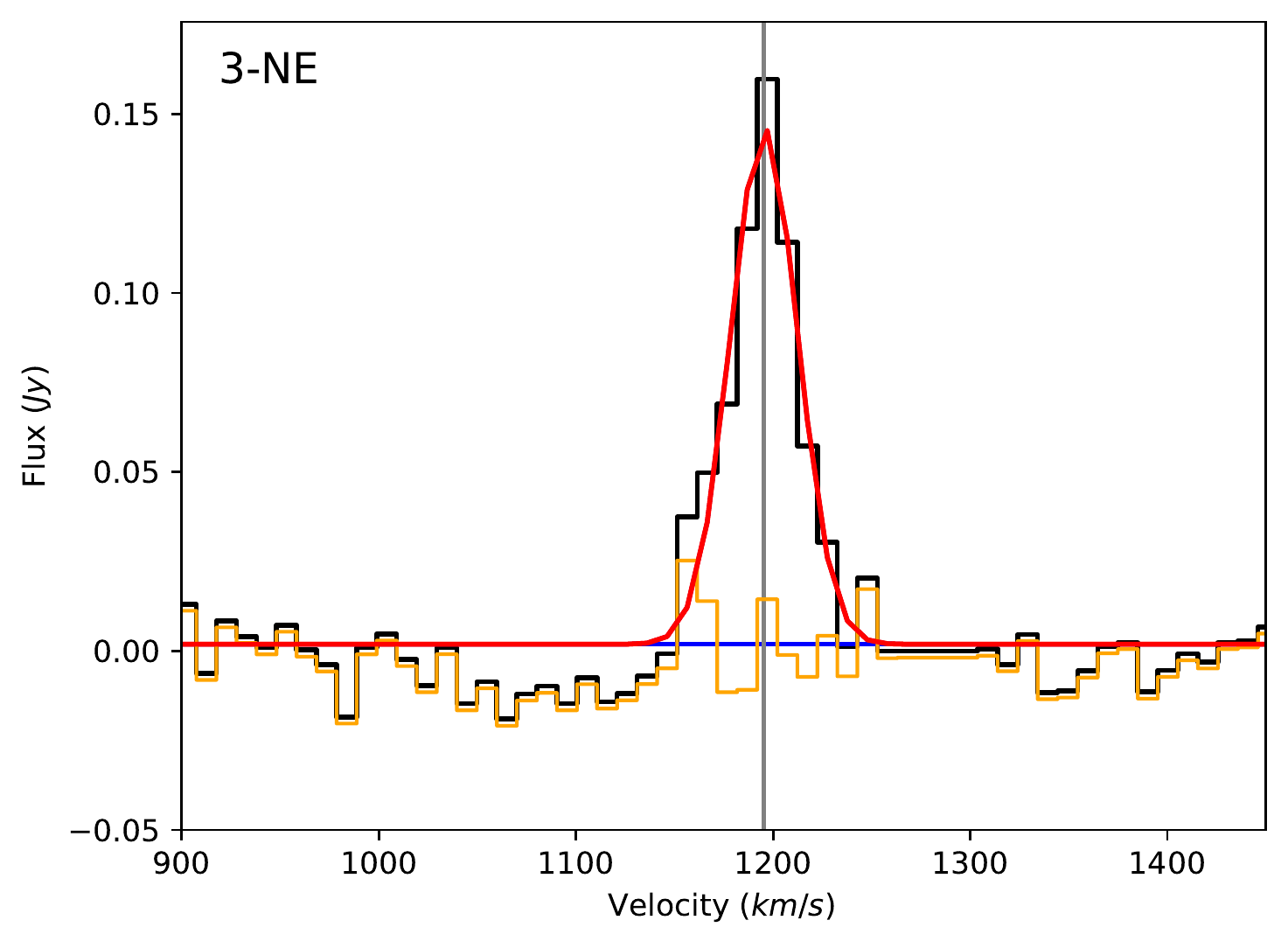}
\includegraphics[width=4.5cm]{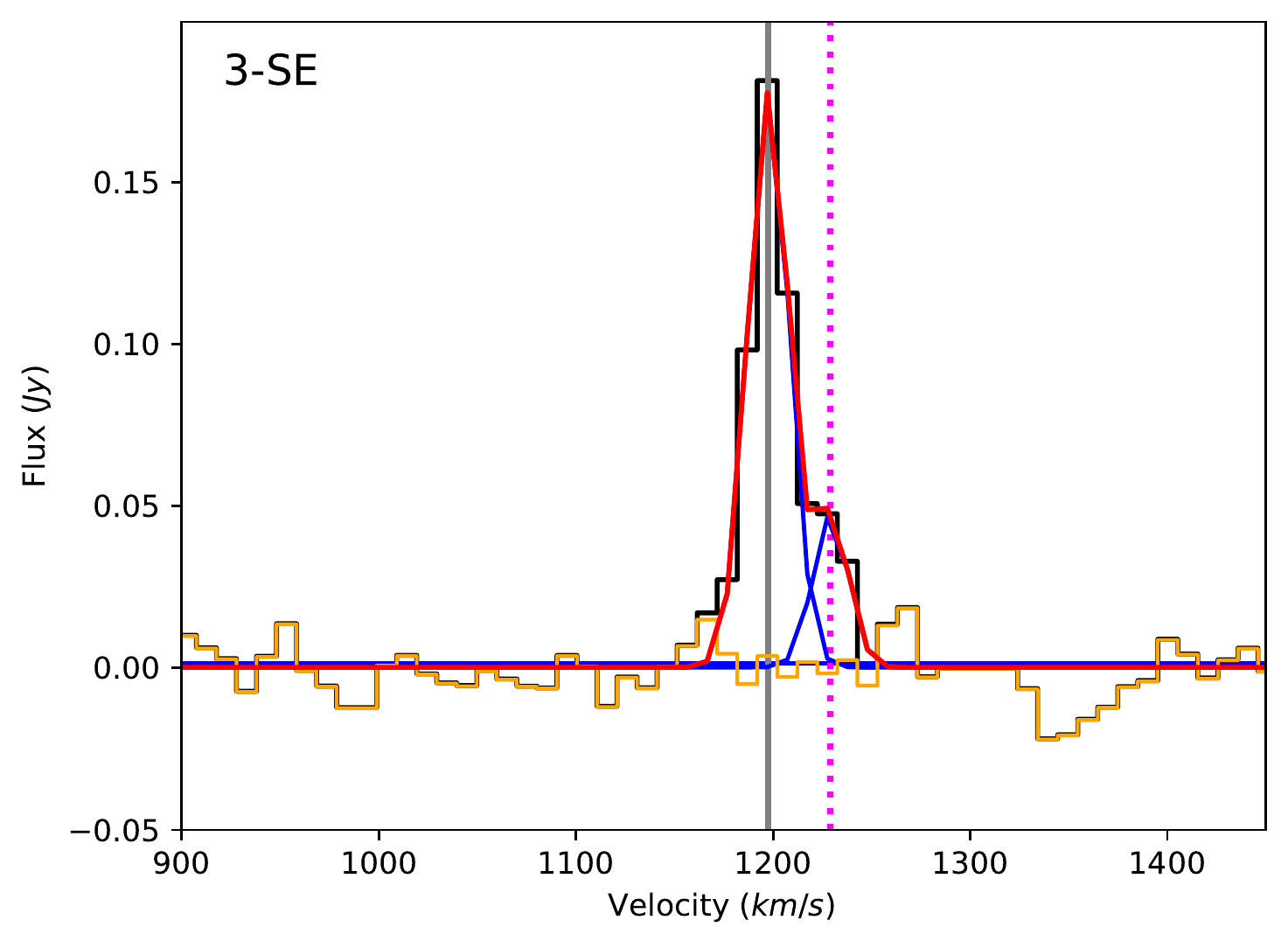}
\includegraphics[width=4.5cm]{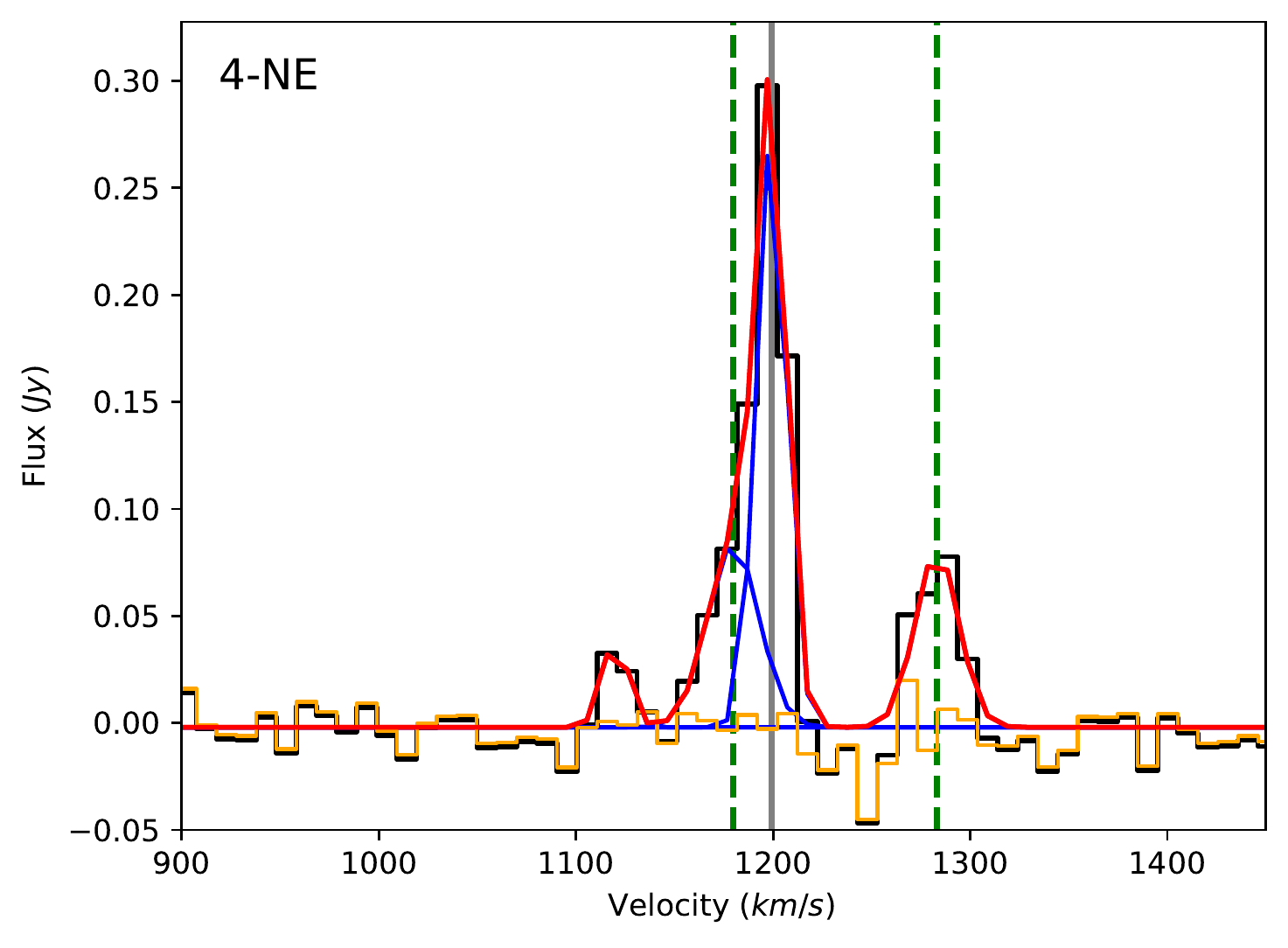}
\includegraphics[width=4.5cm]{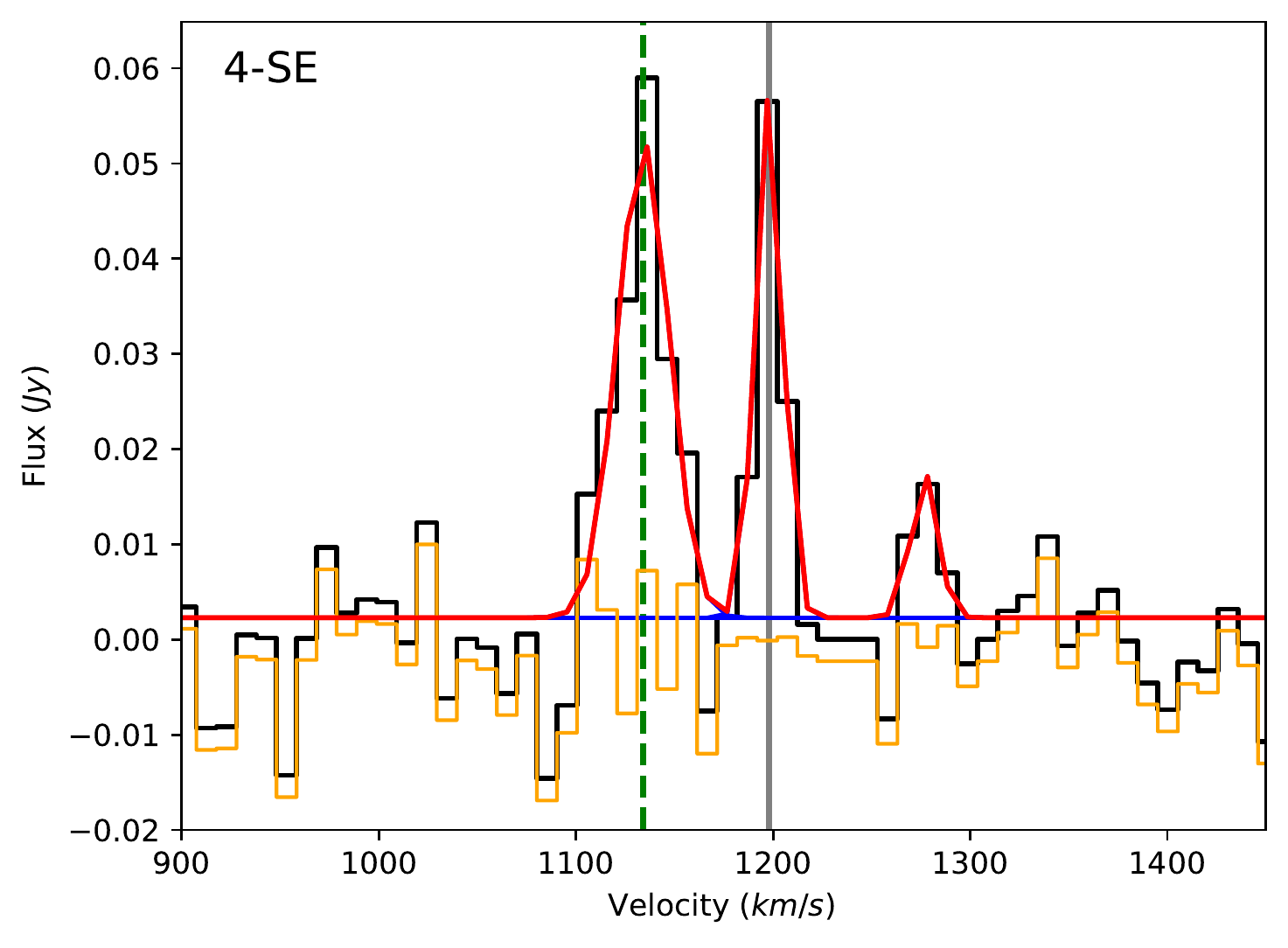}
\includegraphics[width=4.5cm]{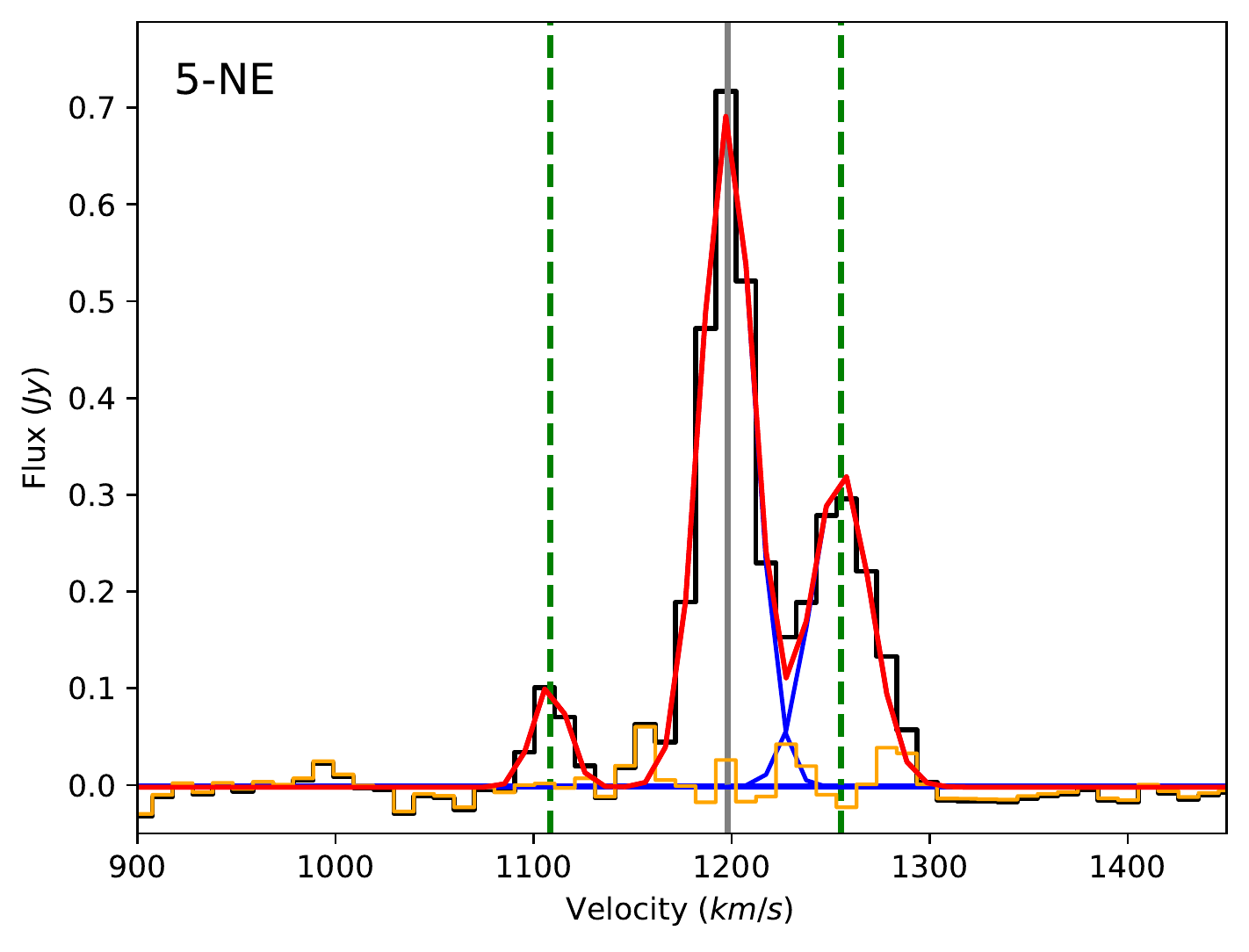}
\includegraphics[width=4.5cm]{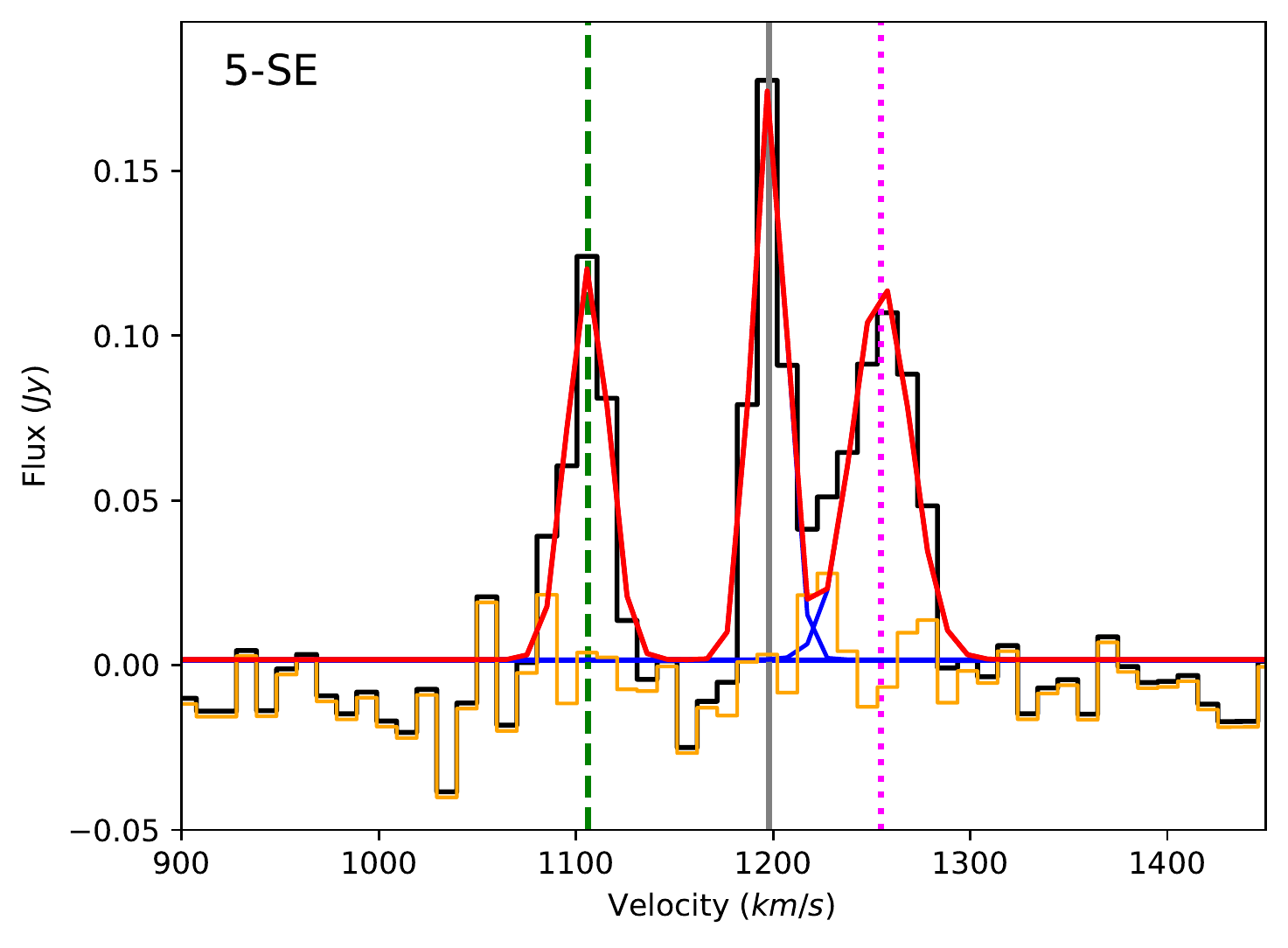}
\par}
\caption{Molecular CO(2-1) emission line fitting for the stacked spectra in each slice. From left to right panels: eastern 1, 2, 3, 4 \& 5 regions. Grey vertical solid line corresponds with the systemic component, the green dashed and dotted pink lines are those components identified as outflows and inflows, respectively. Orange solid line corresponds with the fit residuals.}
\label{line_fits_co1}
\end{figure*}

\begin{figure*}
\centering
\par{
\includegraphics[width=4.5cm]{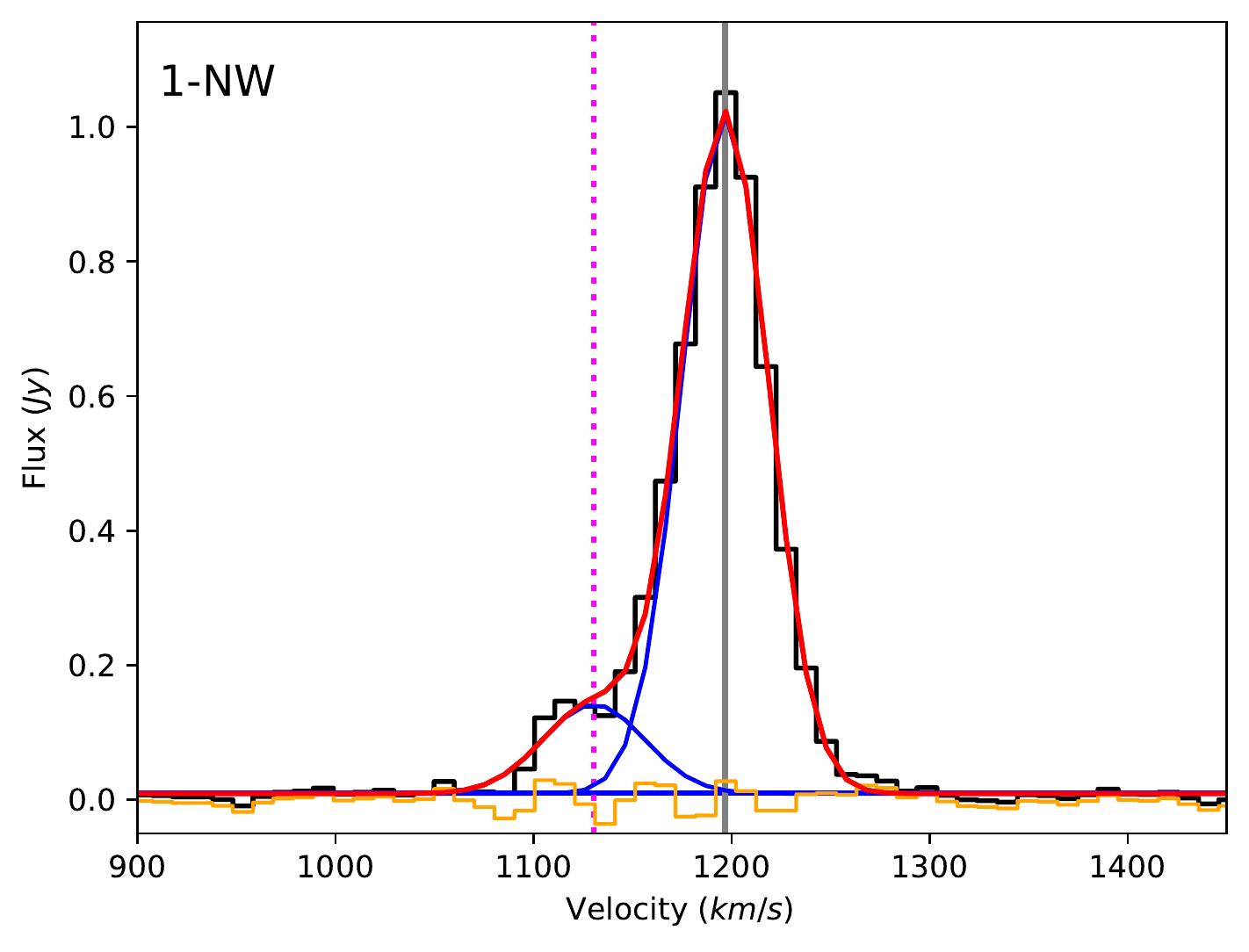}
\includegraphics[width=4.5cm]{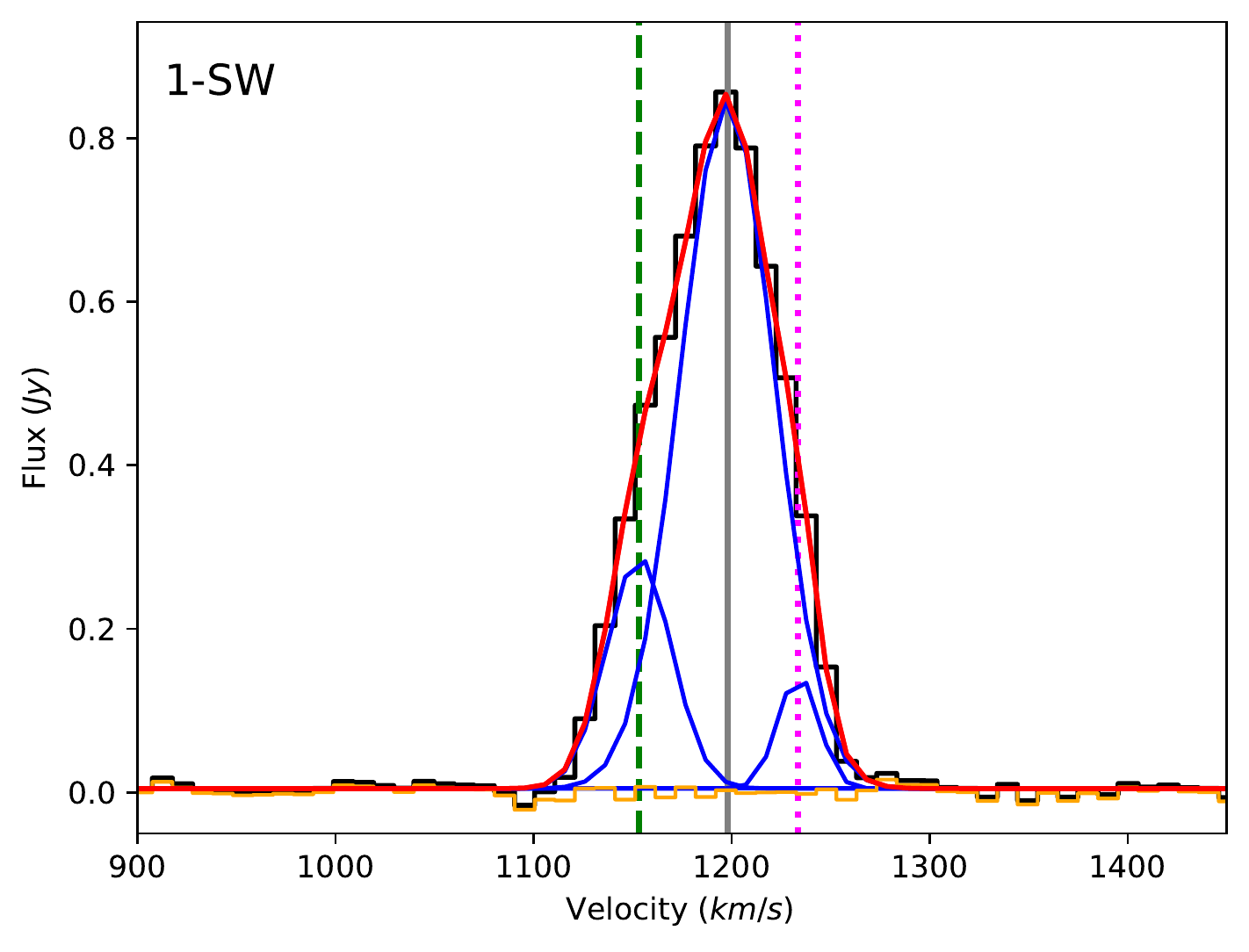}
\includegraphics[width=4.5cm]{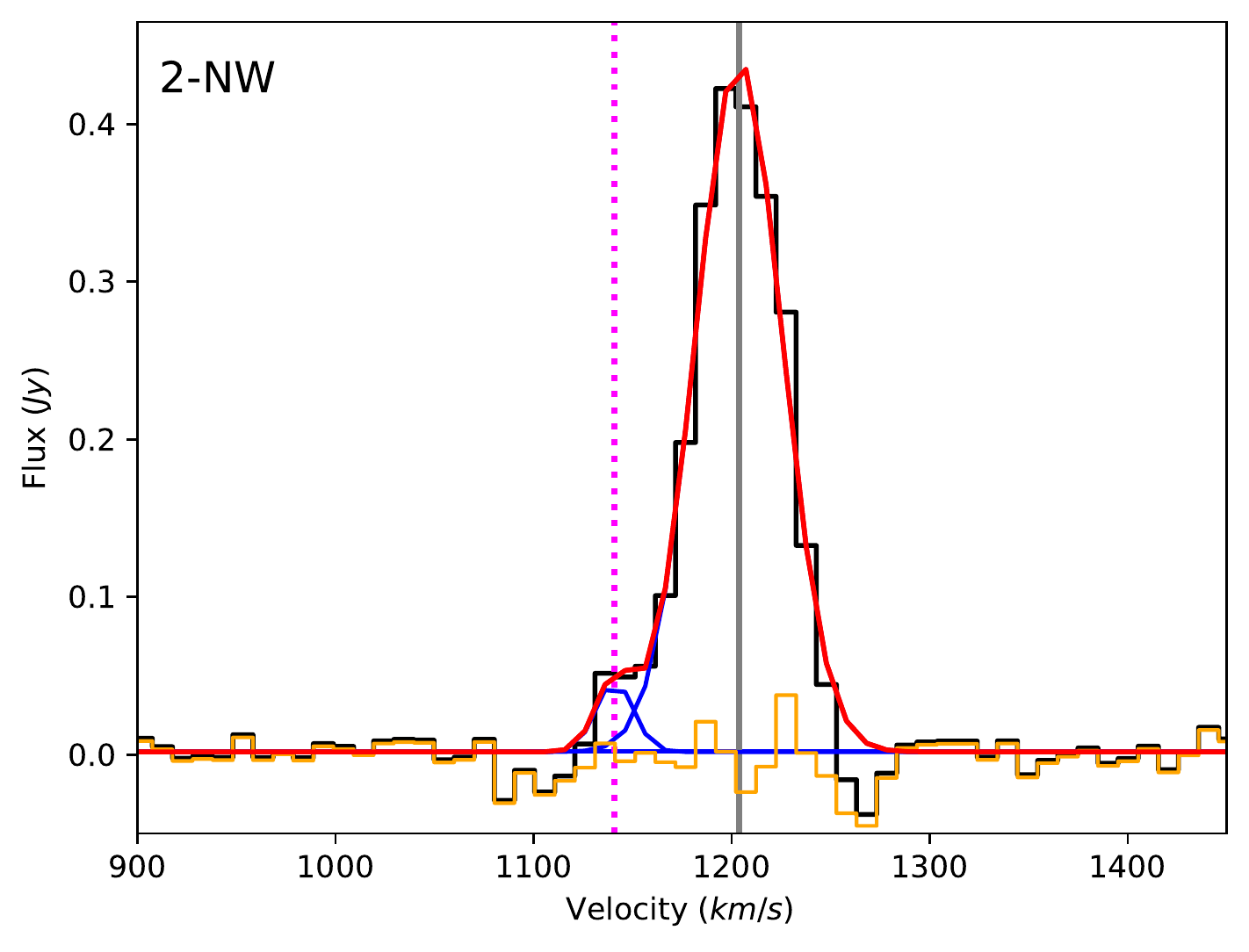}
\includegraphics[width=4.5cm]{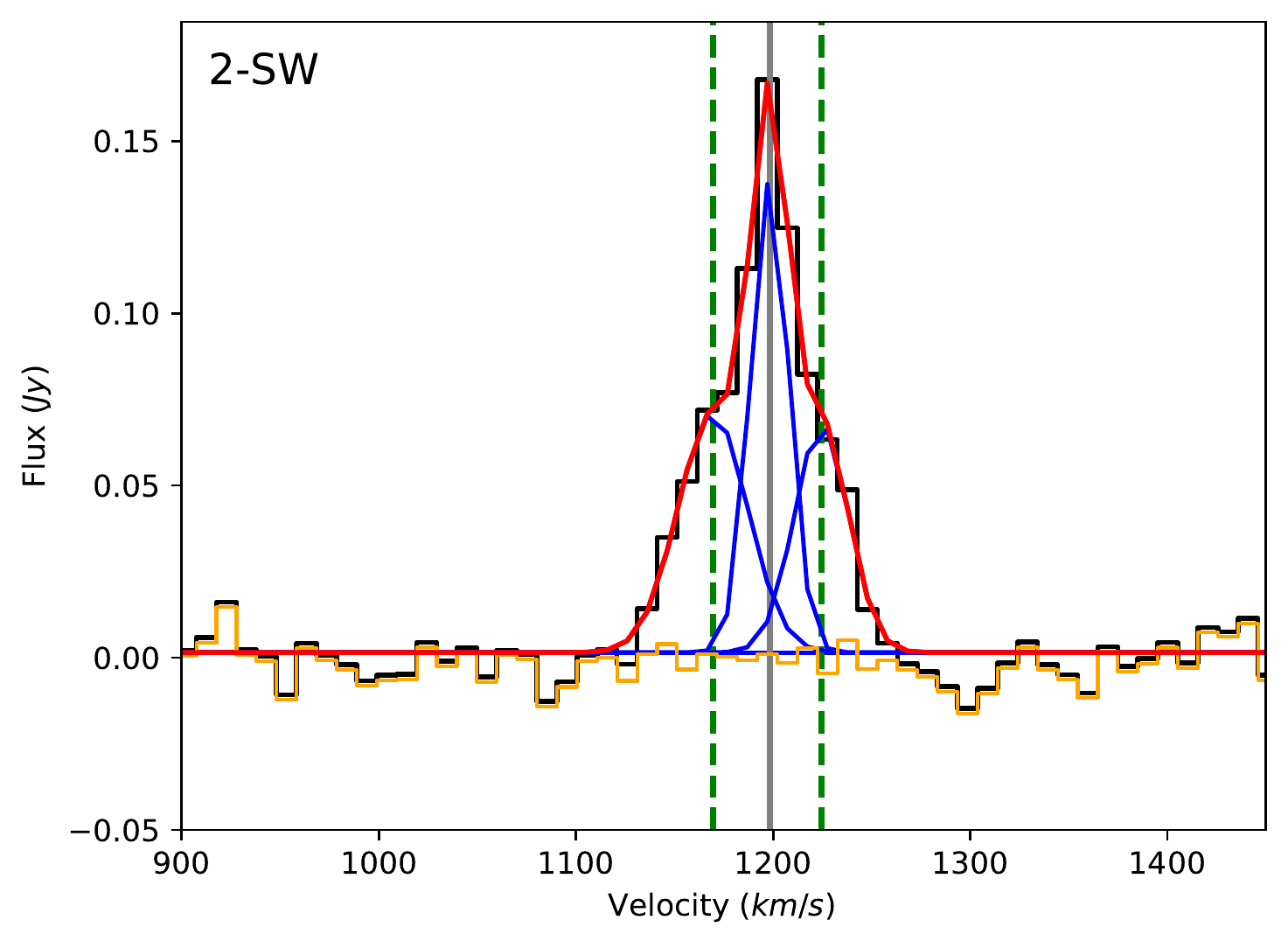}
\includegraphics[width=4.5cm]{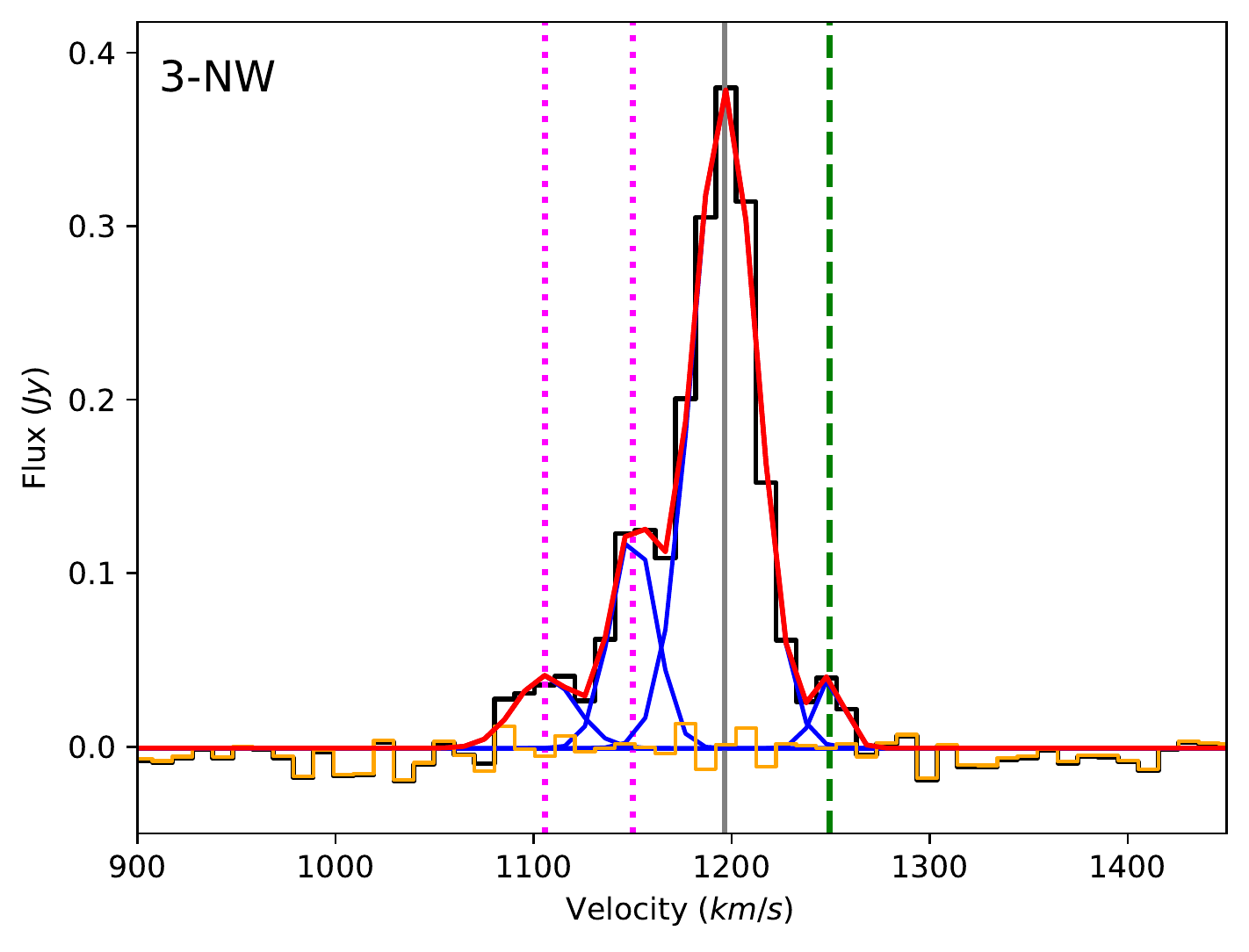}
\includegraphics[width=4.5cm]{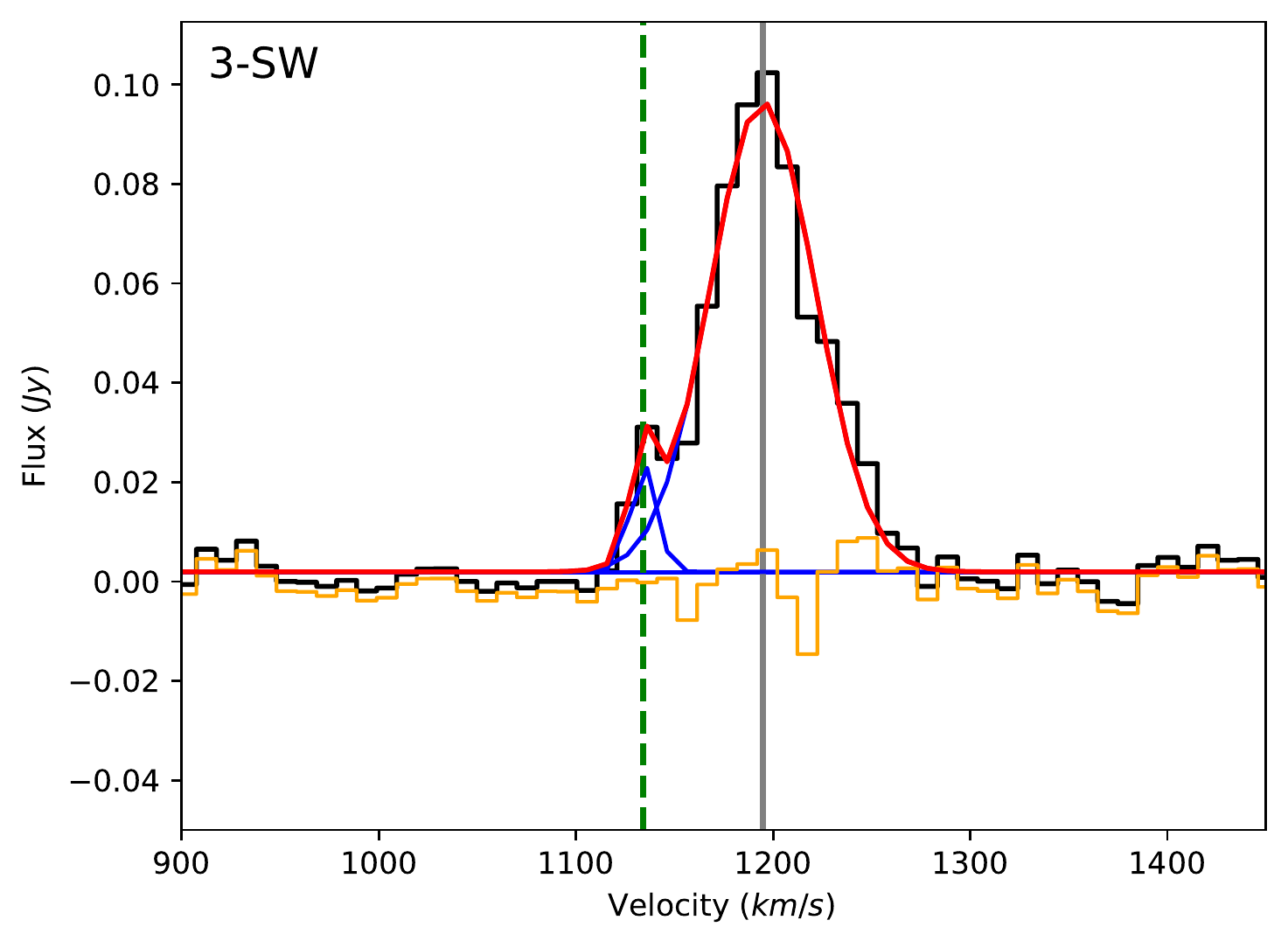}
\includegraphics[width=4.5cm]{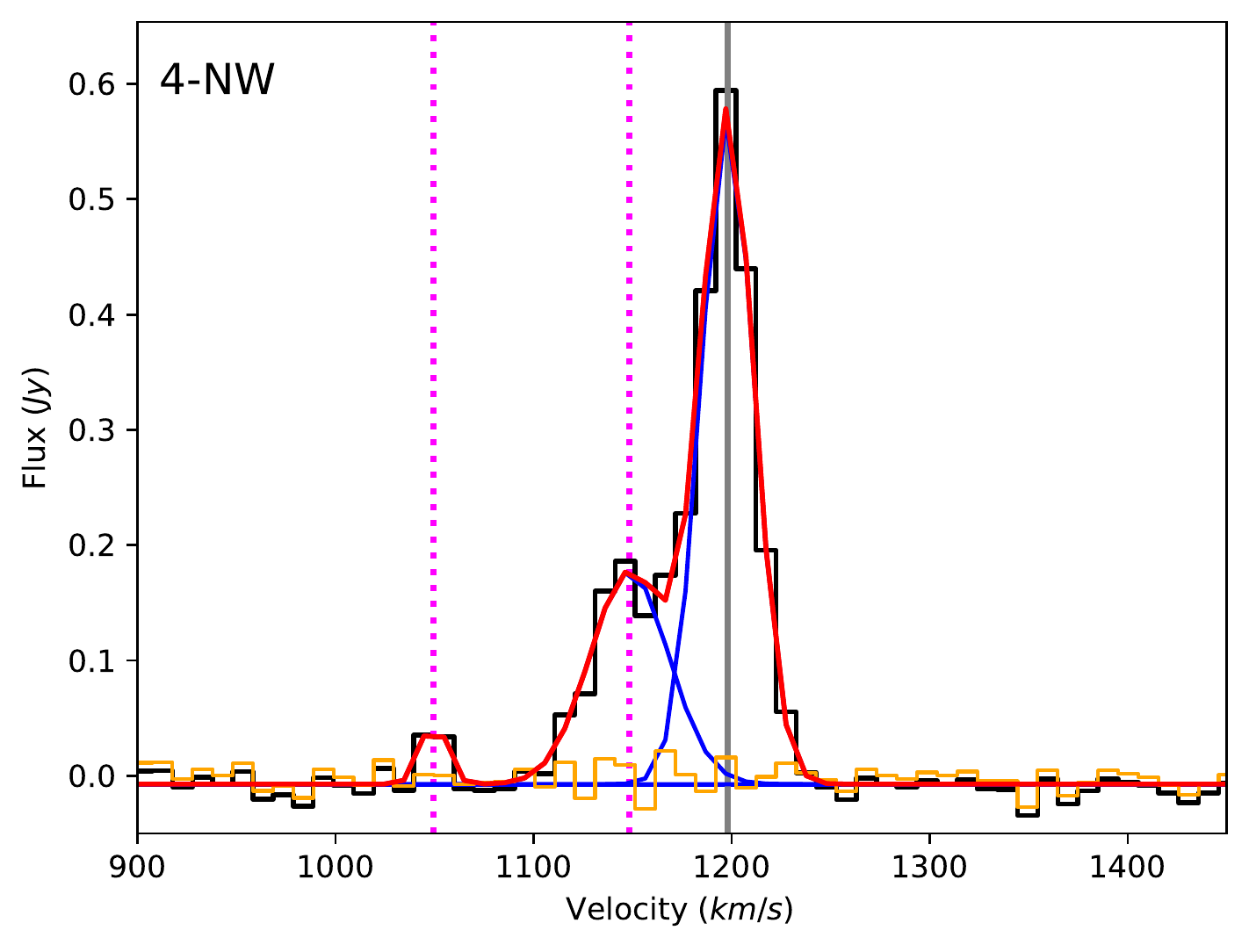}
\includegraphics[width=4.5cm]{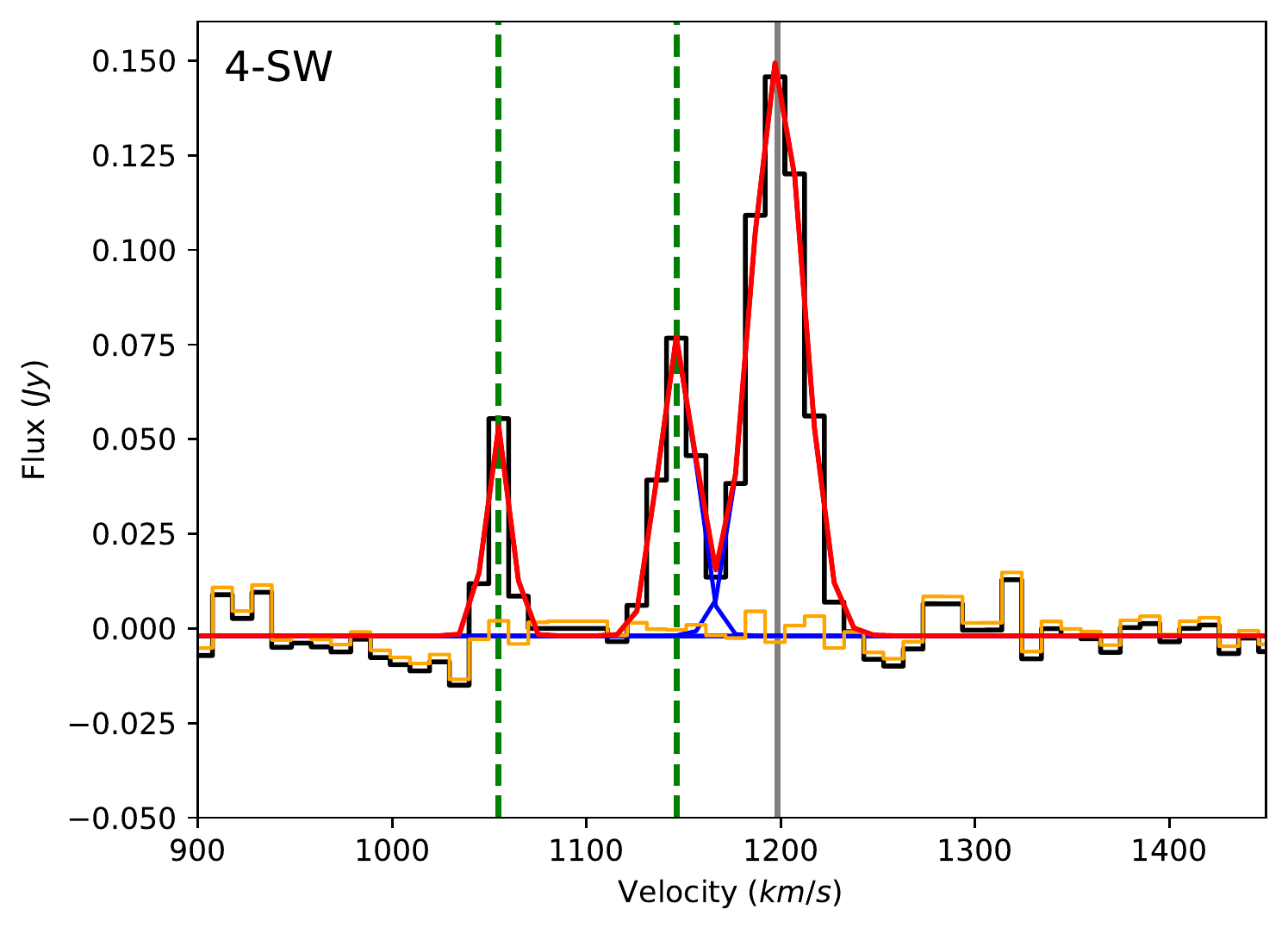}
\includegraphics[width=4.5cm]{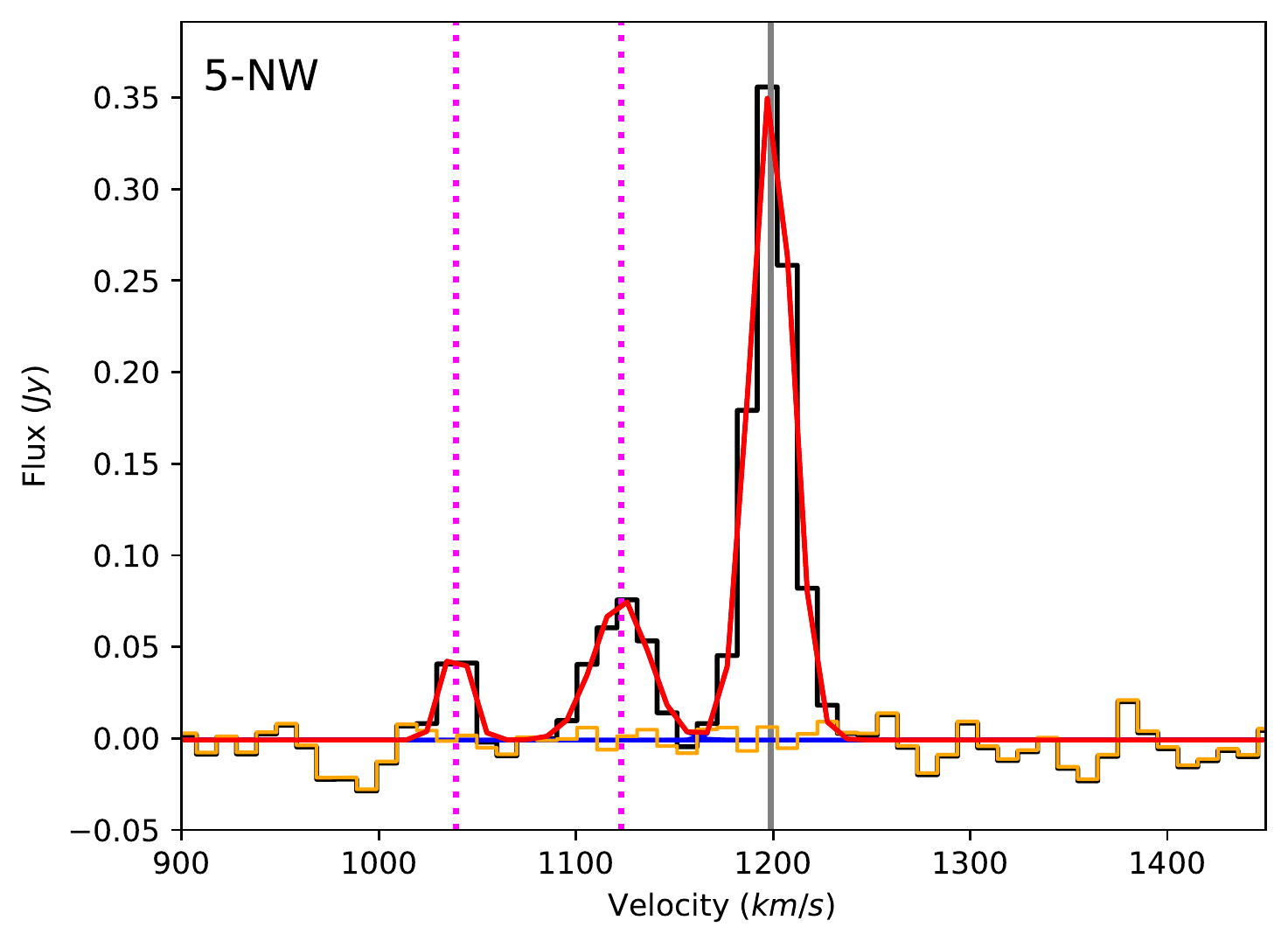}
\includegraphics[width=4.5cm]{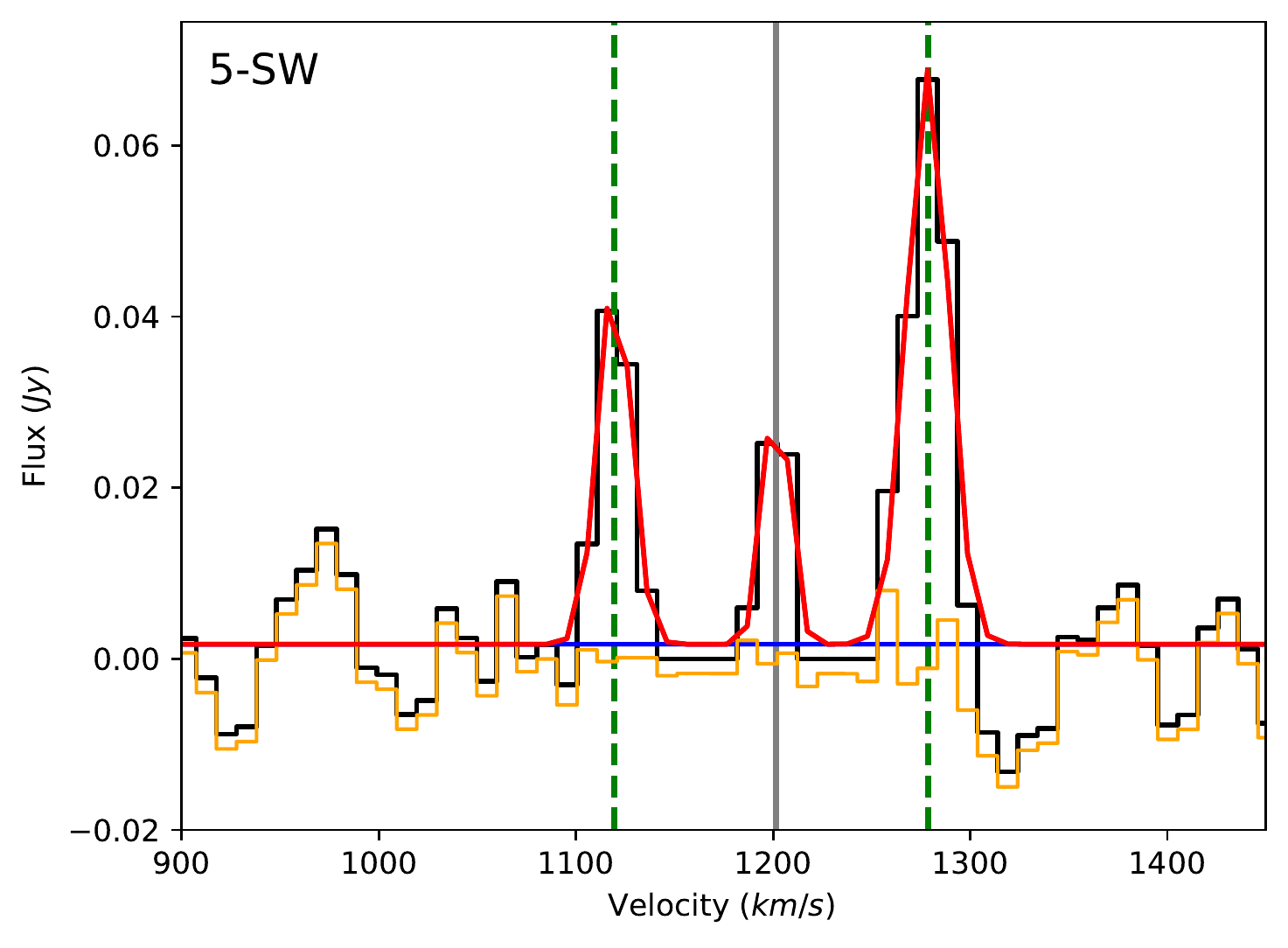}
\par}
\caption{Molecular CO(2-1) emission line fitting for the stacked spectra in each slice. From left to right panels: western 1, 2, 3, 4 \& 5.}
\label{line_fits_co2}
\end{figure*}

To derive the outflow properties we defined various slices along the outflow at both sides of the nucleus. As described in Section \ref{Regions}, to maximize the S/N of the outflow component we use the stacking technique.   Using the large scale stellar bar orientation (see eg. \citealt{Mulchaey97}), we divide the slices selected along the outflow, in northern and southern regions. Figure \ref{stacking_slices} shows those regions where the contribution of from local inflows are not expected (black boxes) and those with inflows movements according to bar models (red boxes).  In the case of the CO(2-1) line profiles, we indicate the outflow and inflow components (see Fig.  \ref{line_fits_co1} and \ref{line_fits_co2}).

In the case of the [O\,{\sc iii}] emission profile it is easy to identify the galaxy disk rotating component, which is the narrow component in the line profiles (see Section \ref{individual_o3}). However, the molecular gas emission is more difficult to distinguish. When possible (high signal-to-noise ratio) we use the derived rotating disk model to identify the rotation component. Otherwise, we select the brightest line which we associated with the galaxy disk. The only exceptions are those cases that present a large offset ($>$100~km~s$^{-1}$) with respect to the immediate previous region. In this situation we look for continuity in the galaxy disk velocities. Note that we check that the identified rotation disk component is in agreement within the errors with those of the stellar kinematics derived from the MUSE data (see Section \ref{stellar}). Once we have identified where the rotation disk components are we shift the velocities to a common rest-frame wavelength. 

Finally, to rule out any possible contribution of the instrumental profile wings at the levels of the measured broad emission lines (i.e. outflow components), we compare the observed emission line profile with the instrumental one. The instrumental profile is defined as the instrument response to an observed unresolved line (i.e. much narrower than the velocity resolution), which depends on the employed instrumentation. This profile can be slightly asymmetric and/or present broad weak wings. To characterize it we use VLT/MUSE data of the planetary nebula NGC\,7009, which is an unresolved source (the velocity extent of the emission line components is $\sim$60~km~s$^{-1}$; see \citealt{Walsh18}), to obtain the [O\,{\sc iii}] emission line profile in various regions{\footnote{Optical integral field spectroscopy of NGC\,7009 was taken using VLT/MUSE, which was observed as part of the program 60.A-9347.  We downloaded the fully reduced and calibrated science data cube from the ESO data archive. This MUSE data cube, previously presented in \citet{Walsh18}, was observed under similar seeing conditions ($\sim$0.5\,\arcsec) than the one used in this work for NGC\,5643.}}. Note that the VLT/MUSE spectral resolution is $\sim$75~km~s$^{-1}$. Since the various regions produce practically the same profile, we use the aperture extracted in the center of the nebula.  
We find that the MUSE/VLT instrumental profile at $\lambda$5007$\AA$~is slightly asymmetry, but the detected broad lines in our work cannot be explained by an instrumental effect. This confirms that the Gaussian approach is a good approximation to the instrumental profile and provides reliable flux estimations.

\begin{figure*}
\centering
\par{
\includegraphics[width=4.5cm]{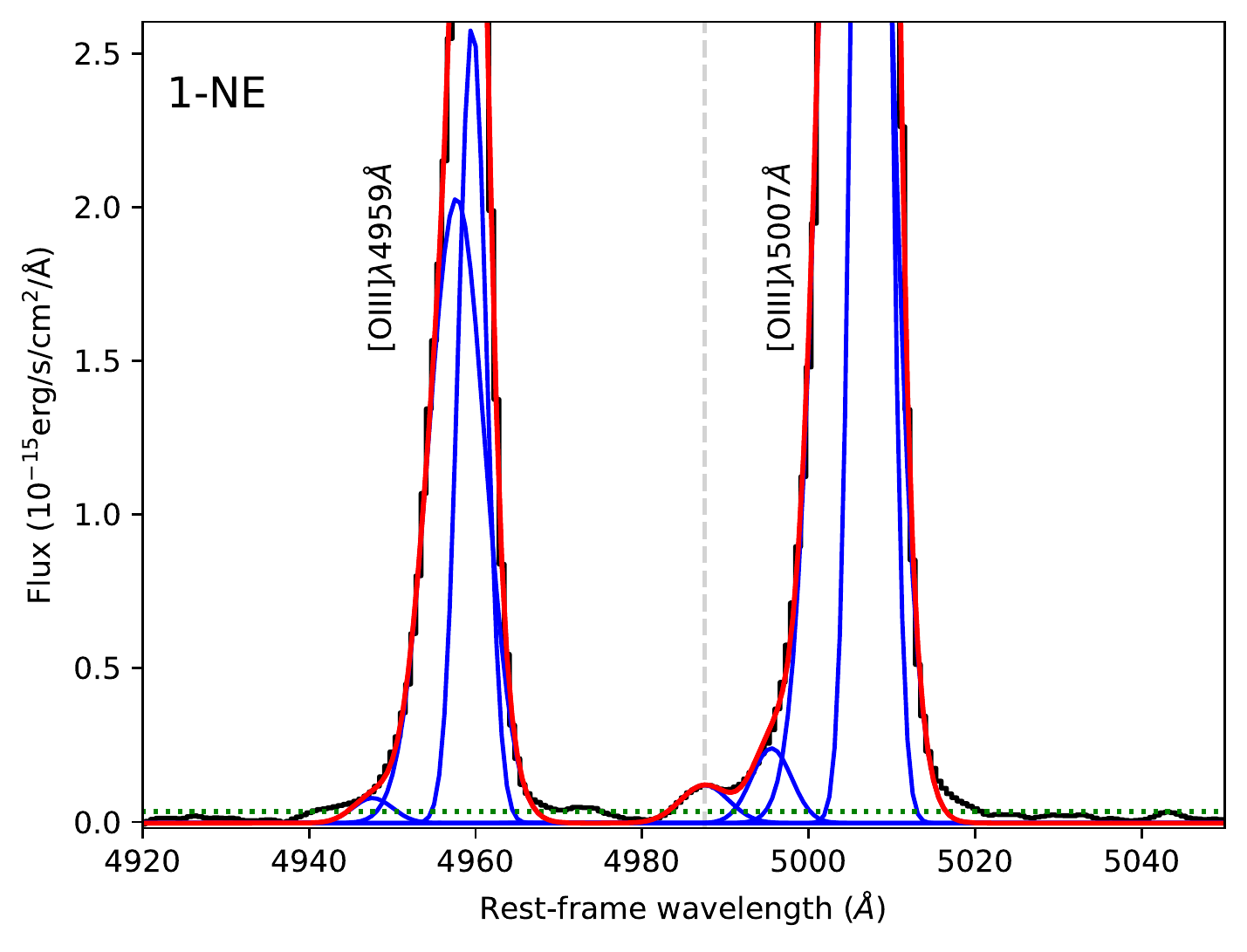}
\includegraphics[width=4.5cm]{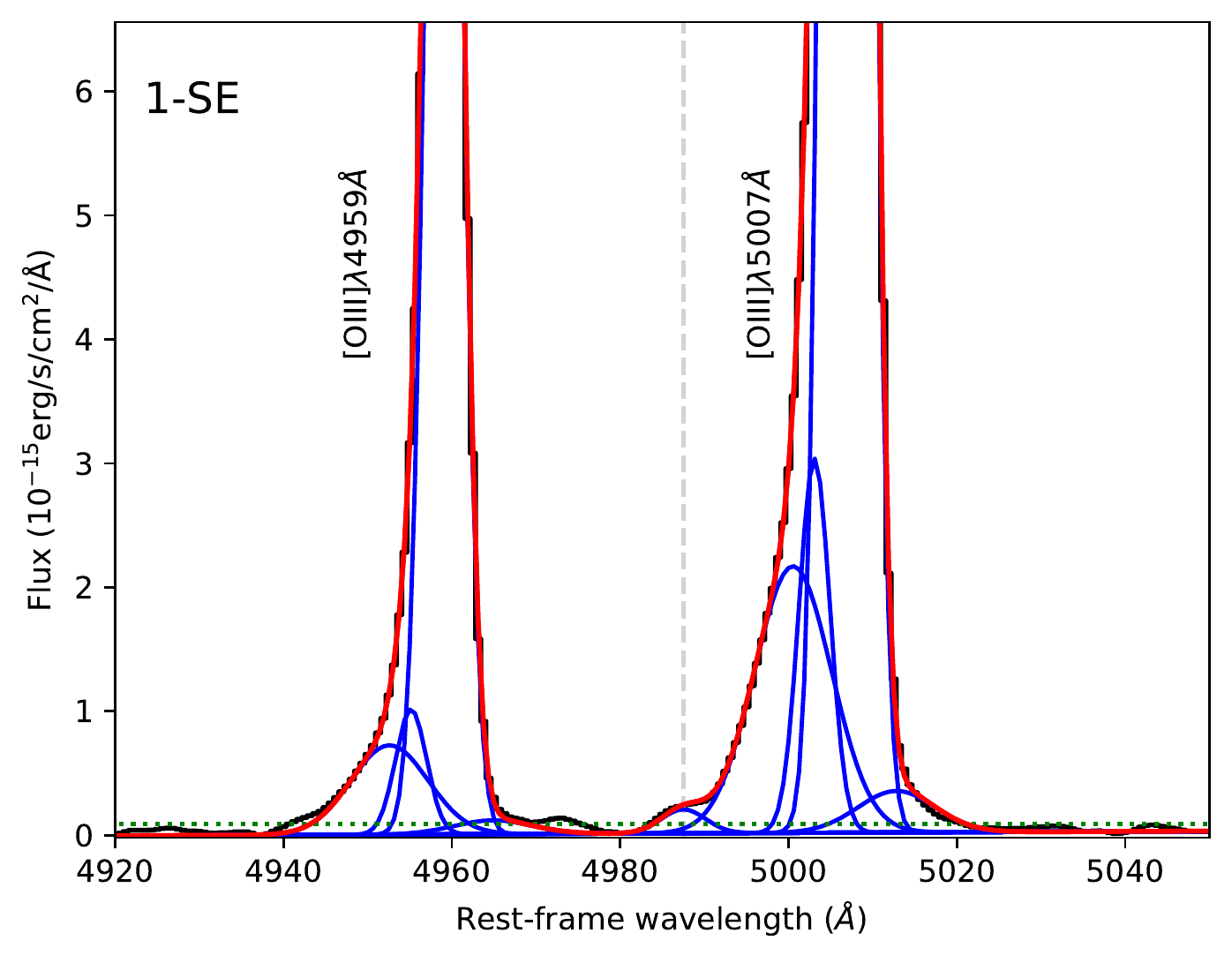}
\includegraphics[width=4.5cm]{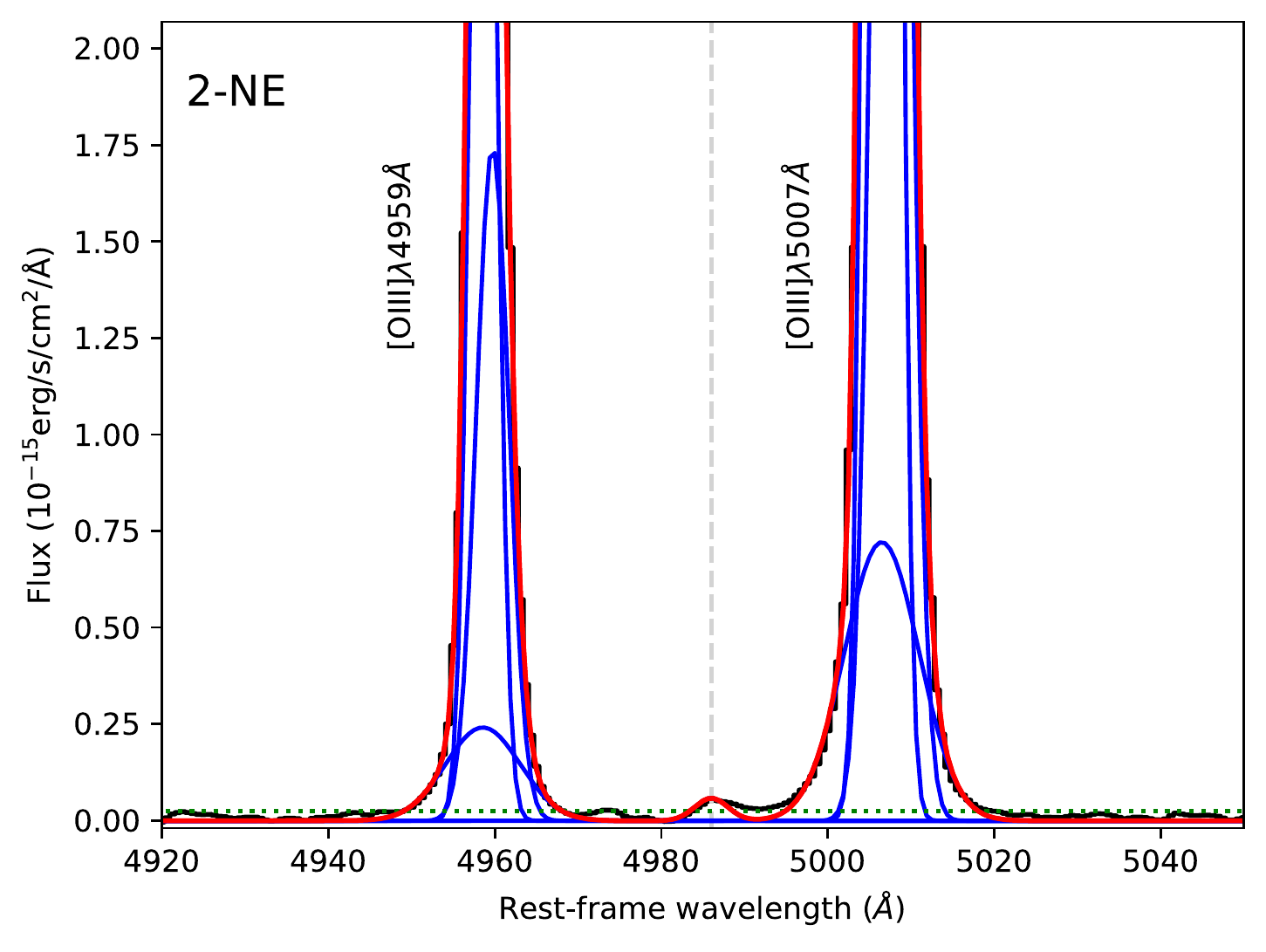}
\includegraphics[width=4.5cm]{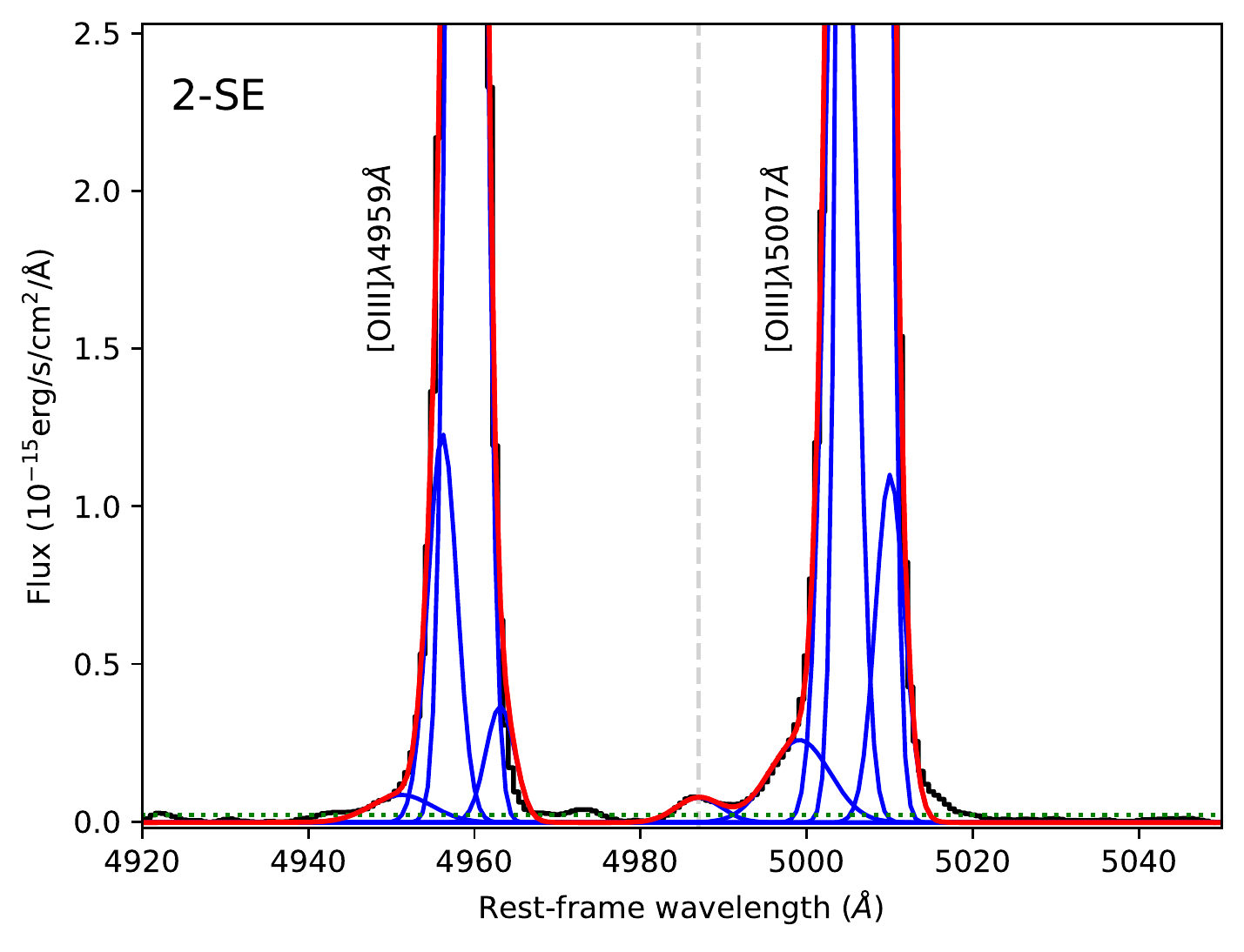}
\includegraphics[width=4.5cm]{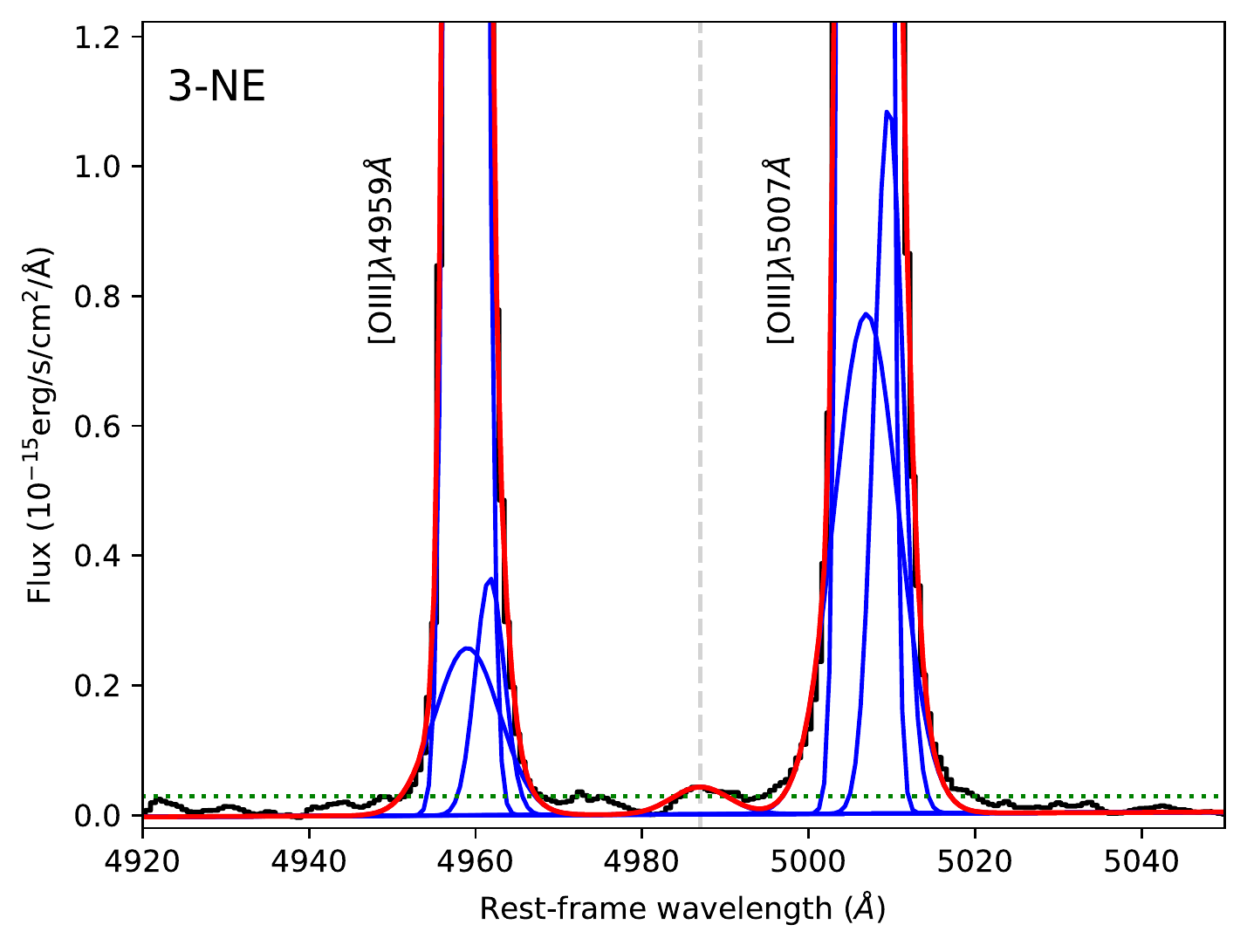}
\includegraphics[width=4.5cm]{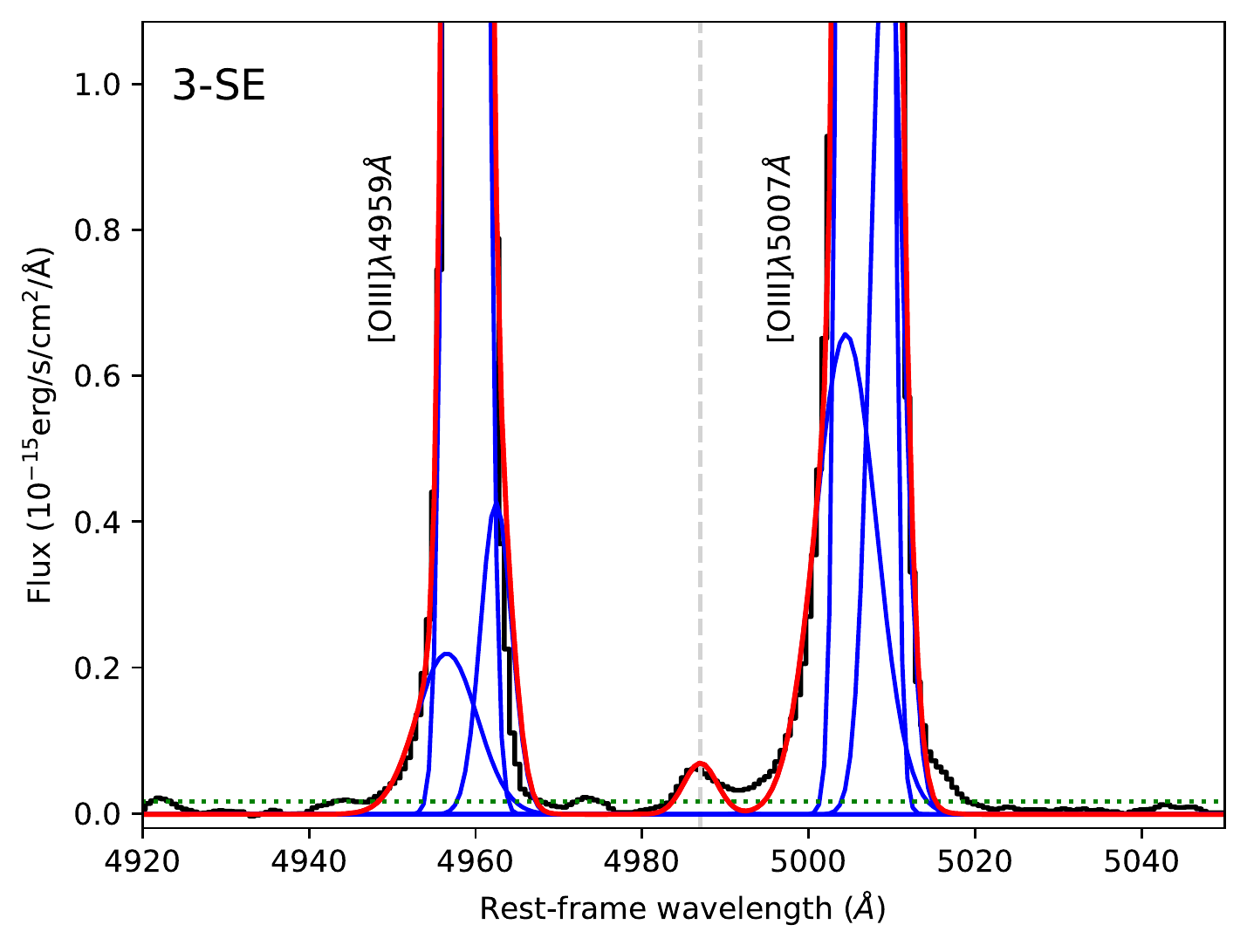}
\includegraphics[width=4.5cm]{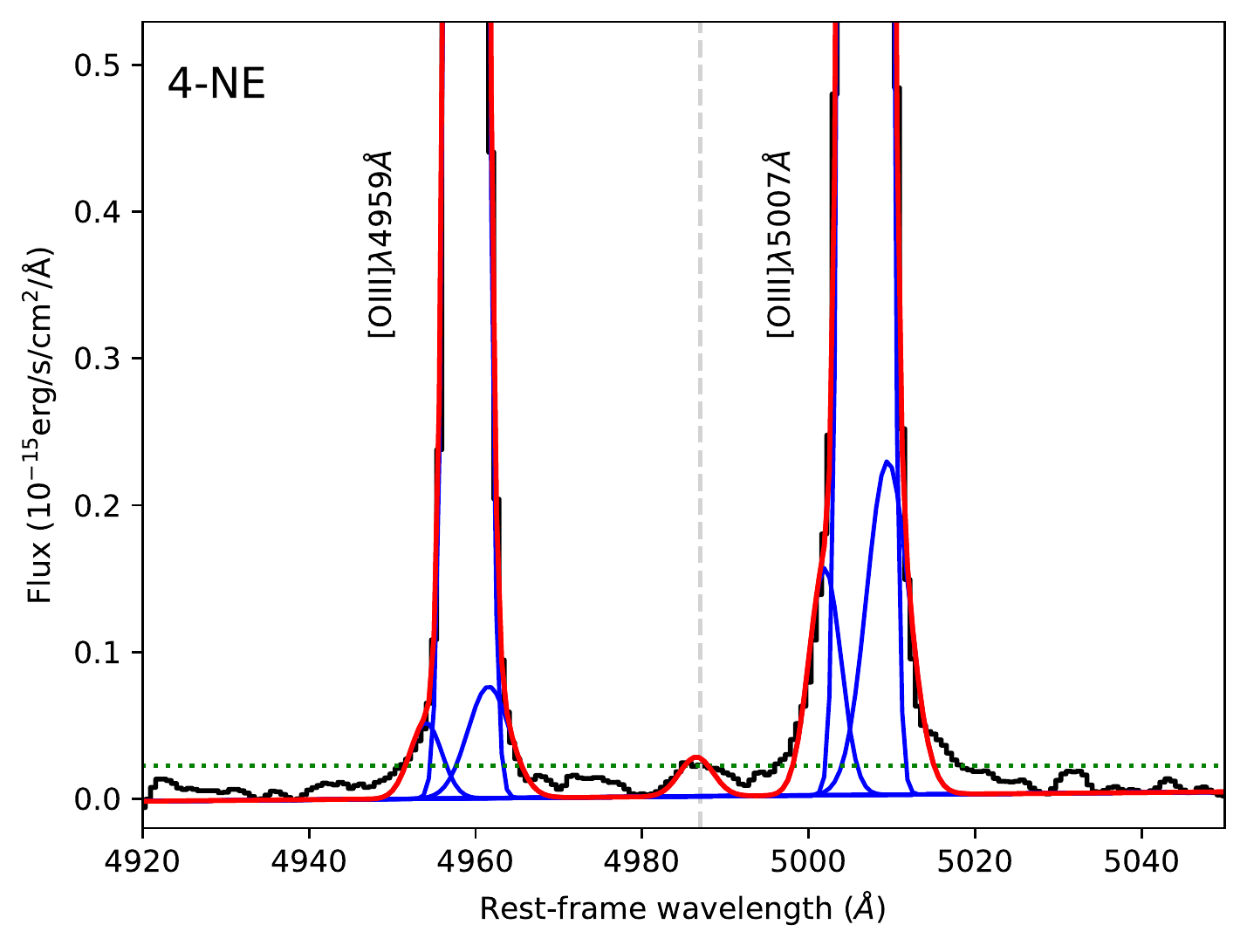}
\includegraphics[width=4.5cm]{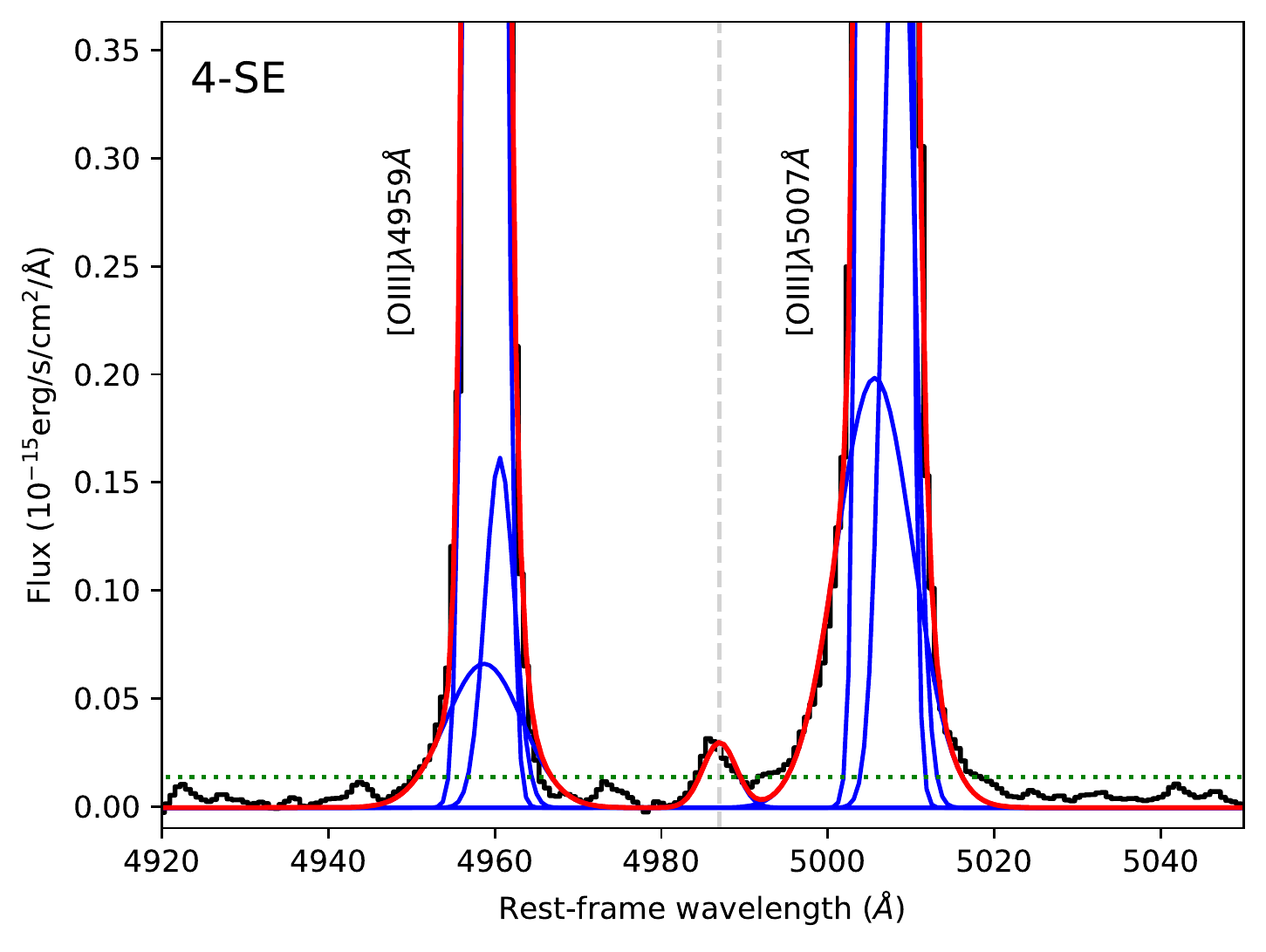}
\includegraphics[width=4.5cm]{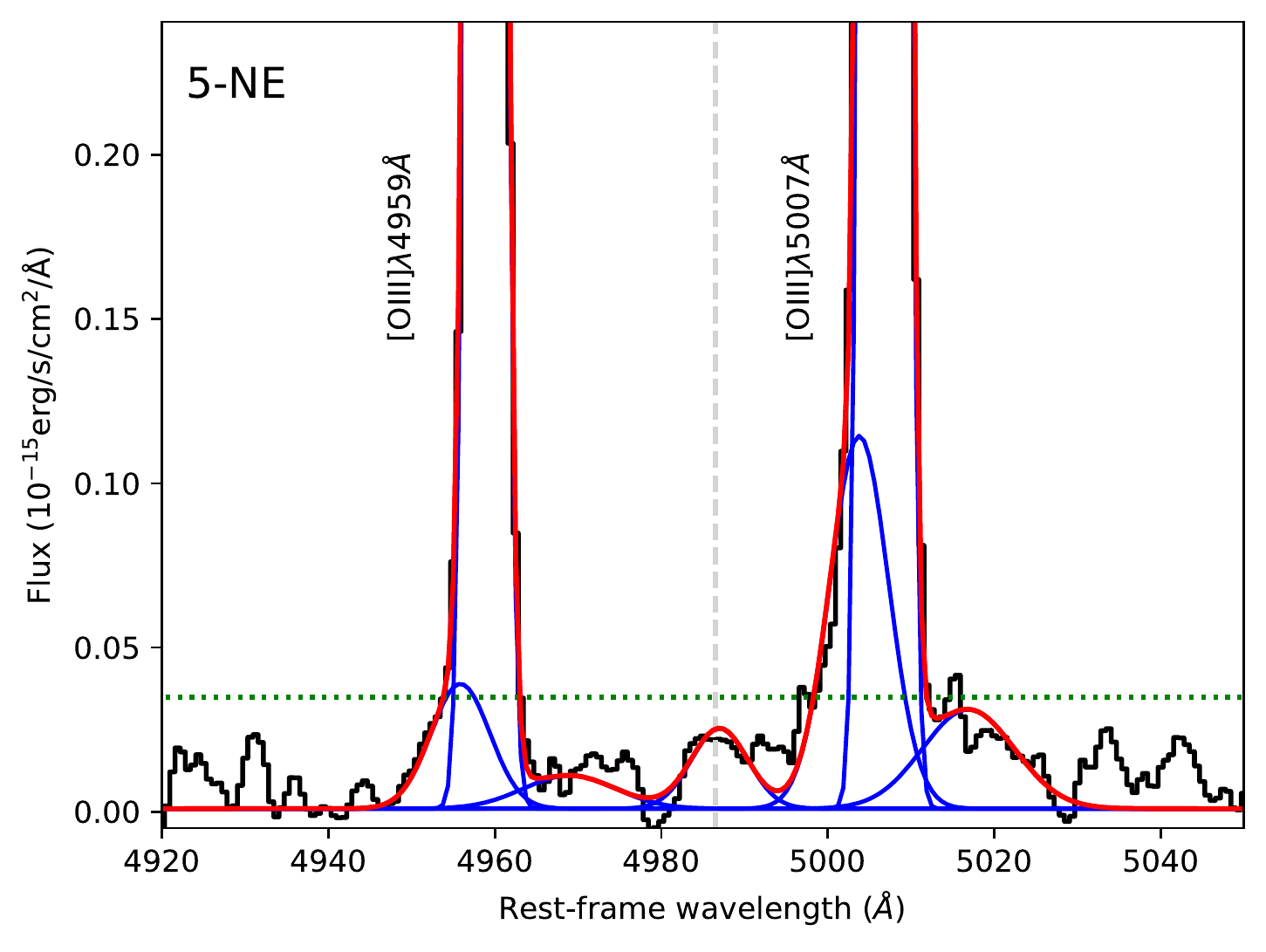}
\includegraphics[width=4.5cm]{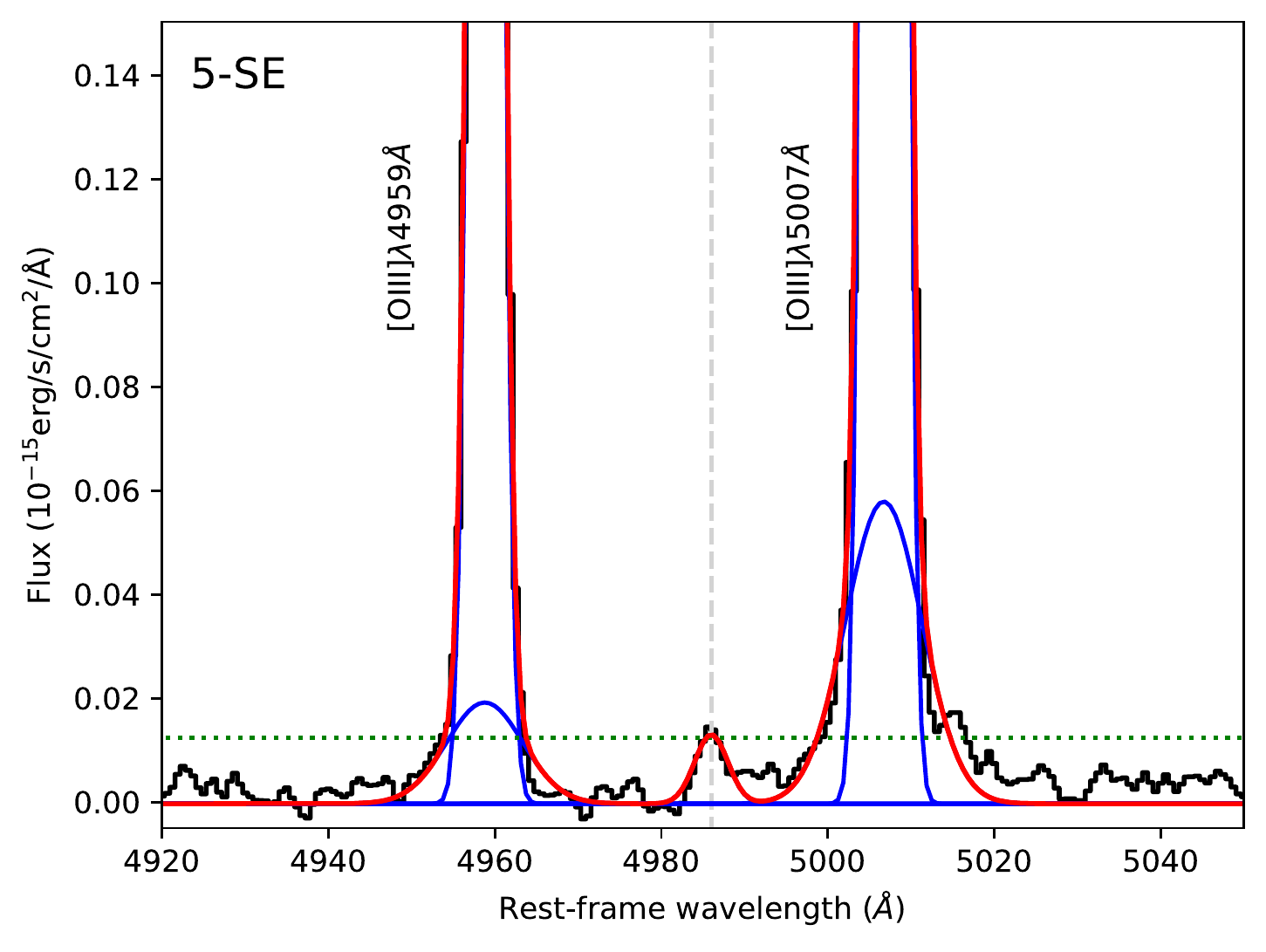}
\par}
\caption{Same as Fig. \ref{line_fits_co1} but for the [O\,{\sc iii}] emission line.}
\label{line_fits_o31}
\end{figure*}

\begin{figure*}
\centering
\par{
\includegraphics[width=4.5cm]{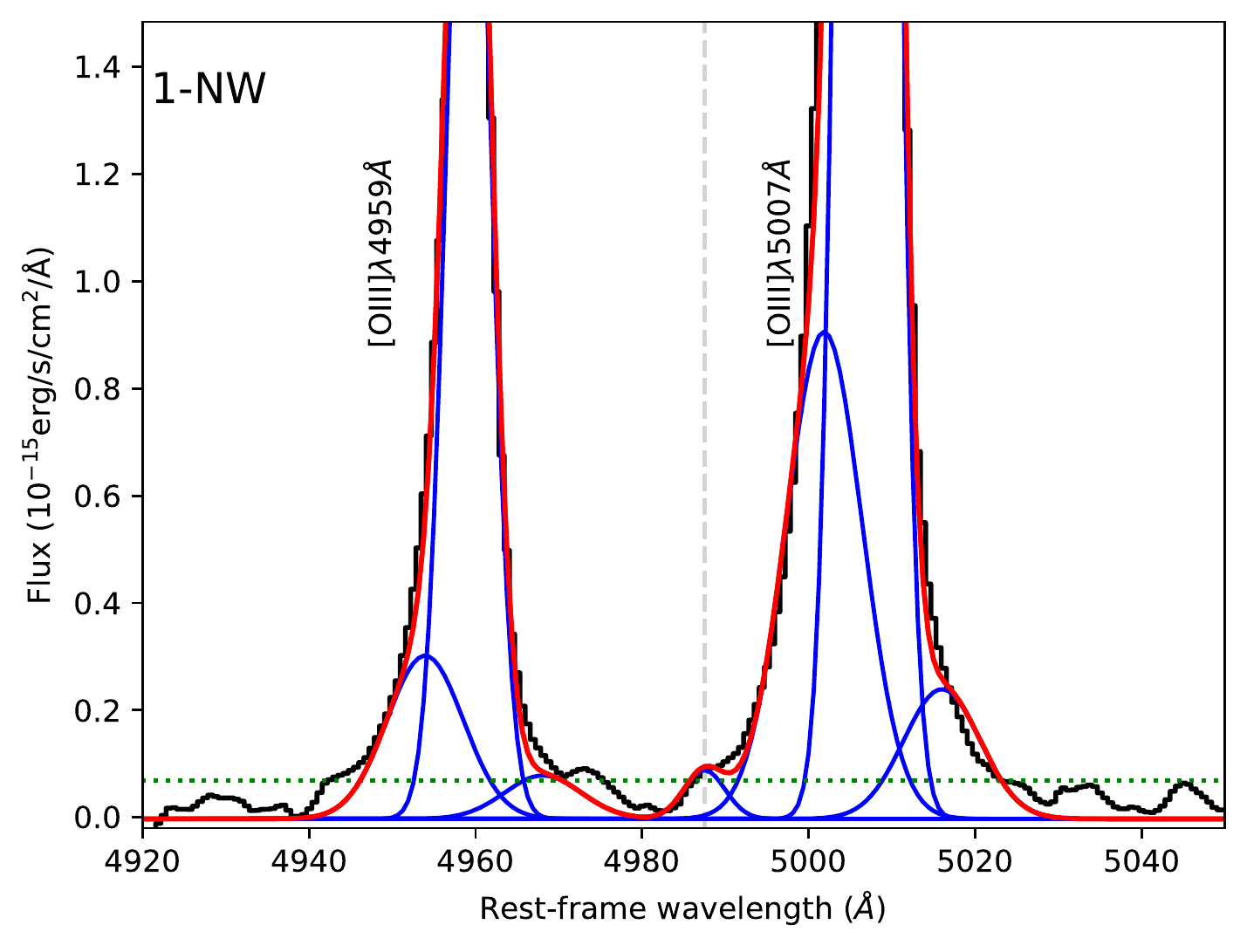}
\includegraphics[width=4.5cm]{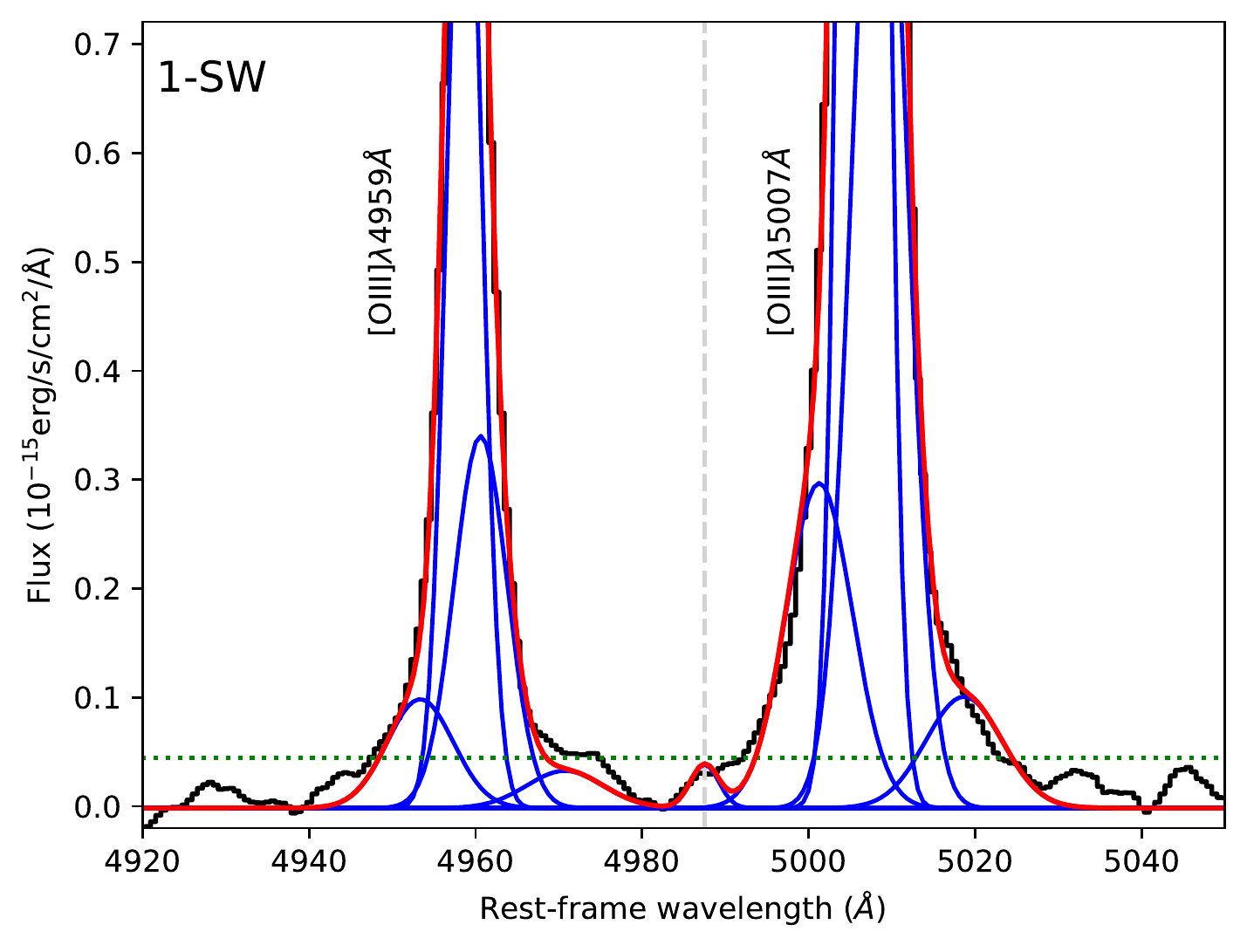}
\includegraphics[width=4.5cm]{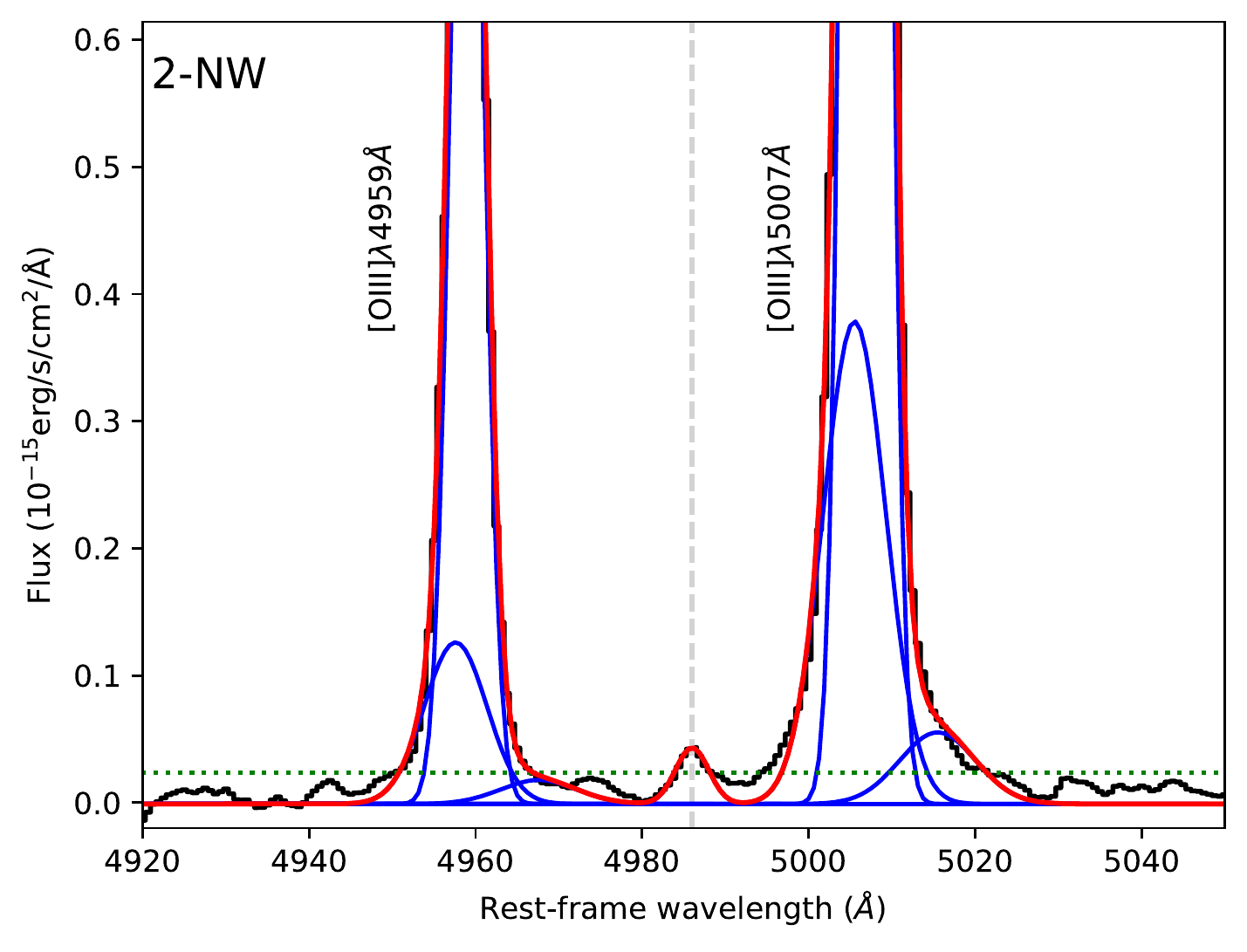}
\includegraphics[width=4.5cm]{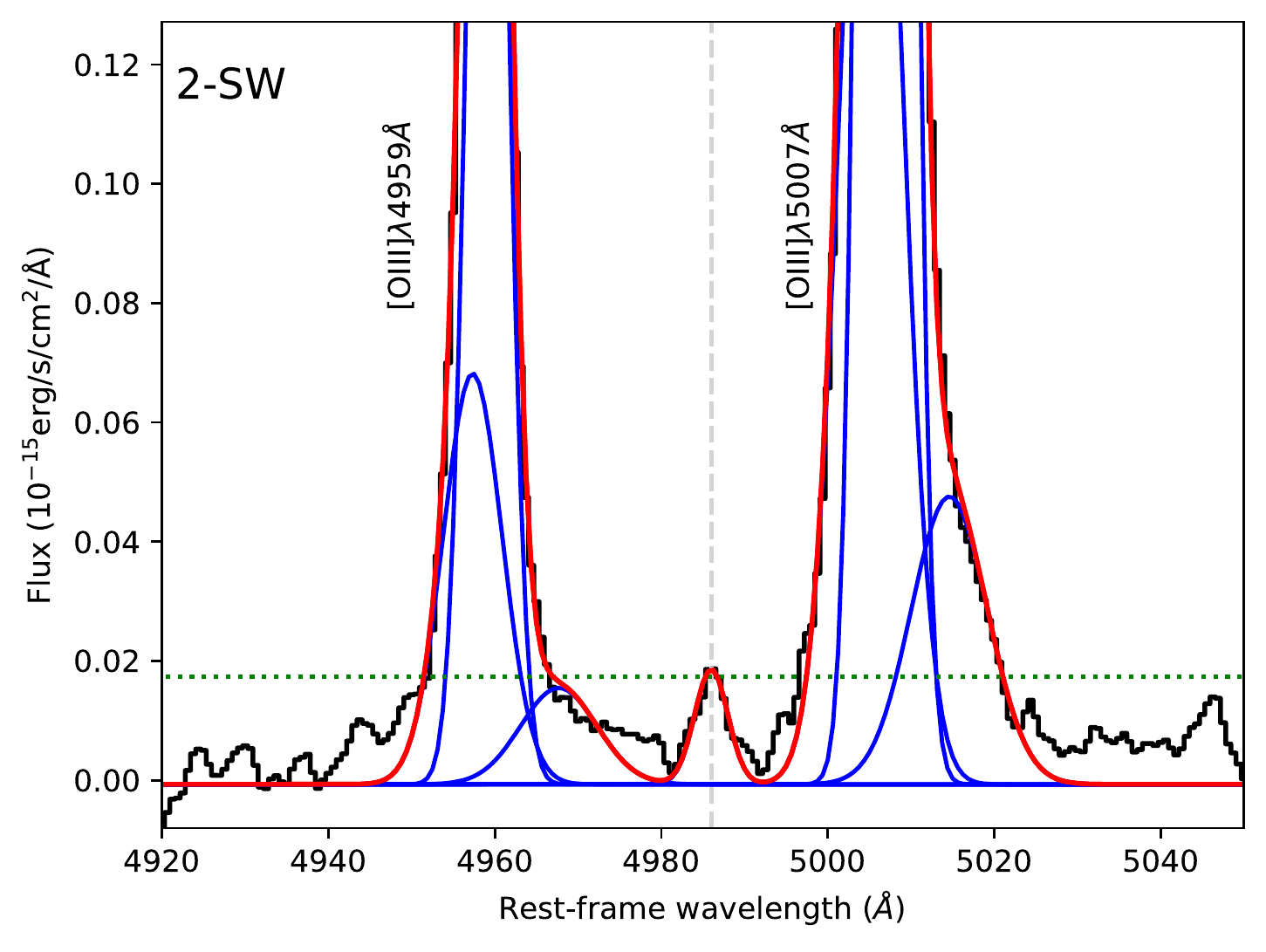}
\includegraphics[width=4.5cm]{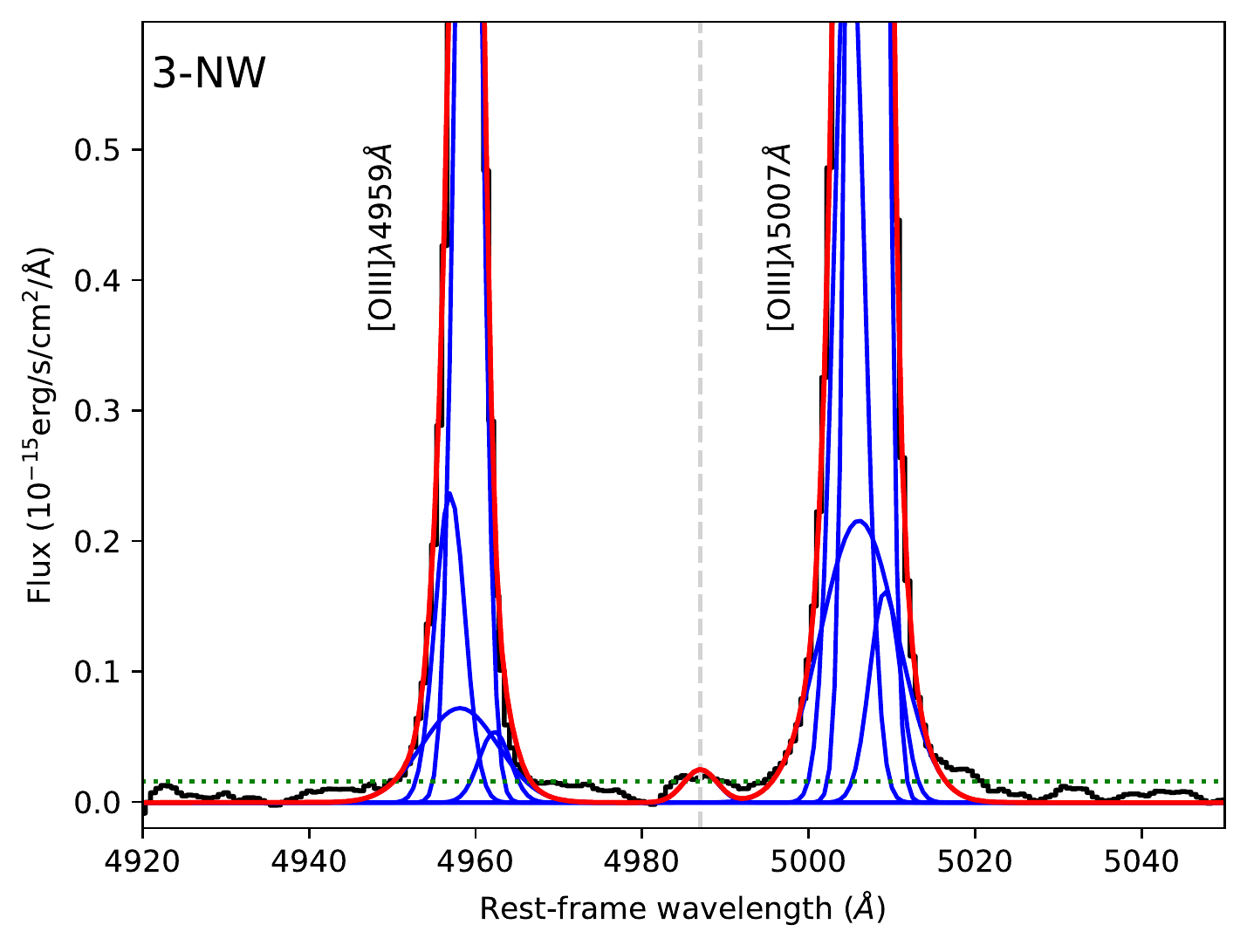}
\includegraphics[width=4.5cm]{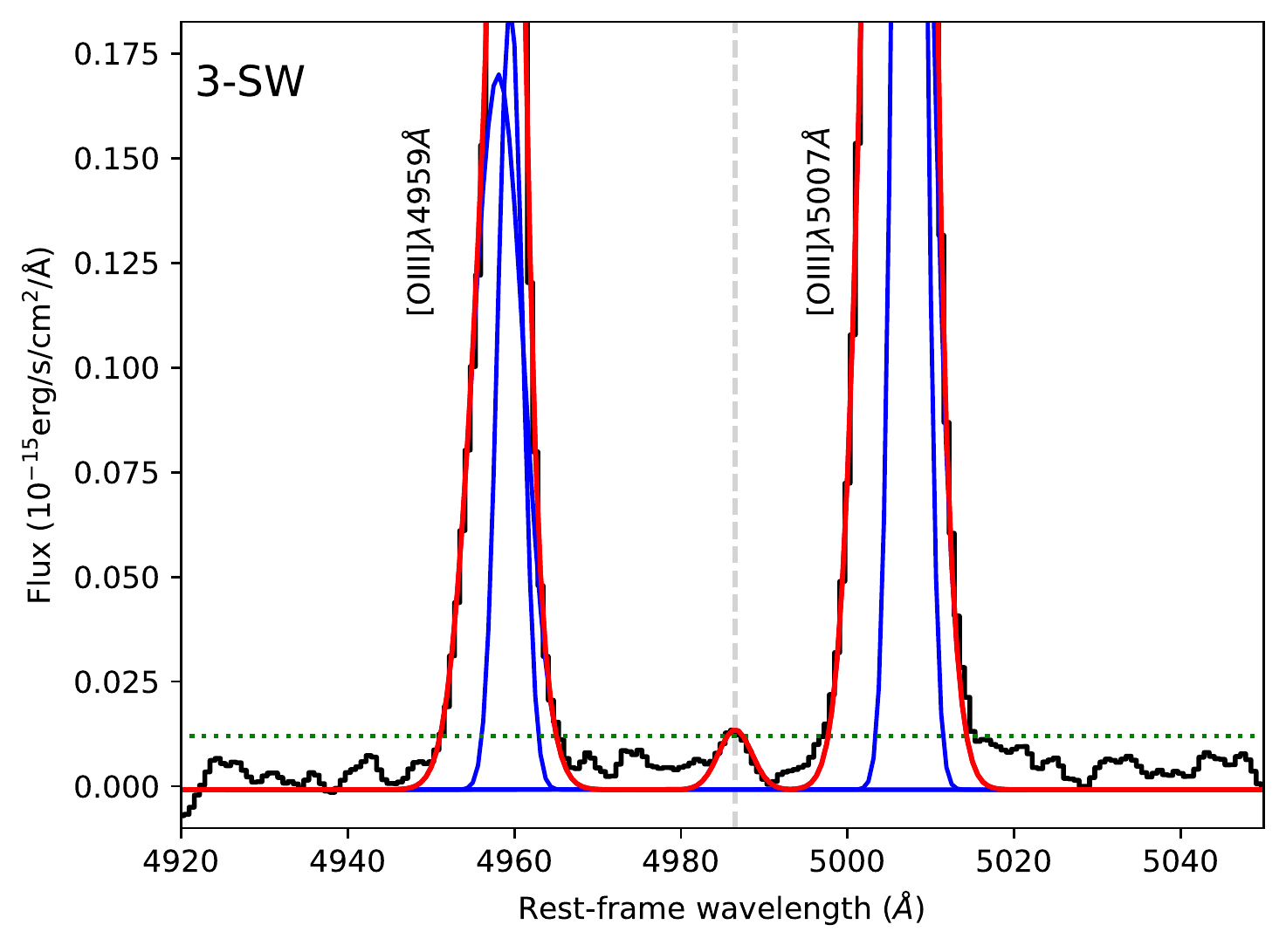}
\includegraphics[width=4.5cm]{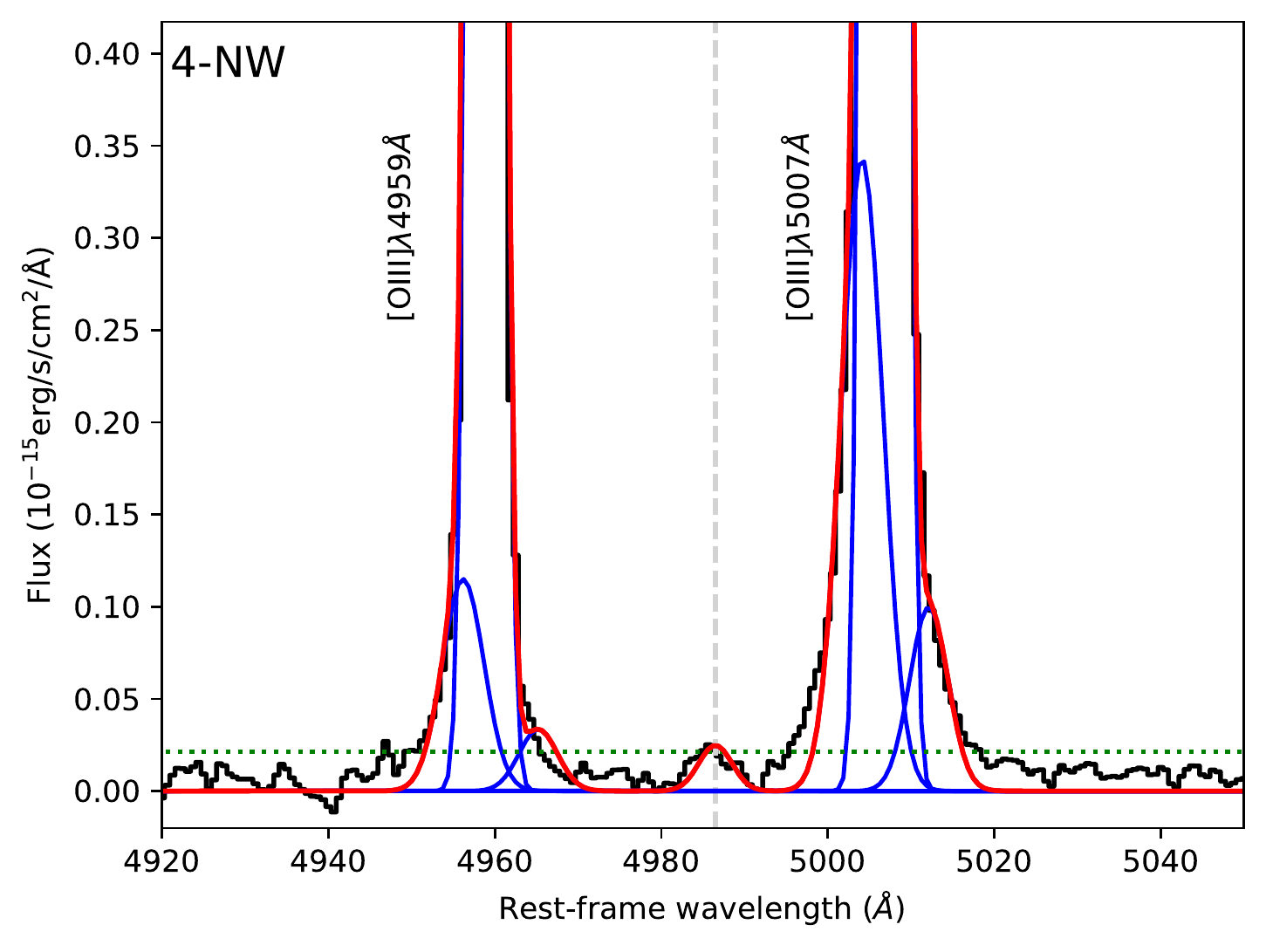}
\includegraphics[width=4.5cm]{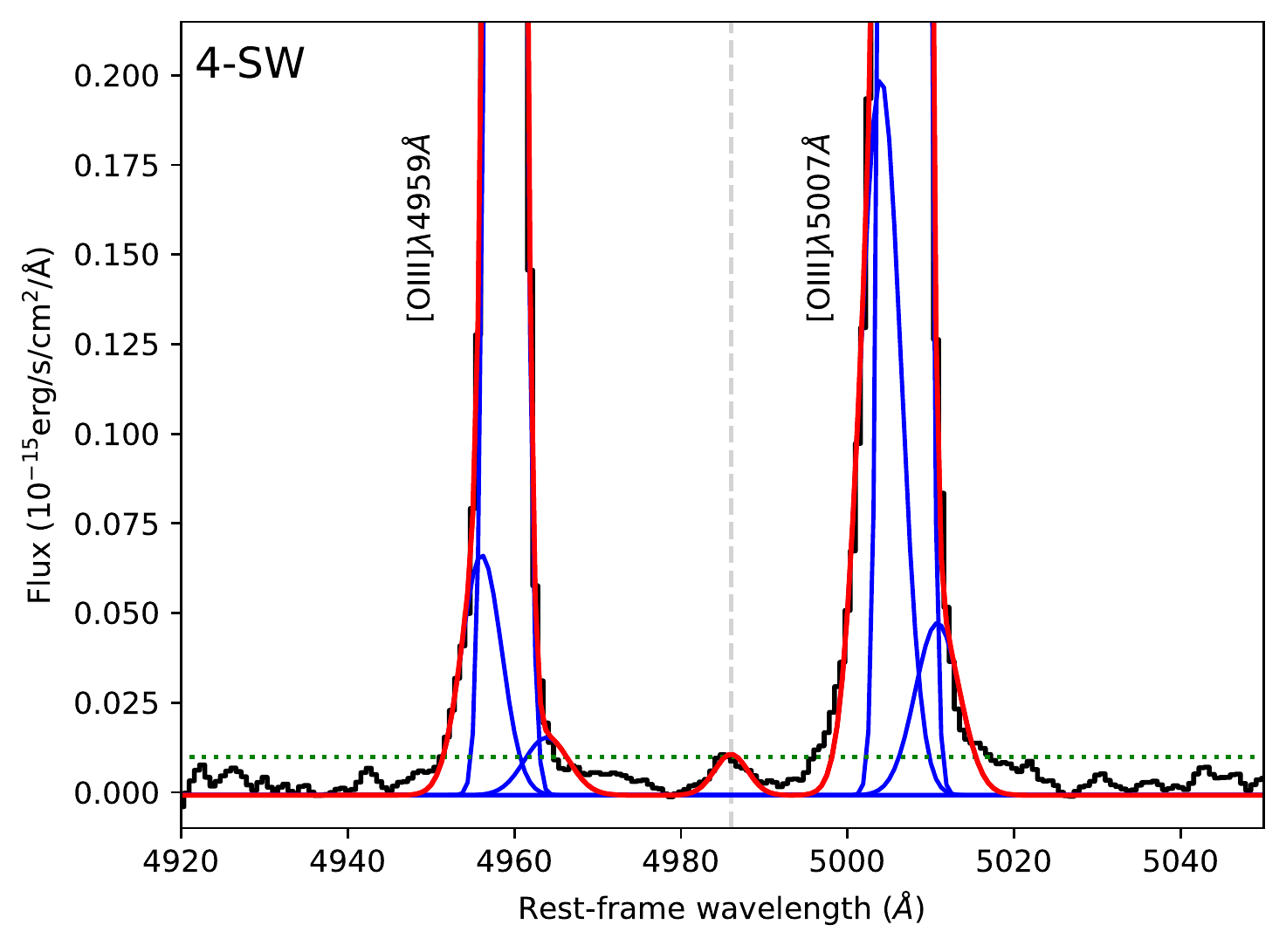}
\includegraphics[width=4.5cm]{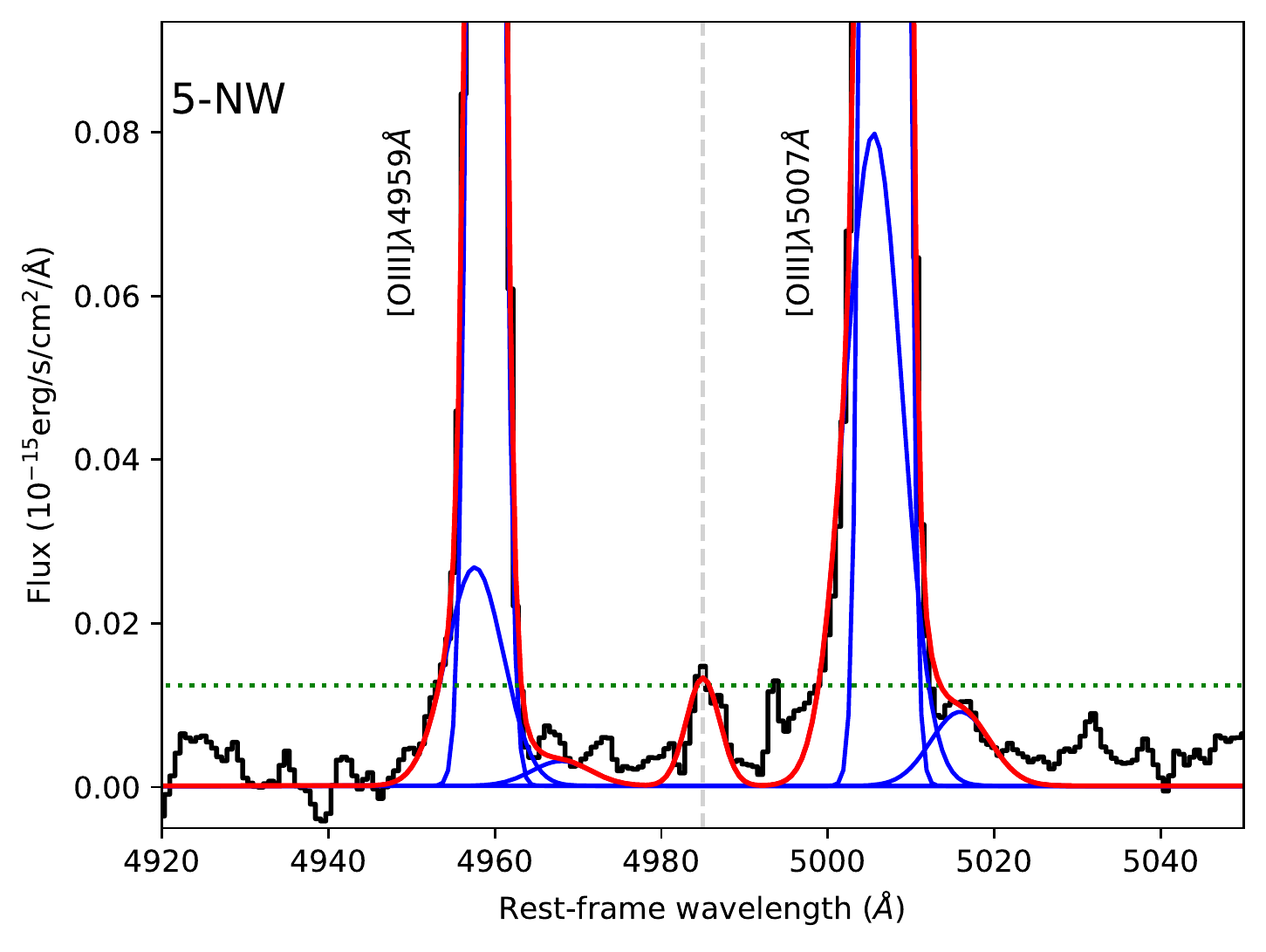}
\includegraphics[width=4.5cm]{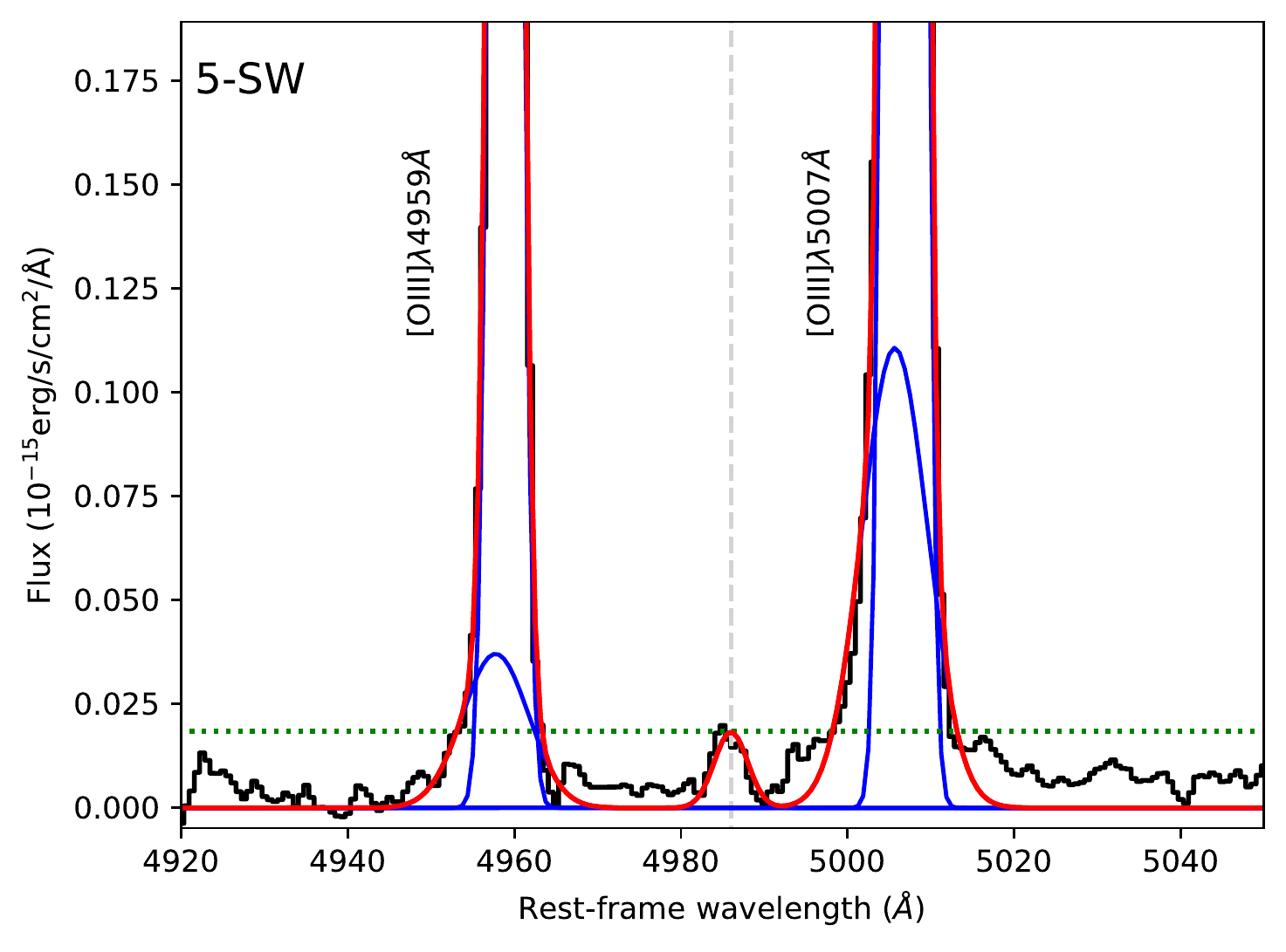}
\par}
\caption{Same as Fig. \ref{line_fits_co2} but for the [O\,{\sc iii}] emission line.}
\label{line_fits_o32}
\end{figure*}

Figures \ref{line_fits_co1},  \ref{line_fits_co2}, \ref{line_fits_o31} and \ref{line_fits_o32} show the molecular and ionized emission lines profiles for each slice. We measured the various component following the methodology presented in Section \ref{data_analysis}. Tables \ref{tab_o3} and \ref{tab_co} report the ionized and molecular outflow measurements, respectively.

\begin{table*}[ht]
\small 
\centering
\begin{tabular}{lcccccccrcc}
\hline
Slice &    Comp.& $\lambda_{out}$ & FWHM$_{out}$ & $\lambda_{gal}$& Flux$_{obs}^{out}$                   &  Flux$_{unred}^{out}$                      &  H$\alpha$/H$\beta$ & V$_{central}^{out}$& [S\,II]$_{ratio}^{out}$ & n$_e^{out}$\\
        
 &    & ($\AA$) & ($\AA$) & ($\AA$) & (erg~s$^{-1}$~cm$^{-2}$)          &  (erg~s$^{-1}$~cm$^{-2}$)                     & & (km~s$^{-1}$)& & (cm$^{-3}$)\\  

(1) &(2)& (3) & (4) & (5) & (6)          &  (7)                     & (8) & (9)& & (10)\\           
\hline
1ne & a &   4995.54 & 5.67& 5007.53    &   (1.47$\pm$0.03)$\times$10$^{-15}$ &   (1.67$\pm$0.03)$\times$10$^{-15}$   &  2.97               & -721     & 1.05              & 401 \\ 
1ne & b &   5005.64 & 8.02 & 5007.53    &   (5.18$\pm$0.01)$\times$10$^{-14}$ &   (5.90$\pm$0.01)$\times$10$^{-14}$   &  2.97               & -114     & 1.05              & 401 \\   
2ne & a &   5007.66 & 4.62 & 5006.56    &   (2.56$\pm$0.02)$\times$10$^{-14}$ &   (4.07$\pm$0.03)$\times$10$^{-14}$   &  3.26               &  66     & 1.20              & 202 \\    
2ne & b &   5006.49 & 10.55 & 5006.56    &   (8.11$\pm$0.21)$\times$10$^{-15}$ &   (1.29$\pm$0.03)$\times$10$^{-14}$   &  3.26               & -5     & 1.20              & 202 \\
3ne & a &   5009.63 & 4.11 & 5006.80   &   (4.77$\pm$0.34)$\times$10$^{-15}$ &   (5.64$\pm$0.40)$\times$10$^{-15}$   &  3.00               &  170     & 1.10              & 328 \\
3ne & b &   5006.93 & 9.15 & 5006.80   &   (7.50$\pm$0.57)$\times$10$^{-15}$ &   (8.87$\pm$0.67)$\times$10$^{-15}$   &  3.00               &  8     & 1.10              & 328 \\
4ne & a &   5001.95 & 4.52 & 5006.84    &   (7.44$\pm$0.32)$\times$10$^{-16}$ &   (8.16$\pm$0.35)$\times$10$^{-16}$   &  2.94               &  -294     & 1.15              & 258 \\  
4ne & b &   5009.50 & 5.92 & 5006.84    &   (1.43$\pm$0.05)$\times$10$^{-15}$ &   (1.57$\pm$0.05)$\times$10$^{-15}$  &  2.94               &  160    & 1.15              & 258 \\
5ne & a &   5003.79 & 8.34 & 5006.85    &   (1.01$\pm$0.05)$\times$10$^{-15}$ &   (1.14$\pm$0.05)$\times$10$^{-15}$  &  2.96               &  -184    & 1.53              & $<$50 \\   
\hline
1se & a &   5003.08 & 4.32 & 5007.01    &   (1.39$\pm$0.01)$\times$10$^{-14}$ &   (3.79$\pm$0.03)$\times$10$^{-14}$   &  3.81               & -236     & 1.06              & 384 \\
1se & b &   5012.82 & 10.79 & 5007.01   &   (3.85$\pm$0.10)$\times$10$^{-15}$ &   (1.05$\pm$0.03)$\times$10$^{-14}$   &  3.81               & 350     & 1.06              & 384 \\   
1se & c &   5000.53 & 10.81& 5007.01    &   (2.47$\pm$0.01)$\times$10$^{-14}$ &   (6.76$\pm$0.01)$\times$10$^{-14}$   &  3.81               & -390     & 1.06              & 384 \\ 
2se & a &   5004.07 & 4.10 & 5006.97    &   (1.61$\pm$0.01)$\times$10$^{-14}$ &   (1.61$\pm$0.01)$\times$10$^{-14}$   &  2.86               & -174     & 1.18              & 231 \\
2se & b &   5010.04 & 4.11 & 5006.97    &   (4.82$\pm$0.10)$\times$10$^{-15}$ &   (4.82$\pm$0.10)$\times$10$^{-15}$   &  2.86               & 185     & 1.18              & 231 \\  
2se & c &   4999.13 & 8.65 & 5006.97    &   (2.41$\pm$0.02)$\times$10$^{-15}$ &   (2.41$\pm$0.02)$\times$10$^{-15}$   &  2.86               & -471     & 1.18              & 231 \\
3se & a &   5009.39 & 4.38 & 5006.81    &   (5.96$\pm$0.25)$\times$10$^{-15}$ &   (2.72$\pm$0.11)$\times$10$^{-14}$   &  4.41               &  155     & 1.29              & 120 \\  
3se & b &   5004.46 & 8.59 & 5006.81    &   (6.02$\pm$0.19)$\times$10$^{-15}$ &   (2.75$\pm$0.08)$\times$10$^{-14}$   &  4.41               &  -141    & 1.29              & 120 \\
4se & a &   5008.52 & 4.09 & 5006.79    &   (2.11$\pm$0.19)$\times$10$^{-15}$ &   (4.31$\pm$0.39)$\times$10$^{-15}$  &  3.50               & 104     & 1.22              & 184 \\
4se & b &   5005.64 & 10.80  & 5006.79  &   (2.29$\pm$0.20)$\times$10$^{-15}$ &   (4.67$\pm$0.40)$\times$10$^{-15}$  &  3.50               &  -69     & 1.22              & 184 \\    
5se & a &   5006.72 & 10.81 & 5006.84   &   (6.70$\pm$0.45)$\times$10$^{-16}$ &   (8.01$\pm$0.54)$\times$10$^{-16}$  &  3.01               &  -7    & 1.11              & 310 \\
\hline
1nw & a &   5016.01 & 10.81& 5007.09    &   (2.78$\pm$0.05)$\times$10$^{-15}$ &   (2.32$\pm$0.04)$\times$10$^{-14}$   &  5.23               & 536     & 1.06              & 387 \\ 
1nw & b &   5001.86 & 10.80 & 5007.09    &   (1.05$\pm$0.01)$\times$10$^{-14}$ &   (8.71$\pm$0.08)$\times$10$^{-14}$   &  5.23               & -315     & 1.06              & 387 \\
2nw & a &   5005.51 & 8.84 & 5006.91    &   (3.57$\pm$0.03)$\times$10$^{-15}$ &   (2.52$\pm$0.02)$\times$10$^{-14}$  &  4.99              & -84     & 1.11              & 316 \\
2nw & a &   5015.50 & 10.79 & 5006.91    &   (6.48$\pm$0.08)$\times$10$^{-16}$ &   (4.58$\pm$0.06)$\times$10$^{-15}$  &  4.99              & 516     & 1.11              & 316 \\
3nw & a &   5004.86 & 4.21 & 5007.07    &   (3.17$\pm$0.03)$\times$10$^{-15}$ &   (1.16$\pm$0.01)$\times$10$^{-14}$   &  4.14               &  -133     & 1.14             & 275 \\
3nw & b &   5009.19 & 4.38 & 5007.07    &   (7.58$\pm$0.33)$\times$10$^{-16}$ &   (2.77$\pm$0.01)$\times$10$^{-15}$   &  4.14               &  127     & 1.14             & 275 \\   
3nw & c &   5006.04 & 10.80 & 5007.07    &   (2.49$\pm$0.04)$\times$10$^{-15}$ &   (9.09$\pm$0.01)$\times$10$^{-15}$   &  4.14              &  -62     & 1.14             & 275 \\    
4nw & a &   5012.14 & 5.45 & 5006.86    &   (1.94$\pm$0.23)$\times$10$^{-15}$ &   (2.84$\pm$0.33)$\times$10$^{-15}$   &  3.34               &  318     & 1.22              & 180 \\   
4nw & b &   5004.12 & 5.92 & 5006.86    &   (2.16$\pm$0.13)$\times$10$^{-15}$ &   (3.73$\pm$0.21)$\times$10$^{-15}$  &  3.34               &  -165     & 1.22              & 180 \\
5nw & a &   5005.50 & 8.05 & 5006.87    &   (6.83$\pm$0.17)$\times$10$^{-16}$ &   (7.70$\pm$0.20)$\times$10$^{-16}$  &  2.96               &  82    & 1.43              & $<$50 \\
\hline
1sw & a &   5008.53 & 7.48 & 5006.58    &   (8.13$\pm$0.03)$\times$10$^{-15}$ &   (1.51$\pm$0.01)$\times$10$^{-13}$   &  6.58               & 117     & 1.07              & 377 \\
1sw & b &   5001.27 & 9.36 & 5006.58    &   (2.97$\pm$0.03)$\times$10$^{-15}$ &   (5.54$\pm$0.05)$\times$10$^{-14}$   &  6.58               & -319     & 1.07              & 377 \\
1sw & c &   5018.67 & 10.79 & 5006.58    &   (1.18$\pm$0.01)$\times$10$^{-15}$ &   (1.93$\pm$0.02)$\times$10$^{-15}$   &  6.58               & 726     & 1.07              & 377 \\
2sw & a &   5005.26 & 8.34 & 5007.06    &   (1.82$\pm$0.01)$\times$10$^{-15}$ &   (1.51$\pm$0.01)$\times$10$^{-14}$   &  5.22               & -108     & 1.33              & 84 \\
2sw & b &   5014.61 & 10.80 & 5007.06    &   (5.55$\pm$0.05)$\times$10$^{-16}$ &   (4.58$\pm$0.04)$\times$10$^{-15}$   &  5.22               & 454     & 1.33              & 84 \\
3sw & a &   5005.98 & 7.21 & 5007.42    &   (3.92$\pm$0.01)$\times$10$^{-15}$ &   (6.81$\pm$0.01)$\times$10$^{-15}$   &  3.35               &  -86     & 1.30              & 109 \\  
4sw & a &   5003.98 & 5.69 & 5006.86    &   (1.21$\pm$0.09)$\times$10$^{-15}$ &   (2.13$\pm$0.16)$\times$10$^{-15}$  &  3.36               & -173     & 1.31              & 102 \\
4sw & b &   5010.79 & 6.27  & 5006.86    &   (3.21$\pm$0.71)$\times$10$^{-16}$ &   (5.64$\pm$1.24)$\times$10$^{-16}$  &  3.36               & 236     & 1.31              & 102 \\    
5sw & a &   5005.67 & 9.30  & 5006.86    &   (1.10$\pm$0.03)$\times$10$^{-15}$ &   (1.13$\pm$0.03)$\times$10$^{-15}$  &  2.89               &  -72   & 1.51              & $<$50 \\    
\hline
\end{tabular}						 
\caption{Summary of the ionized outflow measurements. (a), (b) and (c) correspond to the various fitted outflowing components.} %Note that when we use only two Gaussians for the fit, the component b is not present in the Table.} 
\label{tab_o3}
\end{table*}

\begin{table*}[ht]
\small 
\centering
\begin{tabular}{llcccc}
\hline
Slice &    Comp.& V$_{non-rot}$ & FWHM$_{non-rot}$ & V$_{gal}$& Flux$_{non-rot}$    \\%& V$_{max}$\\
      &         & (km~s$^{-1}$)          & (km~s$^{-1}$)           & (km~s$^{-1}$)     & (Jy~km~$s^{-1}$) \\%& (km~s^{-1})\\
(1)      &(2)         &(3)          &(4)           &(5)     &(6) \\%&(7)\\
      
\hline
1ne & a$_{out}$ &   1242 & 27 & 1198   &   6.71$\pm$0.02 \\%& 57 \\
1ne & b$_{out}$ &   1280 & 40 & 1198   &   5.12$\pm$0.02 \\%& 102 \\ 

2ne & ...       & ...       & ...   & ...       &   ...           \\%& ...    \\

3ne & ...       & ...       & ...   & ...       &   ...          \\% & ...    \\

4ne & a$_{out}$ &   1180 & 31 & 1199   &   2.80$\pm$0.02 \\% & -35 \\
4ne & b$_{out}$ &   1283 & 26 & 1199   &   2.28$\pm$0.07 \\%& 97 \\

5ne & a$_{out}$ &   1108 & 21 & 1198   &   2.37$\pm$0.03 \\%& -100 \\
5ne & b$_{out}$ &   1255 & 35 & 1198   &   12.18$\pm$0.13 \\%& 75 \\
\hline
1se & a$_{in}$  &   1165 & 41 & 1198   &   10.20$\pm$0.03 \\%& {\textit{-55}} \\
1se & b$_{out}$ &   1237 & 43 & 1198   &   16.26$\pm$0.05 \\%& 59 \\

2se & a$_{in}$  &   1267 & 23 & 1198   &   0.81$\pm$0.03 \\%& {\textit{80}} \\

3se & a$_{in}$  &   1229 & 21 & 1198   &   1.06$\pm$0.02 \\%& {\textit{42}} \\

4se & a$_{out}$ &   1134 & 31 & 1198   &   1.63$\pm$0.04 \\%& -79 \\

5se & a$_{out}$ &   1106 & 25 & 1198   &   3.10$\pm$0.07 \\%& -104 \\
5se & b$_{in}$  &   1255 & 35 & 1198   &   4.26$\pm$0.06 \\%& {\textit{74}} \\ 

\hline
1nw & a$_{in}$  &   1130 & 61 & 1197   &   8.55$\pm$0.13 \\%& {\textit{-97}} \\

2nw & a$_{in}$  &   1141 & 22 & 1204   &   1.05$\pm$0.04 \\%& {\textit{-74}} \\

3nw & a$_{in}$  &   1106 & 36 & 1196   &   1.59$\pm$0.04 \\%& {\textit{-108}} \\
3nw & b$_{in}$  &   1150 & 27 & 1196   &   3.58$\pm$0.01 \\%& {\textit{-59}} \\
3nw & c$_{out}$ &   1249 & 17 & 1196   &   0.75$\pm$0.02 \\%& 62 \\

4nw & a$_{in}$  &   1050 & 15 & 1198   &   0.91$\pm$0.02 \\%& {\textit{-156}} \\
4nw & b$_{in}$  &   1148 & 47 & 1198   &   9.20$\pm$0.12 \\%& {\textit{-73}} \\

5nw & a$_{in}$  &   1039 & 16 & 1199   &   0.94$\pm$0.02 \\%& {\textit{-167}} \\
5nw & b$_{in}$  &   1123 & 33 & 1199   &   2.69$\pm$0.03 \\%& {\textit{-92}} \\
\hline
1sw & a$_{out}$ &   1153 & 39 & 1198   &   11.69$\pm$0.04 \\%& -64 \\
1sw & b$_{in}$  &   1234 & 24 & 1198   &   3.57$\pm$0.03 \\%& {\textit{48}} \\

2sw & a$_{out}$ &   1169 & 42 & 1198   &   3.08$\pm$0.01 \\%& -49 \\
2sw & b$_{out}$ &   1225 & 32 & 1198   &   2.28$\pm$0.02 \\%& 42 \\

3sw & a$_{out}$ &   1134 & 16 & 1195   &   0.36$\pm$0.02 \\%& -68 \\

4sw & a$_{out}$ &   1055 & 15 & 1198   &   0.88$\pm$0.03 \\%& -151 \\
4sw & b$_{out}$ &   1147 & 22 & 1198   &   1.84$\pm$0.01 \\%& -62 \\

5sw & a$_{out}$ &   1120 & 20 & 1202   &   0.91$\pm$0.01 \\%& -92 \\
5sw & b$_{out}$ &   1278 & 25 & 1202   &   1.76$\pm$0.03 \\%& 89 \\
\hline
\end{tabular}						 
\caption{Summary of the molecular non-rotational measurements. (a), (b) and (c) correspond to the various fitted non-rotational components. Inflow and outlfow molecular components are labeled as {\textit{in}} and {\textit{out}}.} %The former with the maximum velocity in italic style.} 
\label{tab_co}
\end{table*}

\section{Spiral arms molecular gas mass}
\label{spiral_arms}
\begin{figure*}
\centering
\includegraphics[width=10.15cm, angle=90]{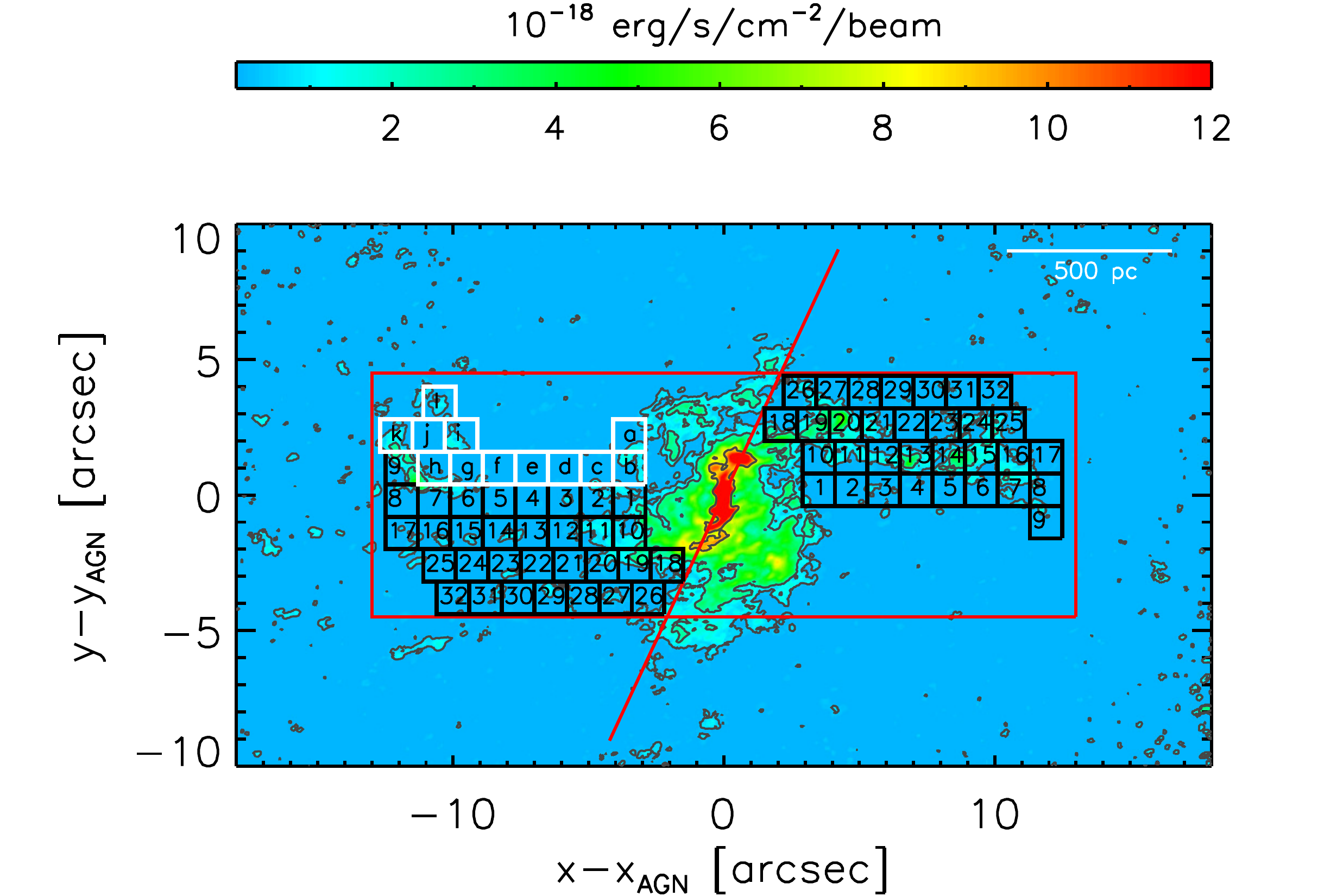}
\caption{ALMA CO(2-1) integrated intensity map of NGC\,5643 produced from the natural-weight data cube. The map is shown in a linear color scale. Black contours corresponds to the CO(2-1) emission map, which are shown in a logarithmic scale with the first contour at 8$\sigma$ and the last contour at 2.2$\times$10$^{-17}$~erg/s/cm$^{-2}$/beam. The various boxes correspond with the regions selected to measure the two main spiral arms (see Section \ref{wind_entrained}). Black boxes indicate those regions selected that are symmetric at both sides of the nucleus. White boxes correspond to the supplementary regions to take into account additional individual CO(2-1) clumps. The red diagonal solid line indicates the separation axis between both arms. North is up and east is left and offsets are measured relative to the AGN.}
\label{fig11}
\end{figure*}

To estimate the impact of the outflow wind in the destruction of the molecular gas we measure the total gas mass in both spiral arms. To do so, we first use the fully reduced optical HST/F606W image of NGC\,5643 from the ESA Hubble Legacy Archive to define the region of the two main spiral arms. Then, using the ALMA CO(2-1) integrated intensity map we extract several regions following both spiral arms (see Fig. \ref{fig11}). We use two different approaches to measure their CO(2-1): a) for simplicity, we assume symmetry in both spiral arms (see black regions in Fig. \ref{fig11}); and b) we complement the previous with the remainder CO(2-1) clumps found by visual inspection (see white regions in Fig. \ref{fig11}). For both approaches, we find that the molecular gas mass of the eastern spiral arm is M$_{east}\sim$(1.5-2.0)$\times$10$^7$~M$_\sun$. These values correspond with 50-70\% the content of the western one (see Section \ref{wind_entrained}).

\section{Observed outflow properties}
\label{appendixC}

\begin{figure*}
\centering
\par{
\includegraphics[width=9.0cm]{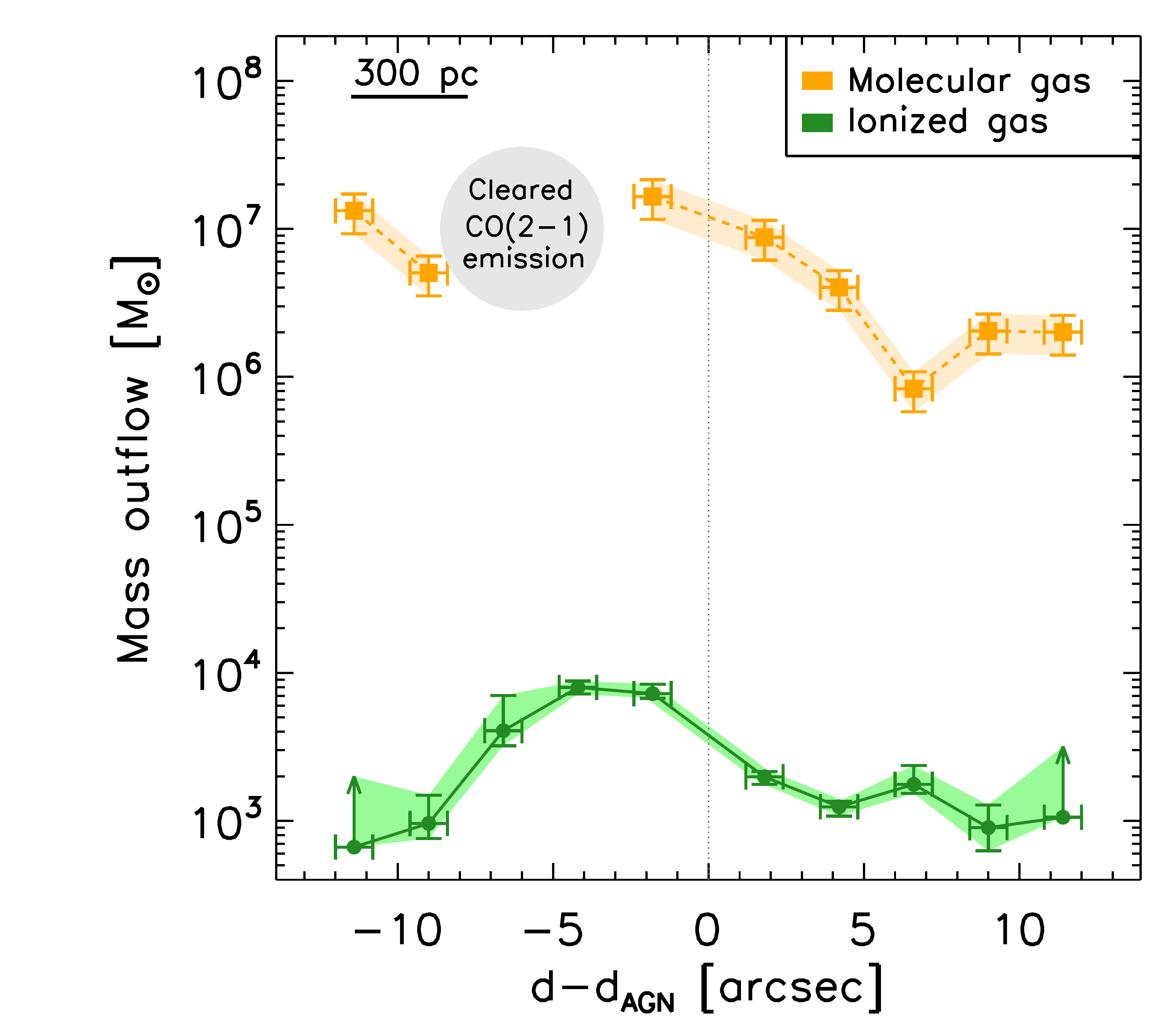}
\includegraphics[width=9.0cm]{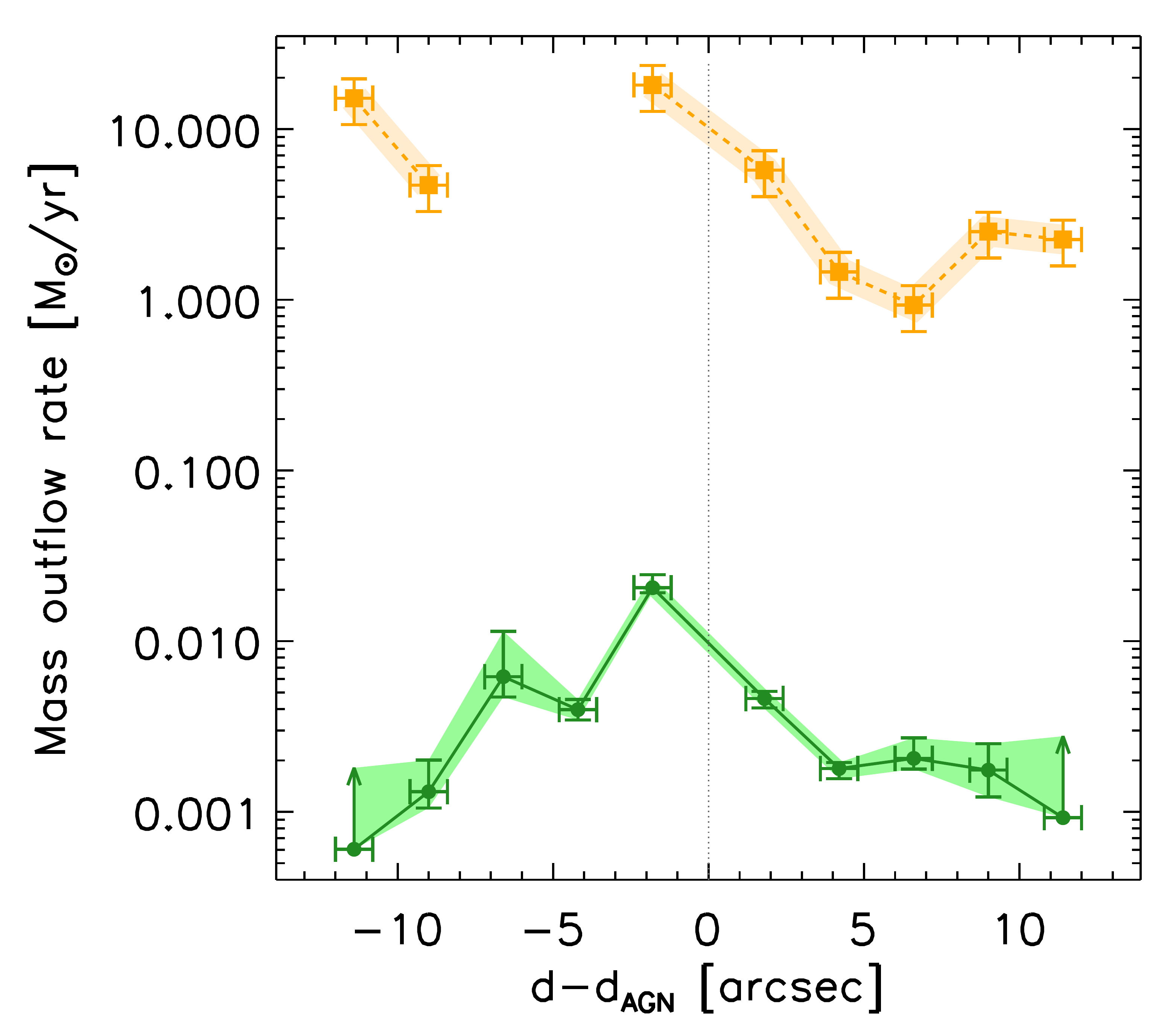}
\includegraphics[width=9.0cm]{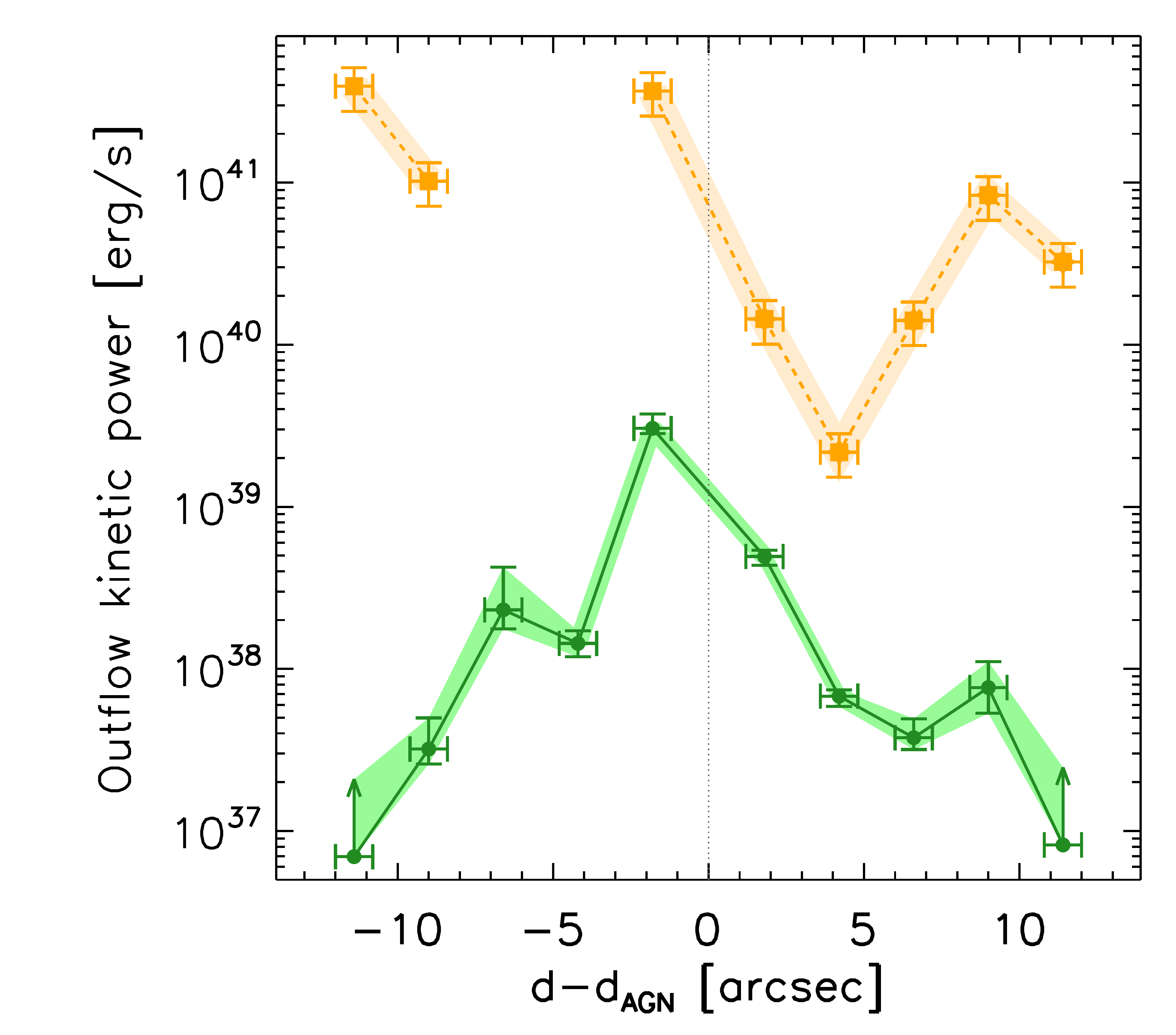}
\includegraphics[width=9.0cm]{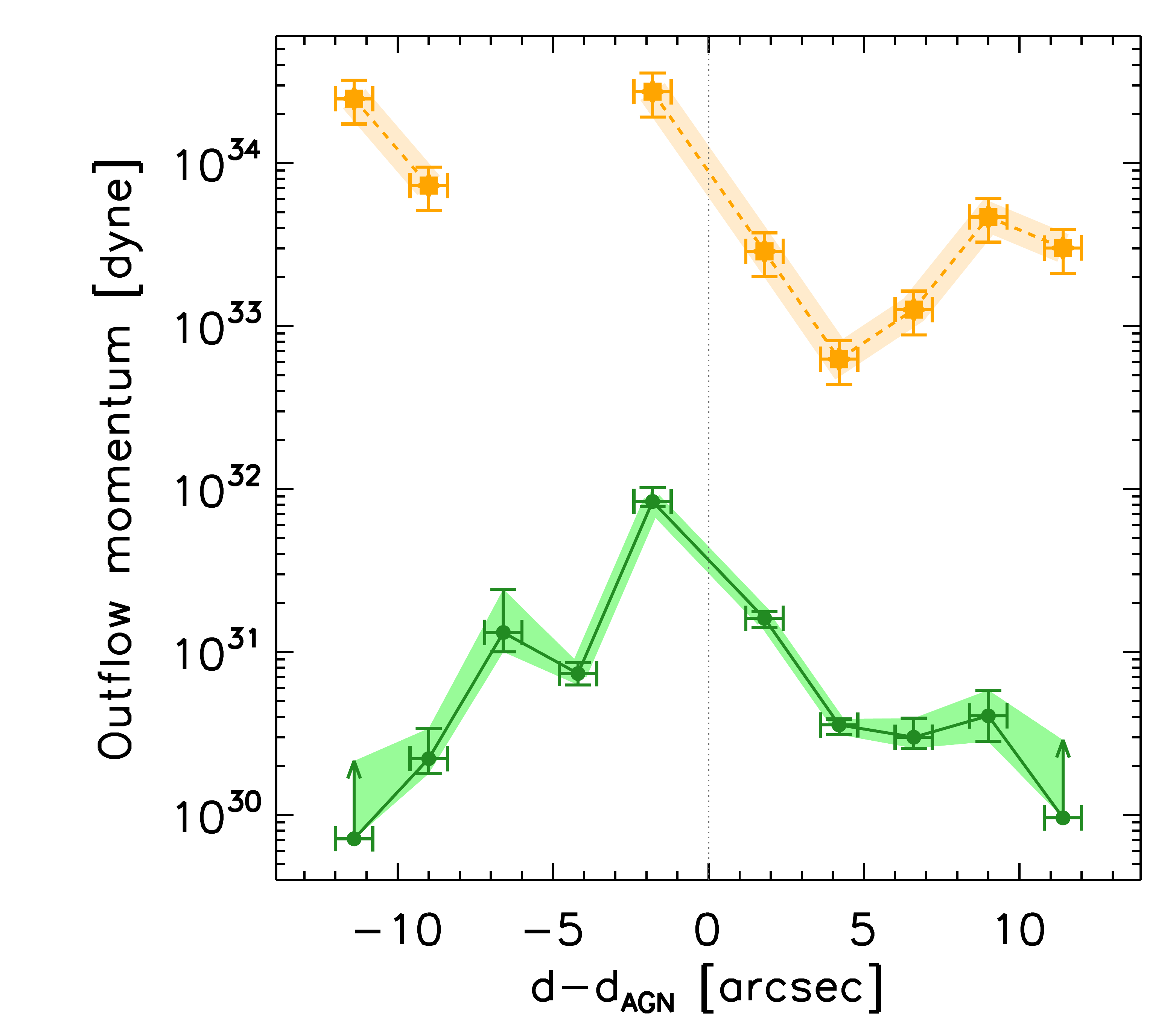}
\par}
\caption{Observed molecular and ionized gas properties. Top left panel: mass outflow profile. Top right panel: mass outflow rate profile. Bottom left panel: outflow kinematic power profile. Bottom right panel: outflow momentum profile. Orange squares and green circles correspond to the molecular and ionized outflow gas phases. Note that the two regions to the east of the AGN (negative d-d$_{AGN}$ values) with no molecular outflow derived properties correspond to regions with cleared CO(2-1) emission.}
\label{c1}
\end{figure*}

In Section \ref{Velocity_profiles} we present the outflow properties of NGC\,5643 using extinction corrected fluxes. Here we also derived the various outflow properties profiles (see Fig. \ref{c1}) without taking into account the extinction correction. As expected, these profiles show a good agreement with the morphology present in the observed [O\,{\sc iii}] emission map. In Table \ref{tabc1} we report the main derived observed outflow properties at both sides of the nucleus.

\begin{table*}
	%\small 
	\centering
	\begin{tabular}{lccc}
		\hline
		Integrated outflow properties		&Eastern &Western & Total\\
		&side    &side	  &\\
		(1)&(2)&(3)& (4)\\	
		
		\hline
		Ionized gas outflow mass  (M$_{\odot}$)			&  (2.1$\pm_{0.1}^{0.3}$)$\times$10$^4$ & (0.7$\pm0.1$)$\times$10$^4$&(2.8$\pm_{0.2}^{0.4}$)$\times$10$^4$\\
		 \noalign{\smallskip}
		Ionized gas outflow mass rate  (M$_{\odot}$ yr$^{-1}$)	& 0.03$\pm0.01$				&	0.01$\pm0.01$    &0.04$\pm0.01$\\
		 \noalign{\smallskip}
		Ionized gas outflow kinematic power (erg s$^{-1}$)& (3.5$\pm_{0.2}^{0.7}$)$\times$10$^{39}$&(6.8$\pm0.6$)$\times$10$^{38}$ &(4.1$\pm_{0.3}^{0.7}$)$\times$10$^{39}$ \\
		 \noalign{\smallskip}
		Ionized gas outflow momentum (dyne)& 
		(1.1$\pm_{0.1}^{0.2}$)$\times$10$^{32}$&(2.8$\pm0.3$)$\times$10$^{31}$ &(1.3$\pm_{0.1}^{0.2}$)$\times$10$^{32}$ \\
		 \noalign{\smallskip}
		\hline
	\end{tabular}						 
	\caption{Summary of the observed ionized outflow properties.} 
	\label{tabc1}
\end{table*}

\section{CO-to-H$_2$ conversion factor}
\label{appendixD}
\begin{figure*}[ht!]
\centering
\par{
\includegraphics[width=8.0cm]{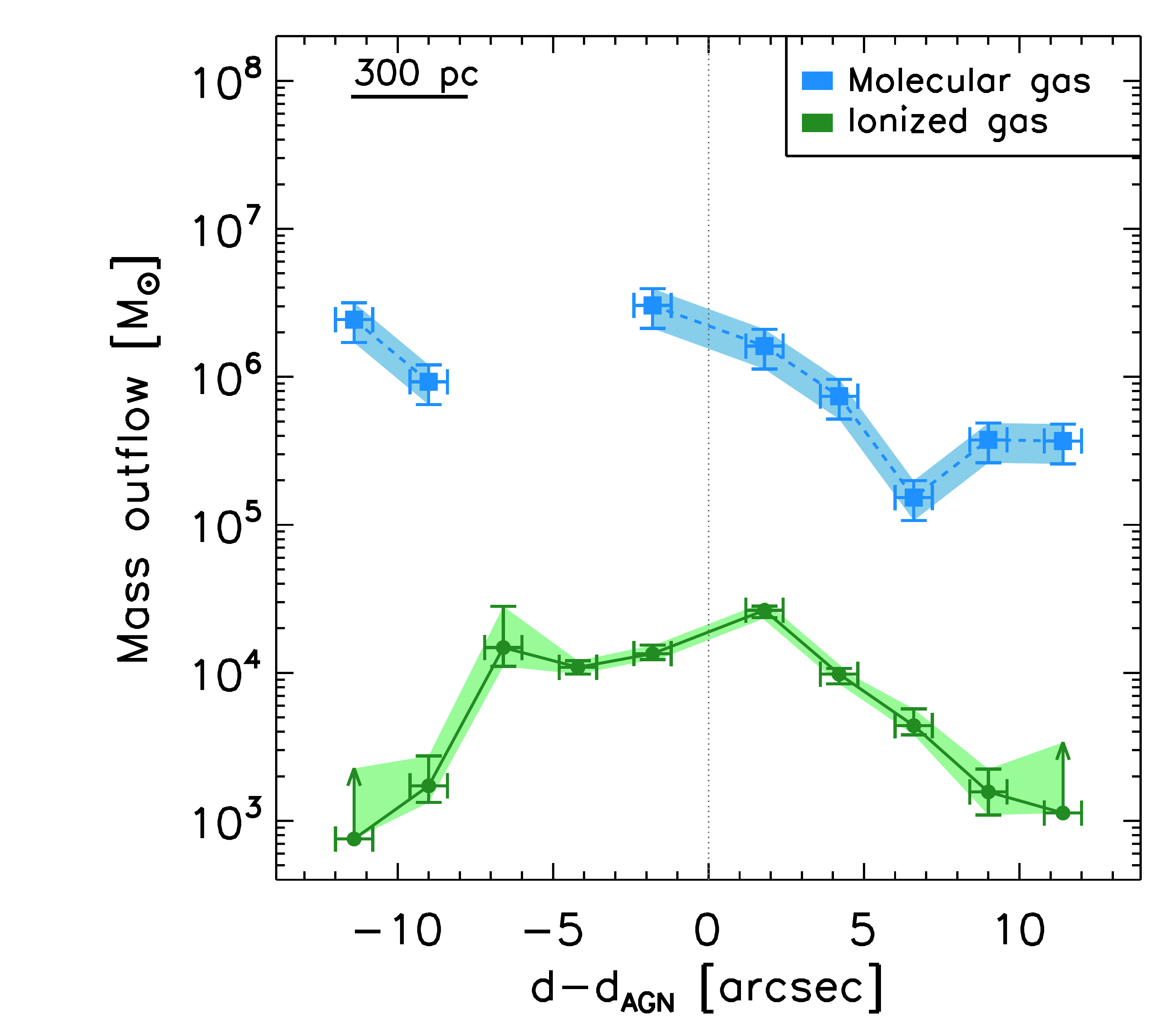}
\includegraphics[width=8.0cm]{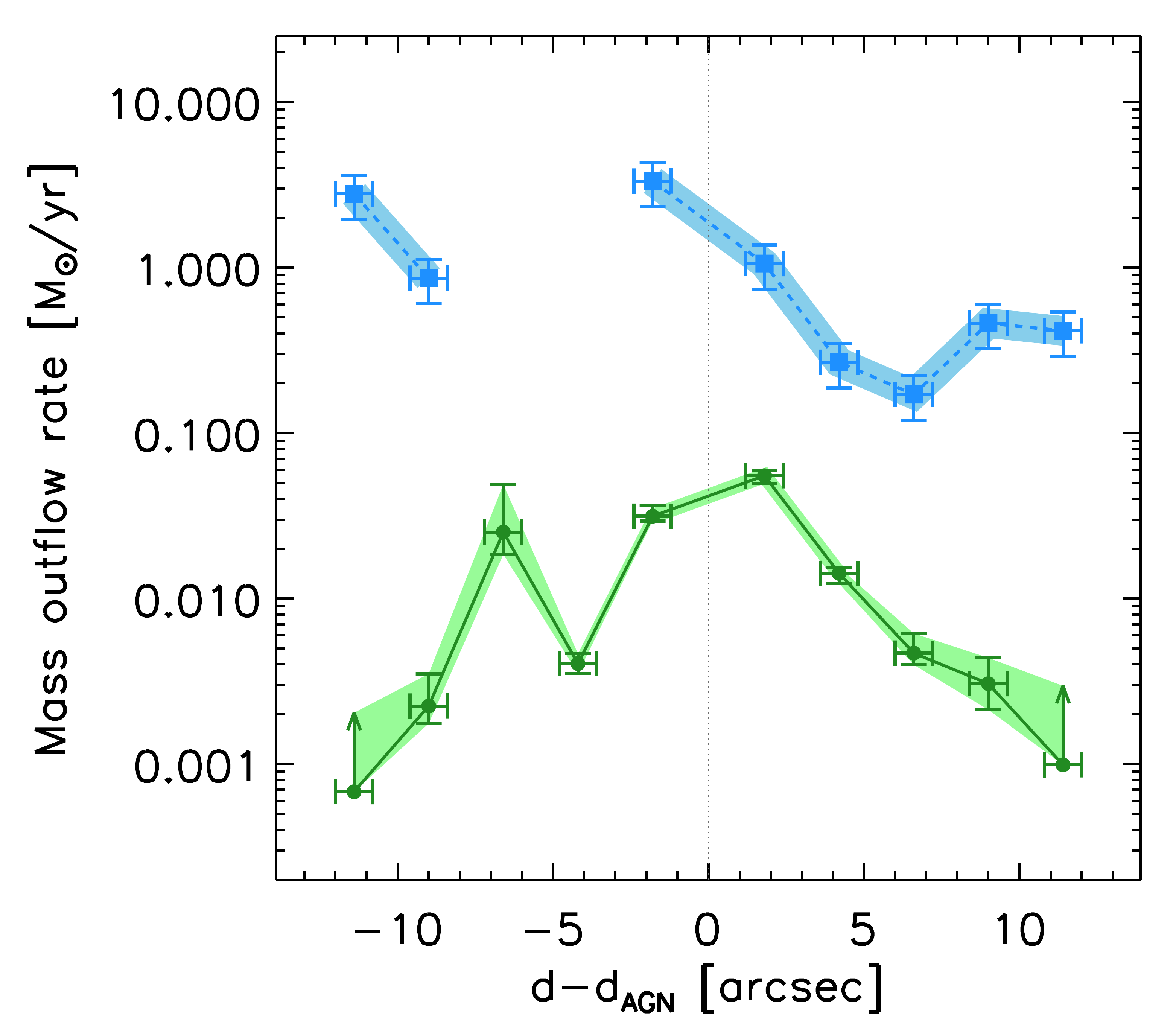}
\includegraphics[width=8.0cm]{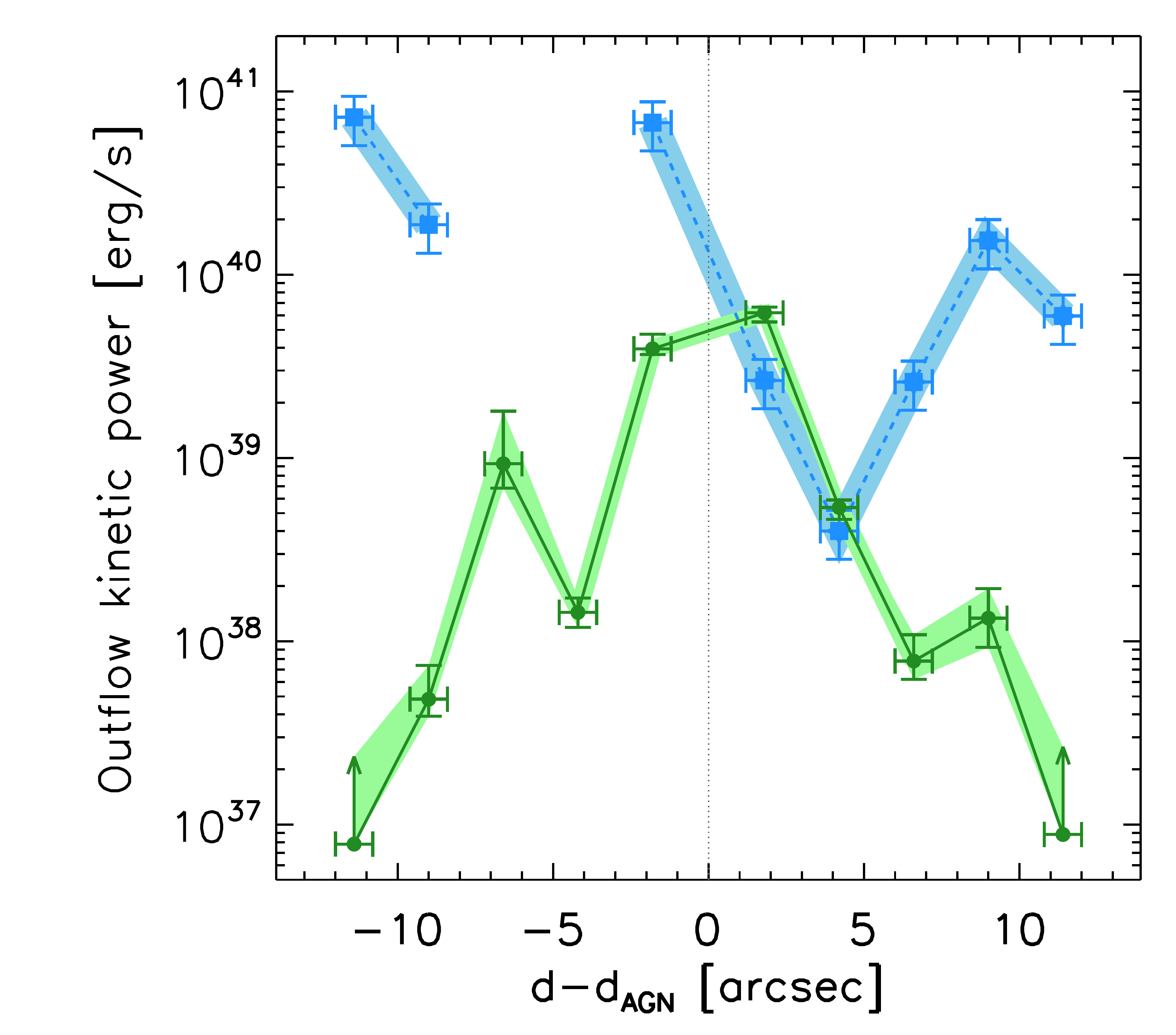}
\includegraphics[width=8.0cm]{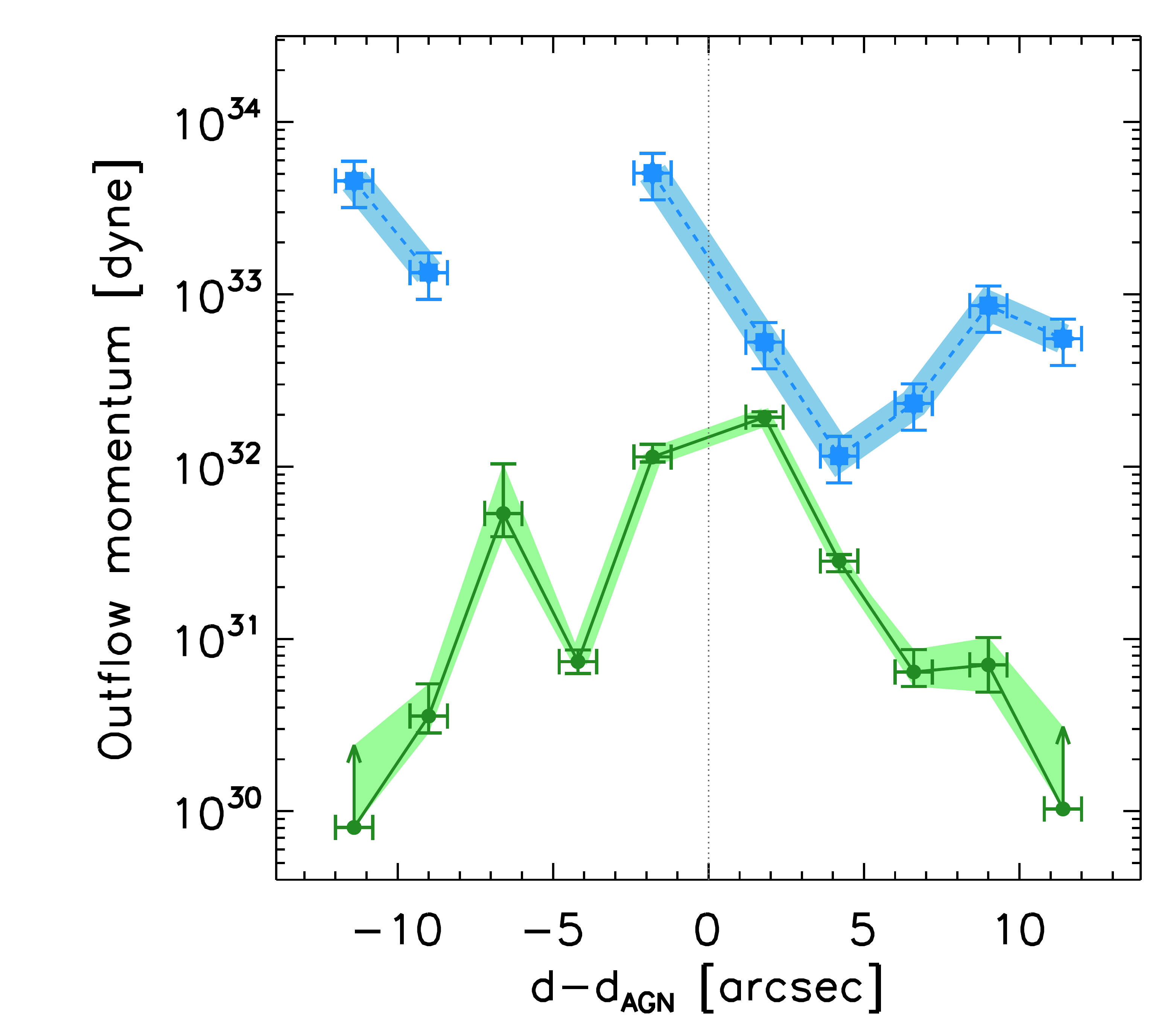}
\par}
\caption{Spatially resolved properties of the ionized (green circles) and molecular (blue squares) phases of the outflow:  outflowing mass (top left), 
mass outflow rate (top right), outflow kinetic power (center left), outflow momentum (center right). Blue squares and green circles correspond to the molecular and ionized outflow gas, respectively. Note that the two regions to the east of the AGN (negative d-d$_{AGN}$ values) with no molecular outflow derived properties correspond to regions with cleared CO(2-1) emission.}
\label{figd1}
\end{figure*}

\begin{figure*}
\centering
\par{
\includegraphics[width=8.6cm]{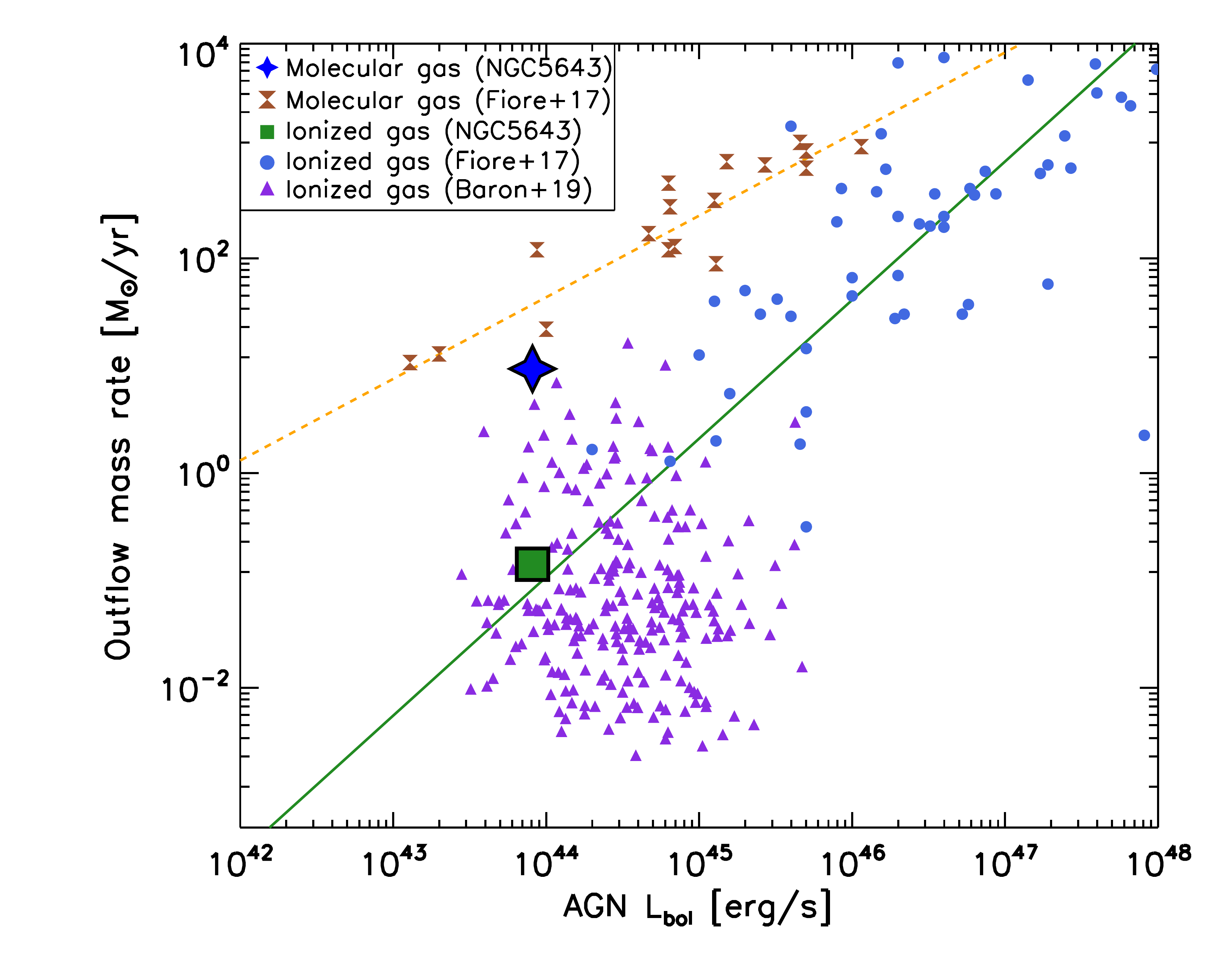}
\includegraphics[width=8.6cm]{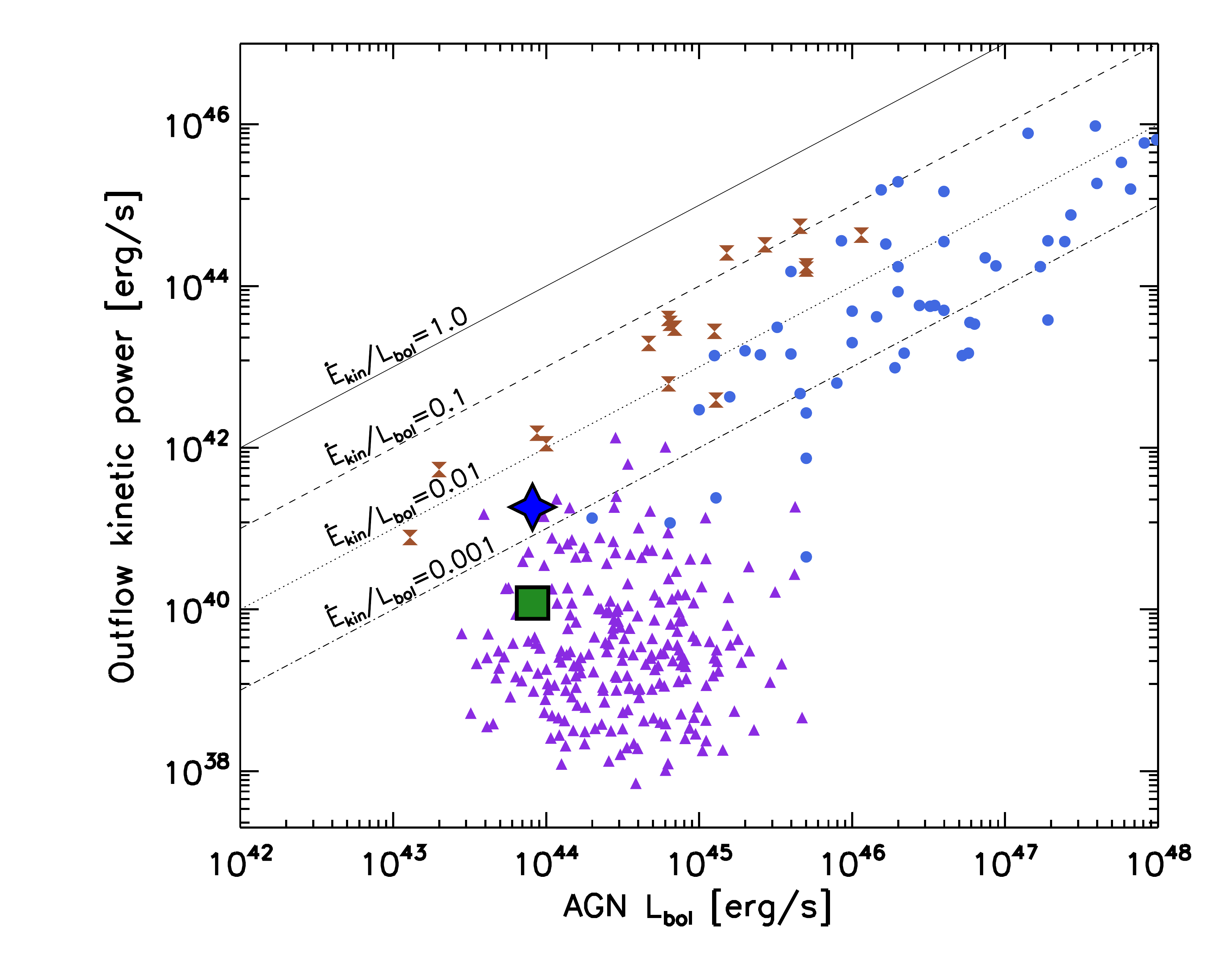}
\includegraphics[width=8.6cm]{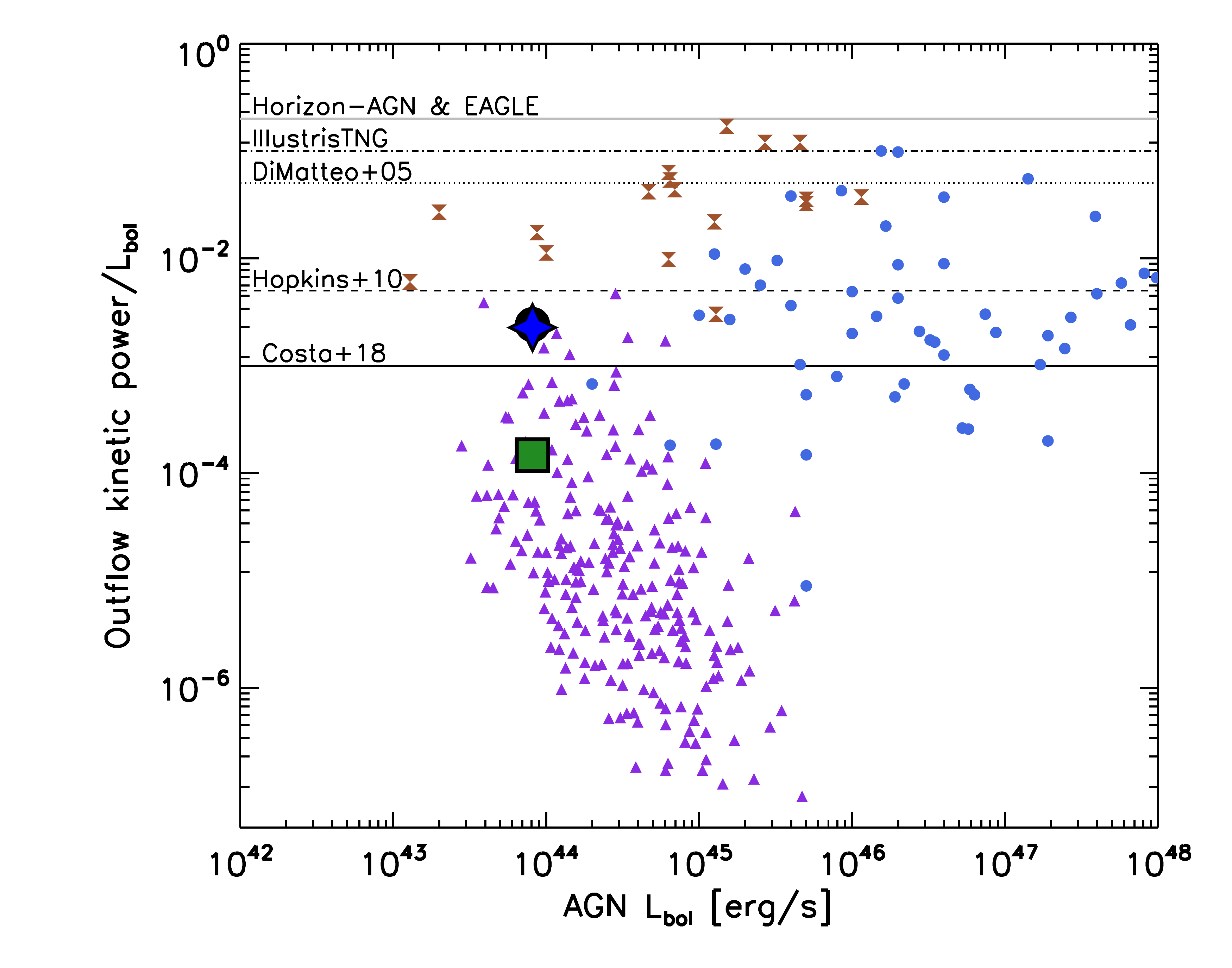}
\par}
\caption{Top left panel: outflow mass rate as a function of the AGN luminosity. The dashed orange and solid green lines are the best fit correlations derived by \citet{Fiore17} for the molecular and ionized gas, respectively. Top right panel: same as  top left panel, but for the outflow kinetic power. Solid, dashed, dotted and dash-dotted lines represent  $\dot{E}_{kin}=$1.0,0.1,0.01,0.001~L$_{bol}$. Bottom panel: kinetic coupling efficiencies. The various horizontal lines correspond to theoretical values (\citealt{Costa18,DiMatteo05,Dubois14,Hopkins10,Schaye15,Weinberger17}). Blue stars and green squares represent the values derived in this work for the molecular and ionized phase of NGC\,5643, respectively. The black circle is the total kinetic coupling efficiencies for both (ionzed and molecular) gas phases.  Brown hourglass and blue circles are from \citet{Fiore17} for the molecular and ionized phase, respectively, and the purple triangles from \citet{Baron19} for ionized outflows. Note that we have consistently applied the same methodology as in \citet{Fiore17} for  the total ionized gas mass, as $3\times  M_{\rm out}^{\rm [O\,III]}$.}
\label{figd2}
\end{figure*}

As previously mentioned, the CO-to-H$_2$ conversion factor is poorly constrained for Seyfert galaxies. Therefore, in Section \ref{Velocity_profiles} we present the molecular gas outflow properties based on the Galactic conversion factor ($\alpha_{\rm CO}=$M$_{gas}$/L$_{CO}^{'}$=4.35~M$_{\sun}$[K~km~s$^{-1}$~pc$^2$]$^{-1}$; \citealt{Bolatto13}) and a the CO(1-0)/CO(2-1) brightness temperature ratio of one. However, these values depend strongly on the CO-to-H$_2$ conversion factor used.  Some previous works also used the conversion factor of ULIRGs (see e.g. \citealt{Cicone14,Lutz20}). Therefore, for comparison, we also estimate the molecular gas outflow properties using a lower  CO-to-H$_2$ conversion factor (i.e. ULIRGs factor; $\alpha_{\rm CO}$=0.8~M$_{\sun}$[K~km~s$^{-1}$~pc$^2$]$^{-1}$; \citealt{Bolatto13}). Table \ref{tabd1}  and Fig. \ref{figd1} show the main molecular gas outflow properties on both sides of the nucleus using $\alpha_{\rm CO}$=0.8~M$_{\sun}$[K~km~s$^{-1}$~pc$^2$]$^{-1}$~pc$^2$]$^{-1}$. Moreover, Fig. \ref{figd2} shows the comparison between the total derived properties for NGC\,5643 and the properties for other AGN, which is discussed in Section \ref{integrated_outflow}. We find that the main results of this work do not change significantly by using different CO-to-H$_2$ conversion factors.

\begin{table*}[ht]
%\small 
\centering
\begin{tabular}{lccc}
\hline
Integrated outflow properties		&Eastern &Western & Total\\
			&side    &side	  &\\
 (1)&(2)&(3)& (4)\\	
			
\hline
Molecular gas outflow mass (M$_{\odot}$)			& (6.4$\pm1.9$)$\times$10$^6$ & (3.2$\pm1.0$)$\times$10$^6$&(9.6$\pm2.9$)$\times$10$^6$\\
 \noalign{\smallskip}
Molecular gas outflow mass rate (M$_{\odot}$ yr$^{-1}$)	& 7.0$\pm2.1$				&	2.4$\pm0.7$& 9.4$\pm2.8$\\
 \noalign{\smallskip}
Molecular gas outflow kinematic power (erg s$^{-1}$)& 
(1.6$\pm0.5$)$\times$10$^{41}$&(2.7$\pm0.8$)$\times$10$^{40}$ &(1.9$\pm0.6$)$\times$10$^{41}$ \\
 \noalign{\smallskip}
Molecular gas outflow momentum (dyne)& 
(1.1$\pm0.3$)$\times$10$^{34}$&(2.3$\pm0.7$)$\times$10$^{33}$ &(1.3$\pm0.4$)$\times$10$^{34}$ \\
 \noalign{\smallskip}
\hline
\end{tabular}						 
\caption{Summary of the molecular outflow properties calculated using the ULIRGs CO-to-H$_2$ conversion factor. }
\label{tabd1}
\end{table*}

\end{appendix}

\end{document}